\documentclass{article}

\usepackage{microtype}
\usepackage{graphicx}
\usepackage{subcaption}
\usepackage{booktabs} 

\usepackage{hyperref}

\usepackage[accepted]{icml2026}

\usepackage{amsmath}
\usepackage{amssymb}
\usepackage{mathtools}
\usepackage{amsthm}

\usepackage{soul}

\newcommand{\compilehidecomments}{false} 
\ifthenelse{ \equal{\compilehidecomments}{true} }{%
	\newcommand{\tk}[1]{}
    \newcommand{\Ulises}[1]{}
    \newcommand{\yw}[1]{}
	\newcommand{\es}[1]{}
	\newcommand{\us}[1]{}
}{
	\newcommand{\tk}[1]{{\color{MidnightBlue}  [Tim: #1]}}
    \newcommand{\Ulises}[1]{{\color{Purple}  [Ulises: #1]}}
	\newcommand{\yw}[1]{{\color{Green}  [Yiliu: #1]}}
	\newcommand{\es}[1]{{\color{BrickRed} [Eric: #1]}}
	\newcommand{\us}[1]{{\color{Orange} [Uygar: #1]}}
}
\usepackage[dvipsnames]{xcolor}
\usepackage{wrapfig}

\usepackage[capitalize,noabbrev]{cleveref}

\theoremstyle{plain}

\theoremstyle{definition}

\theoremstyle{remark}

\makeatletter
\newcommand{\longdash}[1][2em]{%
  \makebox[#1]{$\m@th\smash-\mkern-7mu\cleaders\hbox{$\mkern-2mu\smash-\mkern-2mu$}\hfill\mkern-7mu\smash-$}}
\makeatother
\newcommand{\omitskip}{\kern-\arraycolsep}

\DeclareMathOperator*{\argmax}{arg\,max}
\DeclareMathOperator*{\argmin}{arg\,min}

\usepackage[textsize=tiny]{todonotes}

\icmltitlerunning{Identifying Connectivity Distributions from Neural Dynamics Using Flows}

\begin{document}

\twocolumn[
  \icmltitle{Identifying Connectivity Distributions from Neural Dynamics Using Flows}

  \icmlsetsymbol{equal}{*}

  \begin{icmlauthorlist}
    \icmlauthor{Timothy Doyeon Kim}{yyy,comp}
    \icmlauthor{Ulises Pereira-Obilinovic}{yyy}
    \icmlauthor{Yiliu Wang}{yyy,comp}
    \icmlauthor{Eric Shea-Brown}{comp}
    \icmlauthor{Uygar Sümbül}{yyy}
  \end{icmlauthorlist}

  \icmlaffiliation{yyy}{Allen Institute, Seattle, WA, USA}
  \icmlaffiliation{comp}{University of Washington, Seattle, WA, USA}
  \icmlcorrespondingauthor{Timothy Doyeon Kim}{timkimd@uw.edu}
  \icmlcorrespondingauthor{Ulises Pereira-Obilinovic}{ulises.pereira.o@alleninstitute.org}

  \icmlkeywords{computational neuroscience, low-rank recurrent neural networks, identifiability, interpretability}

  \vskip 0.3in
]



\printAffiliationsAndNotice{}  

\begin{abstract}
  Connectivity structure shapes neural computation, but inferring this structure from population recordings is degenerate: multiple connectivity structures can generate identical dynamics. Recent work uses low-rank recurrent neural networks (lrRNNs) to infer low-dimensional latent dynamics and connectivity from observed activity, enabling a mechanistic interpretation of the dynamics. However, standard approaches for training lrRNNs can recover spurious structures irrelevant to the underlying dynamics. We first characterize the identifiability of connectivity structures in lrRNNs and determine conditions under which a unique solution exists. To find such solutions, we develop an inference framework based on maximum entropy and continuous normalizing flows (CNFs), trained via flow matching. Instead of estimating a single connectivity matrix, our method learns a distribution over connection weights that is maximally unbiased over unidentifiable components while matching the observed dynamics. This approach captures complex yet necessary distributions such as heavy-tailed connectivity found in empirical data. We validate our method on synthetic datasets with connectivity structures that generate multistable attractors, limit cycles, and ring attractors, and demonstrate its applicability in recordings from rat frontal cortex during decision-making. Our framework shifts circuit inference from recovering connectivity to identifying which connectivity structures are computationally required, and which are artifacts of underconstrained inference.
\end{abstract}

\section{Introduction}

A central goal of systems neuroscience is understanding how circuit structure generates computation. A recent line of work addresses this challenge by fitting a dynamical system model directly to population recordings, where state variables correspond to individual neurons or cell-type aggregates that interact through learned synaptic couplings, enabling data-driven discovery of circuit mechanisms. These approaches have uncovered attractor dynamics~\cite{finkelstein2021attractor}, characterized multi-regional interactions~\cite{rajan2016recurrent}, explained trial-by-trial variability~\cite{sourmpis2023trial}, characterized non-normal dynamics during decision-making tasks~\cite{pereira2025neural}, and predicted responses to optogenetic perturbations~\cite{sourmpis2024biologically}. 

Despite substantial progress, the problem of inferring connectivity from neural dynamics is underconstrained and degenerate in many cases, both in biological (e.g., \citet{marder2006variability}) and artificial (e.g., \citet{huang2025measuring}) circuits. Furthermore, existing approaches typically return a single point estimate of recurrent weights. Yet the object of interest is often the structure of {\it families} of circuits consistent with data, particularly given that neural recordings provide only a highly subsampled view of underlying networks \cite{qian2024}. Moreover, relying on single estimates is at odds with the observed biological diversity: synaptic connectivity is highly heterogeneous~\citep{oh2014mesoscale,microns2025functional,dorkenwald2024neuronal,lin2024network}. If the inverse problem admits many solutions, inferring a single connectivity matrix discards information about which circuit features are necessary versus arbitrary. These considerations argue for methods that infer distributions over connectivity rather than point estimates, capturing the family of circuits consistent with the data while imposing minimal assumptions.

Classical statistical-physics models recognize this: they frame circuits as distributions over synaptic weights and use mean-field theory \cite{mezard1987spin, helias2020statistical} to link the statistics of connectivity---not a single parameter set---to neural dynamics \cite{amit1985storing,sompolinsky88,brunel2000dynamics, van1998chaotic}. However, the standard assumptions---binary patterns~\citep{amit1985storing}, independent and identically distributed Gaussian weights~\citep{sompolinsky88}, uniform sparsity~\citep{derrida1987exactly}---do not fully reflect the structured, heterogeneous connectivity found in biological circuits. While recent theoretical work extends these calculations to richer, heterogeneous connectivity \cite{Aljadeff2015, marti18, Dahmen2020, di2025extended}, these approaches remain analytically constrained and cannot flexibly match the structured, cell-type–specific, pathway-dependent organization revealed by modern datasets.

This motivates our central question: How can we infer connectivity distributions from data---rich enough to fit complex weight statistics, constrained by neural recordings, and interpretable for generating testable hypotheses?

Here we introduce {\bf Connector} (\underline{Connect}ivity distributions of l\underline{o}w-rank \underline{R}NNs from neural population dynamics), a framework that learns distributions over synaptic connectivity consistent with observed population dynamics. We leverage recent progress in low-rank recurrent neural networks (lrRNNs), which constrain connectivity to be low rank: rather than estimating every pairwise synapse, the model learns a small set of structured factors that generate the connectivity and summarize population interactions. This factorization reduces free parameters, concentrates activity into a low-dimensional subspace, and---most importantly for interpretability---yields latent variables with explicit equations linking connectivity statistics to familiar dynamical motifs~\citep{MASTROGIUSEPPE18,valente22,pals24}, bridging neuronal tuning and synaptic interactions with population-level computations. 

We build on this approach by placing a continuous normalizing flow (CNF; \citet{chen18}) over low-rank factors to flexibly capture complex, potentially multimodal weight statistics. By recasting circuit inference as density estimation over connectivity, this formulation accounts for partial observability by treating recorded neurons as samples from a larger circuit and by matching the model’s mean-field dynamics to the data. Among the possible distributions consistent with observed data, Connector identifies the maximum entropy solution---the one making the fewest assumptions beyond what the data require (in the spirit of, e.g., \citet{schneidman2006weak}). This shifts circuit inference from learning a single parameter set in fixed-size lrRNNs \cite{valente22,pals24} to learning distributions over connectivity that distinguish computationally necessary structures from arbitrary features.\\

{\bf Main Contributions}\begin{itemize}
\item We characterize three sources of degeneracy in identifying connectivity distribution from neural dynamics, and propose choosing the maximum entropy distribution among the set of possible solutions (Section~\ref{main_identify}).
\item We develop an inference framework, Connector, to find such distributions using CNFs trained via flow matching (Section~\ref{connector_main}).
\item We propose a dissimilarity measure for comparing connectivity distributions based on effective connectivities (Section~\ref{main_distance}).
\item We empirically validate our approach on various synthetic datasets and demonstrate its applicability in recordings from rat frontal cortex during decision-making (Section~\ref{experiments}).
\end{itemize}

\section{Background}

\subsection{Low-Rank Recurrent Neural Networks}

Low-rank recurrent neural networks (lrRNNs; \citet{MASTROGIUSEPPE18}) are RNNs that have the form \begin{equation} \label{eq:main_lr-RNN}
\tau\frac{\boldsymbol{h}_{t} - \boldsymbol{h}_{t-1}}{\Delta t} = -\boldsymbol{h}_{t-1} + \frac{1}{K} \boldsymbol{M}\boldsymbol{N}^\top \phi (\boldsymbol{h}_{t-1}) + \boldsymbol{B}\boldsymbol{u}_t + \boldsymbol{d}.
\end{equation} Here, the strength of interaction between the units $\boldsymbol{h} \in \mathbb{R}^{K}$ are represented by the connectivity matrix $\boldsymbol{J}$ that is factorized into $\boldsymbol{J} = \frac{1}{K}\boldsymbol{M}\boldsymbol{N}^\top$, where $\boldsymbol{M}, \boldsymbol{N} \in \mathbb{R}^{K \times R}$, with $R < K$. Each unit $\boldsymbol{h}_i$ may be interpreted as the membrane potential of neuron $i$, with firing rates $\boldsymbol{r}_i = \phi(\boldsymbol{h}_i)$ \cite{hopfield1984neurons}. The activation function $\phi$ is applied element-wise (typically $\tanh$). In this work, we assume that $\phi$ is strictly monotonically increasing. The external input $\boldsymbol{u} \in \mathbb{R}^{K_{in}}$ projects to the network via $\boldsymbol{B} \in \mathbb{R}^{K \times K_{in}}$, with input biases represented by $\boldsymbol{d} \in \mathbb{R}^K$. It can be shown that the dynamics in Equation~(\ref{eq:main_lr-RNN}) is equivalent to the dynamics of the latent variable $\boldsymbol{z}_t \in \mathbb{R}^R$ \begin{equation} \label{eq:main_zlrrnn}
\tau\frac{\boldsymbol{z}_{t} - \boldsymbol{z}_{t-1}}{\Delta t} = -\boldsymbol{z}_{t-1} + \frac{1}{K} \boldsymbol{N}^\top \phi (\boldsymbol{M} \boldsymbol{z}_{t-1} + \boldsymbol{B} \boldsymbol{v}_{t-1} + \boldsymbol{d}),
\end{equation} where \begin{equation} \label{eq:rate_lvm}
\boldsymbol{r}_t = \phi(\boldsymbol{h}_t) = \phi(\boldsymbol{M} \boldsymbol{z}_t + \boldsymbol{B} \boldsymbol{v}_t + \boldsymbol{d}),
\end{equation} and $\tau \frac{\boldsymbol{v}_{t} - \boldsymbol{v}_{t-1}}{\Delta t} = - \boldsymbol{v}_{t-1} + \boldsymbol{u}_t$. Here, $\boldsymbol{v}_t \in \mathbb{R}^{K_{in}}$ can be thought of as a low-pass filtered input $\boldsymbol{u}_t$ \citep{valente22}.

Let $\boldsymbol{m}_i$, $\boldsymbol{n}_i$, $\boldsymbol{b}_i$, and $\boldsymbol{d}_i$ be the $i$-th row of $\boldsymbol{M}$, $\boldsymbol{N}$, $\boldsymbol{B}$, and $\boldsymbol{d}$, respectively. If we define a probability distribution $p(\boldsymbol{m}, \boldsymbol{n}, \boldsymbol{b}, \boldsymbol{d})$, and sample independent and identically distributed (iid) from this distribution: $\boldsymbol{m}_i, \boldsymbol{n}_i, \boldsymbol{b}_i, \boldsymbol{d}_i \overset{\text{iid}}{\sim} p(\boldsymbol{m}, \boldsymbol{n}, \boldsymbol{b}, \boldsymbol{d})$, then, as $K \rightarrow \infty$, we can show that \begin{equation} \label{eq:main_mf-lrrnn}
\tau\frac{\boldsymbol{z}_{t} - \boldsymbol{z}_{t-1}}{\Delta t} = -\boldsymbol{z}_{t-1} + \mathbb{E}[\boldsymbol{n} \phi (\boldsymbol{m}^\top \boldsymbol{z}_{t-1} + \boldsymbol{b}^\top \boldsymbol{v}_{t-1} + \boldsymbol{d})],
\end{equation} by the law of large numbers, giving us the mean-field dynamics of lrRNNs \cite{beiran21}. $\mathbb{E}[\cdot]$ denotes the expectation over that probability distribution. We refer to $p(\boldsymbol{m}, \boldsymbol{n}, \boldsymbol{b}, \boldsymbol{d})$ as the {\it connectivity distribution}. This distribution is $(2R + K_{in} + 1)$-dimensional. Equation~(\ref{eq:main_mf-lrrnn}) is valid regardless of whether or not the latent state is near a fixed point. See Appendix~\ref{background_lrrnn} for detailed derivations.

\subsection{Inferring Low-D Dynamics with lrRNNs}

Existing approaches typically assume a one-to-one correspondence between lrRNN units and recorded neurons~\cite{valente22,pals24,pereira2025neural}, fitting a network of size $K_{obs}$ directly to activity of $K_{obs}$ neurons. Model parameters are typically trained to match predicted and observed firing rates using backpropagation through time (BPTT) \cite{valente22}. Recent work by \citet{pals24} developed an approach that generalizes to lrRNNs with noise in the latent $\boldsymbol{z}$ and also the observed data, successfully training lrRNNs on single-trial spiking neural data via variational sequential Monte Carlo. Most recently, \citet{li2025disentangledlowrankrnnframework} developed a method to disentangle latent dynamics using the relationship between sequential variational autoencoders (VAEs; \citet{kingma14}) and lrRNNs. These approaches infer a single connectivity instance consistent with the data.

\section{Identifiability} \label{main_identify}

The connectivity distribution $p(\boldsymbol{m}, \boldsymbol{n}, \boldsymbol{b}, \boldsymbol{d})$ can be decomposed into $p(\boldsymbol{m}, \boldsymbol{b}, \boldsymbol{d})p(\boldsymbol{n} | \boldsymbol{m}, \boldsymbol{b}, \boldsymbol{d})$. If we want to infer the connectivity distribution from neural population activity, there are three possible sources that can influence its identifiability: {\bf (1)} the identifiability of the latent variable $\boldsymbol{z}$, {\bf (2)} $p(\boldsymbol{m}, \boldsymbol{b}, \boldsymbol{d})$, and {\bf (3)} $p(\boldsymbol{n} | \boldsymbol{m}, \boldsymbol{b}, \boldsymbol{d})$.  Full derivations for statements in this Section are available in Appendix~\ref{identify}.

{\bf (1) Identifiability of $\boldsymbol{z}$}: The first source comes from the identifiability of the latent variable $\boldsymbol{z}$. In lrRNNs, the latent variable $\boldsymbol{z}$ is identifiable only up to linear transformations $\boldsymbol{z}' = \boldsymbol{A}\boldsymbol{z}$ as long as $\boldsymbol{1}^\top_{T} \notin \textrm{rowspan}(\boldsymbol{u}_{1:T})$ and $\boldsymbol{u}_{1:T} \neq \boldsymbol{0}$ (Appendix~\ref{latent_identify}). One corollary of this result is that when we have constant input $\boldsymbol{u}$ (a case often assumed in literature, e.g., \citet{beiran21}), $\boldsymbol{M}$, $\boldsymbol{B}$, and $\boldsymbol{d}$ trade off with each other and are therefore not identifiable.

The linear identifiability of $\boldsymbol{z}$ implies that the connectivity distribution $p(\boldsymbol{m}, \boldsymbol{n}, \boldsymbol{b}, \boldsymbol{d})$ is identifiable only up to a certain transformation. In other words, for a given linear transformation in $\boldsymbol{z} \mapsto \boldsymbol{z}'$, there is a corresponding transformation $p \mapsto p'$ in connectivity distribution. We can show that this transformation is \begin{equation} \label{eq:p_transformation} \begin{split}
p'(\boldsymbol{m}',\boldsymbol{n}',\boldsymbol{b}',\boldsymbol{d}') &= p(\boldsymbol{A}^\top\boldsymbol{m}',\boldsymbol{A}^{-1}\boldsymbol{n}',\boldsymbol{b}',\boldsymbol{d}'),
\end{split}
\end{equation} where $\boldsymbol{A}$ is invertible (Appendix~\ref{id_connectivity}). In all of our experiments in Section~\ref{experiments}, unless mentioned otherwise, the inferred latents $\boldsymbol{z}$ were matched to the ground-truth $\boldsymbol{z}$ by performing linear regression to find $\boldsymbol{A}$ whenever the ground truth is available. The inferred $p(\boldsymbol{m}, \boldsymbol{n}, \boldsymbol{b}, \boldsymbol{d})$ then went through the transformation in Equation~(\ref{eq:p_transformation}).

We discuss identifiability of $\boldsymbol{z}$ in latent variable models (LVMs) more generally (e.g., linear, switching-linear, general multilayer perceptron-based dynamical models, and LVMs that do not assume dynamics) in Appendix~\ref{identify}.

{\bf (2) $p(\boldsymbol{m}, \boldsymbol{b}, \boldsymbol{d})$}: Suppose that $\boldsymbol{z}_{1:T}$ is fixed. Then $p(\boldsymbol{m}, \boldsymbol{b}, \boldsymbol{d})$ is identifiable in the limit of infinite data (i.e., $K_{obs} \rightarrow \infty$ and $T \geq R + K_{in}$) as long as $\textrm{rank}(\tilde{\boldsymbol{z}}_{1:T}) = R + K_{in}$ and $\boldsymbol{1}^\top_T \notin \textrm{rowspan}(\tilde{\boldsymbol{z}}_{1:T})$, where $\tilde{\boldsymbol{z}}_{1:T} = \begin{bmatrix}
    \boldsymbol{z}_{1:T} \\ \boldsymbol{v}_{1:T}
\end{bmatrix} \in \mathbb{R}^{(R + K_{in}) \times T}$. Loosely, in other words, notice that due to Equation~(\ref{eq:rate_lvm}), the distribution of the observed neural firing rate at time $t$, $p(\boldsymbol{r}_t)$ must be the projection of $p(\boldsymbol{m}, \boldsymbol{b}, \boldsymbol{d})$ onto the vector $\tilde{\boldsymbol{z}}_t$, which is then transformed by $\phi$, a strictly monotonically increasing function. Thus $p(\boldsymbol{m}, \boldsymbol{b}, \boldsymbol{d})$ is recoverable as long as $\tilde{\boldsymbol{z}}$ occupies the full latent space and any linear combination of the elements of $\tilde{\boldsymbol{z}}$ is not constant in time (in the limit $K_{obs} \rightarrow \infty$) (Appendix~\ref{identify}).

{\bf (3) $p(\boldsymbol{n} | \boldsymbol{m}, \boldsymbol{b}, \boldsymbol{d})$}: Note that Equation~(\ref{eq:main_mf-lrrnn}) is equivalent to \begin{equation} \label{eq:main_mf-lr-RNN-final}\begin{split}
&\tau\frac{\boldsymbol{z}_{t} - \boldsymbol{z}_{t-1}}{\Delta t} = -\boldsymbol{z}_{t-1} + \\
&\mathbb{E}[\mathbb{E}[\boldsymbol{n}|\boldsymbol{m}, \boldsymbol{b}, \boldsymbol{d}] \phi (\boldsymbol{m}^\top \boldsymbol{z}_{t-1} + \boldsymbol{b}^\top \boldsymbol{v}_{t-1} + \boldsymbol{d})]
\end{split}
\end{equation} by the law of iterated expectations. Therefore, the mean-field dynamics of lrRNNs in Equation~(\ref{eq:main_mf-lr-RNN-final}) when $K \rightarrow \infty$ depend only on the first moment of $p(\boldsymbol{n}|\boldsymbol{m}, \boldsymbol{b}, \boldsymbol{d})$, i.e., on $\mathbb{E}[\boldsymbol{n}|\boldsymbol{m}, \boldsymbol{b}, \boldsymbol{d}]$. This implies that any probability distribution $p(\boldsymbol{n}|\boldsymbol{m}, \boldsymbol{b}, \boldsymbol{d})$ with the first moment equal to $\boldsymbol{\mu}(\boldsymbol{m}, \boldsymbol{b}, \boldsymbol{d}) = \mathbb{E}[\boldsymbol{n}|\boldsymbol{m}, \boldsymbol{b}, \boldsymbol{d}]$ will have the same dynamics. However, we can define a ``minimally structured'' distribution $p^*(\boldsymbol{n}|\boldsymbol{m}, \boldsymbol{b}, \boldsymbol{d})$ that generates the mean-field dynamics. We can choose this $p^*$ from a family of solutions $p$ by formulating the following optimization problem: \begin{equation} \label{eq:maxent}
\begin{split}
p^*(\boldsymbol{n}|\boldsymbol{m}, \boldsymbol{b}, \boldsymbol{d}) &= \argmax_p  H(p(\boldsymbol{n}|\boldsymbol{m}, \boldsymbol{b}, \boldsymbol{d})) \\
\textrm{subject to } \quad &\mathbb{E}[\boldsymbol{n}|\boldsymbol{m}, \boldsymbol{b}, \boldsymbol{d}] = \boldsymbol{\mu}(\boldsymbol{m}, \boldsymbol{b}, \boldsymbol{d}),\\
&\mathbb{E}[\mathcal{T}(\boldsymbol{n} | \boldsymbol{m}, \boldsymbol{b}, \boldsymbol{d})] = \boldsymbol{0},\\
&\textrm{supp}(p(\boldsymbol{n}|\boldsymbol{m}, \boldsymbol{b}, \boldsymbol{d})) = \mathfrak{N} \subseteq \mathbb{R}^{R},
\end{split}
\end{equation} where we have defined the differential entropy $H(p({\bf x})) = - \int p({\bf x}) \log p({\bf x}) d{\bf x}$, and let the support of $p(\boldsymbol{n}|\boldsymbol{m}, \boldsymbol{b}, \boldsymbol{d})$ be $\mathfrak{N}$. This problem has a unique $p^*(\boldsymbol{n}|\boldsymbol{m}, \boldsymbol{b}, \boldsymbol{d})$ under standard regularity conditions, when the constraints make the entropy $H$ bounded above and $p^*$ exists (theorem attributed to Ludwig Boltzmann, and more recently used in e.g., \citet{loaiza-ganem17}). Here,  the function $\mathcal{T}(\boldsymbol{n}|\boldsymbol{m}, \boldsymbol{b}, \boldsymbol{d}): \mathfrak{N} \rightarrow \mathbb{R}^s$ represents any additional constraints that we impose. $p^*(\boldsymbol{n}|\boldsymbol{m}, \boldsymbol{b}, \boldsymbol{d})$ is called the maximum entropy distribution, and can be interpreted as the distribution assuming the least about the data given the constraints \cite{jaynes1957}.

In this work, we focus on the simple case $\mathfrak{N} = \mathbb{R}^R$ and choose $\mathcal{T}(\boldsymbol{n}|\boldsymbol{m}, \boldsymbol{b}, \boldsymbol{d})$ such that the first and second moments are constrained to $\boldsymbol{\mu}(\boldsymbol{m}, \boldsymbol{b}, \boldsymbol{d})$ and a positive semidefinite covariance $\boldsymbol{S}$. Under these constraints, the unique maximum entropy distribution is Gaussian, $p^*(\boldsymbol{n}|\boldsymbol{m}, \boldsymbol{b}, \boldsymbol{d})=\mathcal{N}(\boldsymbol{\mu}, \boldsymbol{S})$. While $\boldsymbol{\mu}$ can be inferred directly from neural activity $\boldsymbol{r}^{data}_{1:T}$ and inputs $\boldsymbol{v}_{1:T}$ (Appendix~\ref{inference}), $\boldsymbol{S}$ remains underdetermined by these data alone and is treated as a hyperparameter. Additional constraints---such as causal perturbations or biological structure (e.g., Dale's law)---may help determine $\boldsymbol{S}$ and motivate more restrictive choices of $\mathfrak{N}$ and $\mathcal{T}$, which we leave to future work. More generally, when the maximum entropy distribution cannot be obtained analytically, approaches such as \citet{loaiza-ganem17} may be used to approximate $p^*(\boldsymbol{n}|\boldsymbol{m}, \boldsymbol{b}, \boldsymbol{d})$. In the absence of additional constraints favoring more structured distributions, Gaussian connectivities provide analytical tractability in lrRNNs (Appendix~\ref{beiran_proof}--\ref{mog_proof}) and therefore serve as a natural default. Nevertheless, analytical tractability is not unique to Gaussian connectivities: distributions such as Student-$t$ may also admit tractable analyses, as we show in Appendix~\ref{student-t}.

\section{Connector Framework} \label{connector_main}

Here we describe {\bf Connector}, an approach to learn the connectivity distribution $p(\boldsymbol{m}, \boldsymbol{n}, \boldsymbol{b}, \boldsymbol{d})$ from an lrRNN that has been trained on neural data. Connector presupposes a separate lrRNN training step, performed in advance using methods such as LINT \cite{valente22} or variational sequential Monte Carlo \cite{pals24}. The quality of this fit upper-bounds everything that follows, since a poorly trained lrRNN will yield mean-field dynamics and connectivity that poorly reflect the data. Given the trained lrRNN, we learn each component of  $p(\boldsymbol{m}, \boldsymbol{n}, \boldsymbol{b}, \boldsymbol{d})=p(\boldsymbol{m}, \boldsymbol{b}, \boldsymbol{d})p(\boldsymbol{n}|\boldsymbol{m}, \boldsymbol{b}, \boldsymbol{d})$ in the steps below.

\subsection{Inferring $p(\boldsymbol{m}, \boldsymbol{b}, \boldsymbol{d})$} \label{main_fm}

Suppose we are given neural activity data $\boldsymbol{r}^{data}_{1:T} \in \mathbb{R}^{K_{obs} \times T}$ and, optionally, external input data $\boldsymbol{v}_{1:T} \in \mathbb{R}^{K_{in} \times T}$, together with an lrRNN already trained on them. We take the rows of the learned $\boldsymbol{M}$, $\boldsymbol{B}$, and $\boldsymbol{d}$ to be the data samples for $p(\boldsymbol{m}, \boldsymbol{b}, \boldsymbol{d})$. With the samples $\{\boldsymbol{m}_i, \boldsymbol{b}_i, \boldsymbol{d}_i\}^{K_{obs}}_{i=1}$, there are multiple ways to do density estimation. Here, we use continuous normalizing flows (CNF; \citet{chen18}), a highly flexible model that estimates probability density functions. 

We train our CNF by defining a probability density path and the corresponding vector field that transforms a standard normal distribution to $p(\boldsymbol{m}, \boldsymbol{b}, \boldsymbol{d})$, and by parametrizing the vector field with a neural network (Appendix~\ref{details}). We use the flow matching objective to train this network and infer $p(\boldsymbol{m}, \boldsymbol{b}, \boldsymbol{d})$ \cite{lipman2023flow}. This gives us our inferred $p(\boldsymbol{m}, \boldsymbol{b}, \boldsymbol{d})$, which we can use to sample $\boldsymbol{m}, \boldsymbol{b}, \boldsymbol{d}$ as many times as we like. If we sample $\boldsymbol{m}_i, \boldsymbol{b}_i, \boldsymbol{d}_i \sim p(\boldsymbol{m}, \boldsymbol{b}, \boldsymbol{d})$ for $K$ times (where this $K$ need not be $K_{obs}$), these samples can be used to construct $\boldsymbol{M}$, and $\boldsymbol{B}$, and $\boldsymbol{d}$. How do we infer the remaining $\boldsymbol{N}$ for this new network that has the same latent dynamics $\boldsymbol{z}_{1:T} \in \mathbb{R}^{R \times T}$ as the original network? We show how in Section~\ref{main_inference}.

\begin{figure*}[ht]
\vskip 0.2in
\hspace{1.1in}\includegraphics[width=5.5in]{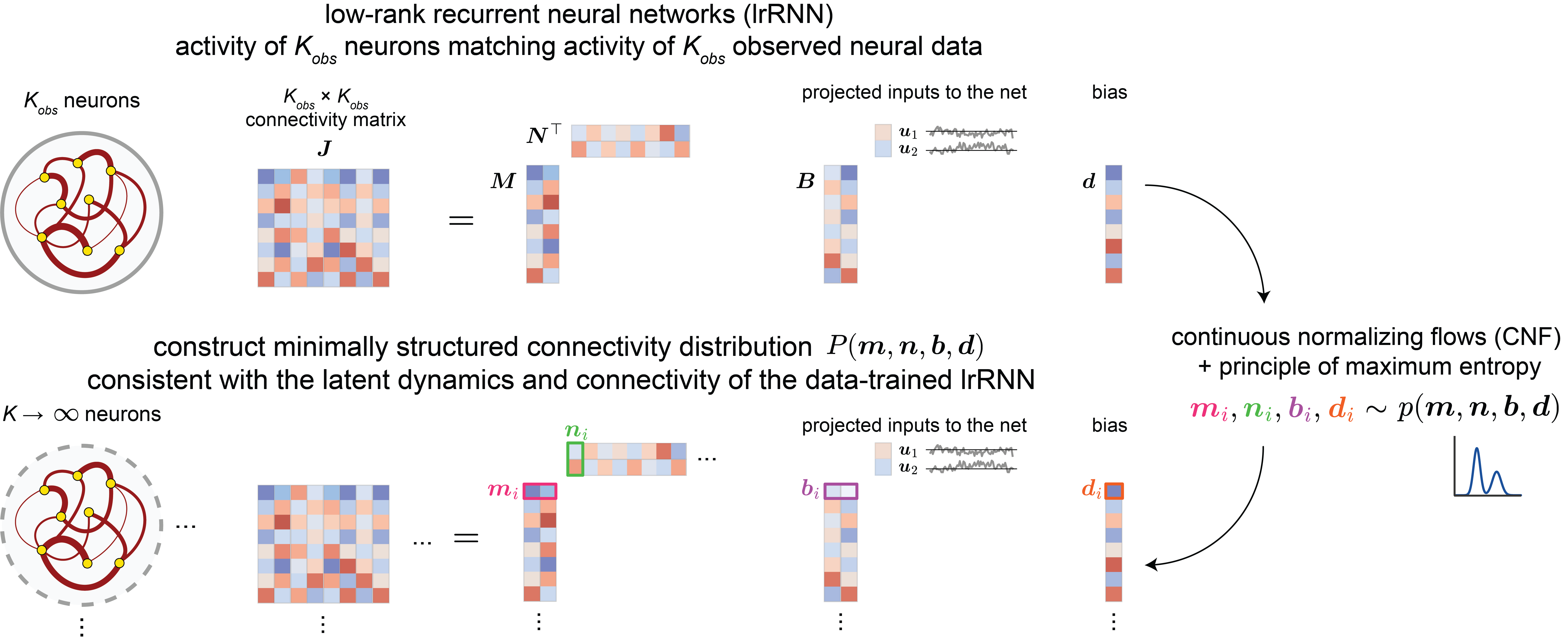}
\caption{Connector is a framework that infers minimally structured distribution over the connection weights of the lrRNN trained on neural data. It constructs, based on maximum entropy and continuous normalizing flows (CNFs), a generative model of the connectivity distribution, and can sample new lrRNNs that have mean-field dynamics that match the latent dynamics learned by the original lrRNN.}
\label{figure-1}
\vskip -0.2in
\end{figure*}

\subsection{Inferring $\boldsymbol{\mu}(\boldsymbol{m}, \boldsymbol{b}, \boldsymbol{d})$} \label{main_inference}

Given the inferred latent trajectories $\boldsymbol{z}_{1:T}$ and loadings $\boldsymbol{M}$, $\boldsymbol{B}$, $\boldsymbol{d}$, Equation~(\ref{eq:main_zlrrnn}) implies that \begin{equation} \label{w_update}
\boldsymbol{w}_t \equiv \boldsymbol{z}_{t+1} + (\alpha - 1) \boldsymbol{z}_t \approx \frac{\alpha}{K}\boldsymbol{N}^\top \boldsymbol{r}_t,
\end{equation} where $\boldsymbol{r}_t$ is given by Equation~(\ref{eq:rate_lvm}). Thus, estimating $\boldsymbol{N}$ reduces to a linear regression problem: we seek a matrix $\boldsymbol{N}$ whose weighted population activity best predicts the latent update $\boldsymbol{w}_t$. Stacking the time points gives $\boldsymbol{w}_{1:(T-1)} \approx \frac{\alpha}{K} \boldsymbol{r}_{1:(T-1)}\boldsymbol{N}$. The solution to this regression problem is typically not unique due to correlations in neural activity, i.e., $\textrm{rank}(\boldsymbol{r}_{1:(T-1)}) = R + K_{in} < K$ for large $K$ (given that $\tilde{\boldsymbol{z}}$ satisfies condition in {\bf (2)} of Section~\ref{main_identify}). We therefore introduce an $\ell_2$ regularization term and estimate $\boldsymbol{N}$ via 
\begin{equation} \label{eq:ridge}
\hat{\boldsymbol{N}} = \argmin_{\boldsymbol{N}} \left\Vert \boldsymbol{w}_{1:(T-1)} - \frac{\alpha}{K} \boldsymbol{r}_{1:(T-1)}\boldsymbol{N} \right\Vert^2_F + c \left\Vert \boldsymbol{N} \right\Vert^2_F.
\end{equation} This yields the closed-form solution \begin{equation} \label{eq:main_nhat}
\hat{\boldsymbol{N}} = \frac{K}{\alpha}\left(\boldsymbol{R} + \frac{cK^2}{\alpha^2} \boldsymbol{I}_K\right)^{-1}\boldsymbol{W},
\end{equation} where $\boldsymbol{R} = \sum_{t=1}^{T-1} \boldsymbol{r}_t \boldsymbol{r}^\top_t = (\boldsymbol{r}_{1:(T-1)}) (\boldsymbol{r}_{1:(T-1)})^\top$ and $\boldsymbol{W} = \sum_{t=1}^{T-1} \boldsymbol{r}_t \boldsymbol{w}^\top_t = (\boldsymbol{r}_{1:(T-1)})(\boldsymbol{w}_{1:(T-1)})^\top$. From a Bayesian perspective, this corresponds to maximum a posteriori estimation under a zero-mean isotropic Gaussian prior on the rows of $\boldsymbol{N}$. For large networks (i.e., large $K$), each row $\hat{\boldsymbol{n}}_i$ depends primarily on the corresponding rows of $\boldsymbol{M}$, $\boldsymbol{B}$, $\boldsymbol{d}$ and on the latent trajectories. We therefore interpret $\hat{\boldsymbol{n}}_i$ as an estimate of the conditional mean $\boldsymbol{\mu}(\boldsymbol{m}_i, \boldsymbol{b}_i, \boldsymbol{d}_i) = \mathbb{E}[\boldsymbol{n}|\boldsymbol{m}_i, \boldsymbol{b}_i, \boldsymbol{d}_i]$ in Equation~(\ref{eq:maxent}), and numerically validate this in Appendix~\ref{num_val}. Unidentifiable components of $\boldsymbol{N}$ lying in the null space of $\boldsymbol{r}_{1:(T-1)}$ are shrunk to zero by the regularization, reflecting the fact that these directions are unconstrained by the data. Because $\boldsymbol{r}_{1:(T-1)}$ is low-rank, the null space of $\boldsymbol{r}_{1:(T-1)}$ is huge. This null space is the unidentifiable space of $\boldsymbol{N}$.

We give a full Bayesian treatment of this formulation in Appendix~\ref{inference}, and consider the more general case where the prior is not isotropic Gaussian and can depend on $\boldsymbol{M}$, $\boldsymbol{B}$, and $\boldsymbol{d}$. In the general case, $\hat{\boldsymbol{N}}$ is the solution to a Sylvester equation. Our results here are consistent with, and extend the linear regression method in Section 6 of \citet{beiran21}, the maximum a posteriori method in \citet{qian2024}, and most recently in \citet{arora2025efficient}.

\subsection{Inferring $p(\boldsymbol{n}|\boldsymbol{m}, \boldsymbol{b},\boldsymbol{d})$} \label{last_step}

Taking the rows of $\hat{\boldsymbol{N}}$ from Equation~(\ref{eq:main_nhat}), and also the rows of $\boldsymbol{M}$, $\boldsymbol{B}$, and $\boldsymbol{d}$ that were used to infer $\hat{\boldsymbol{N}}$, we have $\{\boldsymbol{m}_i, \hat{\boldsymbol{n}}_i, \boldsymbol{b}_i, \boldsymbol{d}_i\}^{K}_{i=1}$, with $K$ not necessarily equal to $K_{obs}$. Since $\hat{\boldsymbol{n}}_i \approx \boldsymbol{\mu}(\boldsymbol{m}_i,\boldsymbol{b}_i, \boldsymbol{d}_i)$ (Section~\ref{main_inference}), and since the maximum entropy $p(\boldsymbol{n}|\boldsymbol{m}, \boldsymbol{b},\boldsymbol{d})$ with fixed covariance $\boldsymbol{S}$ is $\mathcal{N}(\boldsymbol{\mu}(\boldsymbol{m},\boldsymbol{b}, \boldsymbol{d}), \boldsymbol{S})$, $\boldsymbol{n}_i = \hat{\boldsymbol{n}}_i + \boldsymbol{\xi}_i$, with $\boldsymbol{\xi}_i \sim \mathcal{N}(\boldsymbol{0}, \boldsymbol{S})$, are approximately the samples from the maximum entropy distribution $\mathcal{N}(\boldsymbol{\mu}(\boldsymbol{m},\boldsymbol{b}, \boldsymbol{d}), \boldsymbol{S})=p(\boldsymbol{n}|\boldsymbol{m}, \boldsymbol{b},\boldsymbol{d})$.

\subsection{Summary}
Combining steps in Sections~\ref{main_fm}--\ref{last_step} provides an approach to sample $K$ times from our inferred connectivity distribution $p(\boldsymbol{m}, \boldsymbol{n}, \boldsymbol{b}, \boldsymbol{d})$, with the samples being $\{\boldsymbol{m}_i, \boldsymbol{n}_i, \boldsymbol{b}_i, \boldsymbol{d}_i\}^{K}_{i=1}$. We summarize the Connector framework as a schematic in Figure~\ref{figure-1} and as Algorithm~\ref{alg:connector} in the Appendix.

\section{Dissimilarity Between Connectivities} \label{main_distance}

\citet{valente22} compared connectivities via an effective connectivity matrix $\boldsymbol{J}^{\text{eff}} = \frac{1}{K}\boldsymbol{M}\boldsymbol{N}_{\parallel}$,
where $\boldsymbol{N}_{\parallel}$ denotes the projection of $\boldsymbol{N}$ onto the span of $\boldsymbol{m}$, $\boldsymbol{b}$ and $\boldsymbol{d}$. Dissimilarity is defined as $D(\boldsymbol{J}^{\text{eff}, (1)}, \boldsymbol{J}^{\text{eff}, (2)}) = 1-\text{corr}(\text{vec}(\boldsymbol{J}^{\text{eff}, (1)}), \text{vec}(\boldsymbol{J}^{\text{eff}, (2)}))$. This is appropriate for linear $\phi$, but for nonlinear $\phi$, this simplification may fall short as the nonlinearity adds projections in other dimensions (Appendix~\ref{remark}). Therefore, instead, we could define $\boldsymbol{J}_{\textrm{ours}}^{\text{eff}} = \frac{1}{K}\boldsymbol{M}\hat{\boldsymbol{N}}$, where $\hat{\boldsymbol{N}}$ is from Equation~(\ref{eq:main_nhat}).

To compare networks of unequal sizes without one-to-one correspondence between neurons in the two networks, we extend this notion to a dissimilarity $D(p^{(1)}, p^{(2)})$ between connectivity distributions, $p^{(1)}(\boldsymbol{m}, \boldsymbol{n}, \boldsymbol{b}, \boldsymbol{d})$ and $p^{(2)}(\boldsymbol{m}, \boldsymbol{n}, \boldsymbol{b}, \boldsymbol{d})$: \begin{equation}\label{eq:our_distance_measure}
\begin{split}
&D(p^{(1)}, p^{(2)}) = W(p^{(1)}({\boldsymbol{x}}) , p^{(2)}({\boldsymbol{x}})) + \\
&\mathbb{E}_{\boldsymbol{x} \sim p^{(2)}}\left[ \|\mathbb{E}_{\boldsymbol{n} \sim p^{(1)}} [\boldsymbol{n}|\boldsymbol{x}] - \mathbb{E}_{\boldsymbol{n} \sim p^{(2)}}[\boldsymbol{n}|\boldsymbol{x}]\|^2 \right]
\end{split}
\end{equation} where $\boldsymbol{x} = (\boldsymbol{m}, \boldsymbol{b}, \boldsymbol{d})$. The first term $W$ is an optimal transport distance, approximated with the debiased Sinkhorn divergence \cite{feydy2018interpolatingoptimaltransportmmd}, and the second term captures differences in conditional means. Both terms are estimated via sampling (see Appendix~\ref{diss_supp} for details). Intuitively, the two terms quantify differences in the $\{ \boldsymbol{M}, \boldsymbol{B}, \boldsymbol{d} \}$'s and in the effective $\hat{\boldsymbol{N}}$'s, respectively.

\section{Experiments} \label{experiments}

\begin{figure*}[ht]
\vskip 0.2in
\begin{center} 
\centerline{\includegraphics[width=6in]{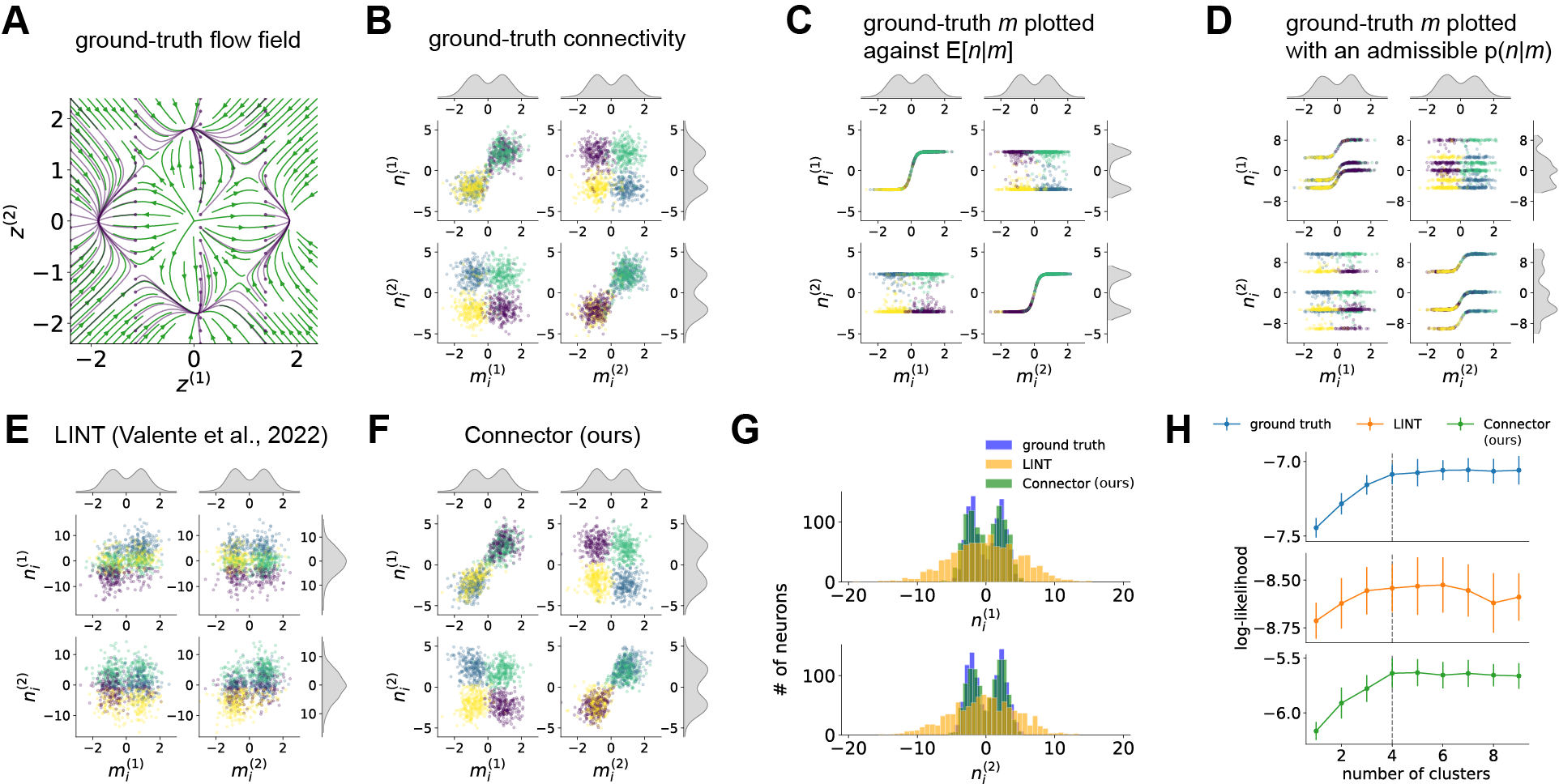}}
\caption{Identifiability of connectivity distributions. ({\bf A}) Flow field of quadstable attractor dynamics generated from a generalized Hopfield network. Purple lines are example latent trajectories with circles indicating initial conditions.
({\bf B}) Connectivity distribution of the generalized Hopfield network is a mixture of four Gaussians. Here, $x$-axis represents the first and second components from the samples $\boldsymbol{m}$ drawn from $p(\boldsymbol{m}, \boldsymbol{n})$, and the $y$-axis represents samples $\boldsymbol{n}$ from $p(\boldsymbol{m}, \boldsymbol{n})$. Color code based on which of the four Gaussians the neuron is drawn from (the neuron's ``cell type''). The same colors are used across {\bf B}--{\bf D} based on the ground-truth cell type.
({\bf C}) Ground-truth $\boldsymbol{m}_i$'s plotted against $\mathbb{E}[\boldsymbol{n}|\boldsymbol{m}_i]$'s.
({\bf D}) Connectivity generated from an arbitrary $p(\boldsymbol{n}|\boldsymbol{m})$ that has the same $\mathbb{E}[\boldsymbol{n}|\boldsymbol{m}]$ as the ground truth connectivity. Connectivity distributions in {\bf B}--{\bf D} are all admissible and generate dynamics nearly identical to the quadstable attractor dynamics in {\bf A}.
({\bf E}) Connectivity inferred from LINT. Color code based on $4$-means clustering.
({\bf F}) Connectivity inferred from Connector (our approach). Color code based on $4$-means clustering. ({\bf G}) The ground-truth and inferred $p(\boldsymbol{n})$ in {\bf B}, {\bf E}, {\bf F}. ({\bf H}) We clustered the neurons based on the learned connectivity using GMM. The 5-fold cross-validated log-likelihood (mean $\pm$ std) was computed to identify the ``elbow''.
}
\label{figure-2}
\end{center}
\vskip -0.2in
\end{figure*}

\subsection{Generalized Hopfield Networks} \label{generalized_hopfield}
 
Here we construct lrRNNs with known ground-truth connectivity to demonstrate how the degeneracies identified in Section~\ref{main_identify}, particularly those arising from $p(\boldsymbol{n}|\boldsymbol{m}, \boldsymbol{b}, \boldsymbol{d})$, make the recovery of connectivity from neural dynamics difficult. We considered Gaussian-mixture lrRNNs \cite{beiran21}, which reduce to classical Hopfield networks in a particular limit. A rank-2 quadstable attractor dynamics (Figure~\ref{figure-2}A) can be generated by arranging four Gaussians in $(\boldsymbol{m}, \boldsymbol{n})$-space as in Figure~\ref{figure-2}B. In this example, we constructed the Gaussian-mixture connectivity distribution such that the conditional covariance of $p(\boldsymbol{n}|\boldsymbol{m})$ is equal to the identity matrix $\boldsymbol{I}_R$. We do not have external inputs and bias, and therefore there are no $\boldsymbol{b}$ and $\boldsymbol{d}$. We sampled 1,000 neurons from this distribution (250 per Gaussian) to generate the network. The four Gaussians can be interpreted as four ``cell types'' in this network \cite{dubreuil22}.

As discussed in Section~\ref{main_identify}, the mean-field dynamics of the quadstable attractor network depend only on the first moment of $p(\boldsymbol{n}|\boldsymbol{m})$: $\boldsymbol{\mu}(\boldsymbol{m}) = \mathbb{E}[\boldsymbol{n}|\boldsymbol{m}]$. Figure~\ref{figure-2}C shows how $\boldsymbol{\mu}$ depends on $\boldsymbol{m}$---as long as $\boldsymbol{\mu}$ is arranged as such, any $p(\boldsymbol{n}|\boldsymbol{m})$ satisfying this arrangement of $\boldsymbol{\mu}$ will have mean-field dynamics identical to the ground-truth dynamics in Figure~\ref{figure-2}A. In Figure~\ref{figure-2}D, we have constructed $p(\boldsymbol{n}|\boldsymbol{m})$ such that the first moment matches Figure~\ref{figure-2}C, and experimentally validated that network with this connectivity structure generates the quadstable attractor dynamics in Figure~\ref{figure-2}A.

Next, using neural activity generated from the ground-truth network, we applied LINT \cite{valente22} to infer connectivity. LINT recovered the quadstable attractors in Figure~\ref{figure-2}A (up to linear transformation of $\boldsymbol{z}$, due to source {\bf (1)} in Section~\ref{main_identify}), and its connectivity matrix correctly approximated $p(\boldsymbol{m})$ of the ground-truth network (Figure~\ref{figure-2}E). The inferred latent trajectories spanned the latent space, satisfying the identifiability condition in {\bf (2)} of Section~\ref{main_identify}. Of note, if we only knew the latent dynamics without the knowledge of how the latents map onto the neural population activity, we would not have been able to identify $p(\boldsymbol{m})$---there are multiple non-unique $p(\boldsymbol{m})$'s that are capable of generating identical latent dynamics (Figure~\ref{supp-figure-2}). To correctly identify $p(\boldsymbol{m})$, we need to know the loading from the latent trajectory to neural activity.

Even when LINT correctly approximates $p(\boldsymbol{m})$, we found that the approximated $p(\boldsymbol{n}|\boldsymbol{m})$ from the connectivity matrix of LINT can be quite different from $p(\boldsymbol{n}|\boldsymbol{m})$ of the ground truth (Figure~\ref{figure-2}G). The recovered approximation of $p(\boldsymbol{n}|\boldsymbol{m})$ depended on how the lrRNN was initialized, and other hyperparameters of the training procedure (Figure~\ref{supp-figure-3}). This suggests that while LINT finds an {\it admissible} solution to the problem of recovering the connectivity distribution from neural population activity, it does not necessarily recover the ground truth connectivity due to the degeneracy in $p(\boldsymbol{n}|\boldsymbol{m})$. Based on the connectivity found by LINT in Figure~\ref{figure-2}E, we grouped neurons into clusters using $k$-means and Gaussian mixture models (GMMs). Clustering inertia (from $k$-means), out-of-sample log-likelihood (from GMM), and silhouette scores failed to recover that four cell types are present in the ground-truth network (Figures~\ref{figure-2}H).

Next, we applied Connector using the LINT-inferred latents and loadings. Connector returns a set of admissible solutions, not just a single solution, that can generate the observed quadstable attractor dynamics. When the conditional covariance $\boldsymbol{S}$ of $p(\boldsymbol{n}|\boldsymbol{m})$ matched the ground-truth conditional covariance of $p(\boldsymbol{n}|\boldsymbol{m})$ (i.e., $\boldsymbol{S} = \boldsymbol{I}_R$), Connector accurately recovered the connectivity distribution $p(\boldsymbol{m}, \boldsymbol{n})$ (Figure~\ref{figure-2}F, G), and correctly identified four cell types via clustering (Figure~\ref{figure-2}H). Results were robust across clustering methods ($k$-means and GMM).

\begin{figure}[h]
\vskip 0.2in
\begin{center} 
\centerline{\includegraphics[width=3.2in]{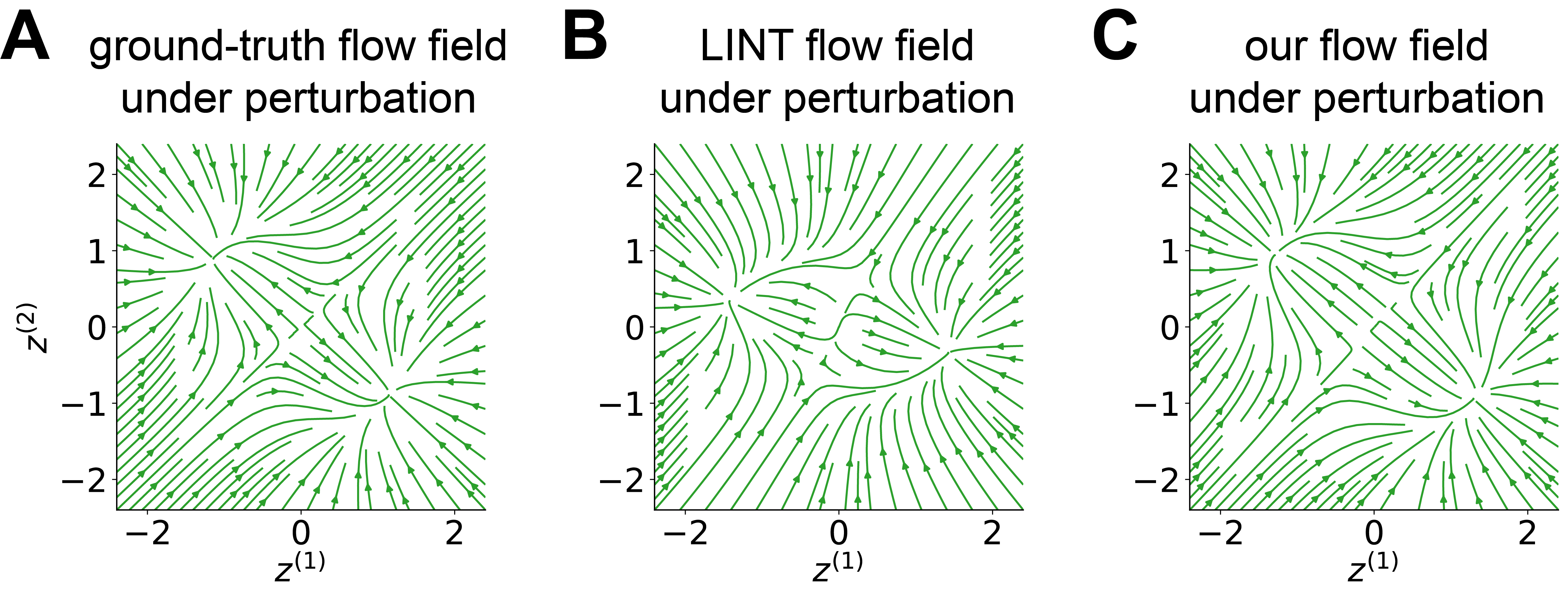}}
\caption{Comparisons of dynamics under perturbation. We clustered the neurons into four different cell types for each model. We then silenced neurons that belonged to one of the four cell types in the ground-truth ({\bf A}), LINT ({\bf B}), and Connector ({\bf C}).
}
\label{supp-figure-5}
\end{center}
\vskip -0.2in
\end{figure}

We further validated these results using {\it in silico} perturbations. When we silence one of the four cell types (see Appendix~\ref{perturb} for how we formalize silencing), the dynamics of the ground-truth network become bistable attractor-like (Figure~\ref{supp-figure-5}A). Silencing one of the Connector-inferred cell types produced similar perturbed dynamics, whereas silencing one of the LINT-based cell types did not (Figure~\ref{supp-figure-5}B--C).

In addition to the quadstable attractors, we performed similar analyses on other dynamical systems, including bistable attractors (Figure~\ref{supp-figure-6}A--F), limit cycles (Figure~\ref{supp-figure-6}G--J), and ring attractor (Figure~\ref{supp-figure-6}K--N). When Connector's conditional covariance $\boldsymbol{S}$ matched the ground-truth conditional covariance, Connector correctly recovered the ground-truth connectivity distribution in various settings, including when the covariance between $\boldsymbol{m}$ and $\boldsymbol{n}$ is non-normal (Figure~\ref{supp-figure-6}). We also performed analyses on dynamical systems generated from heavy-tailed distributions such as Student-$t$ and log-normal distributions (Figure~\ref{supp-figure-17}).

Indeed, when Connector's $\boldsymbol{S}$ is not equal to the ground-truth conditional covariance, the inferred connectivity distribution deviates from the ground truth. Connector {\it does not} eliminate degeneracy---it is still fundamentally difficult to infer connectivity from dynamics alone---but what Connector allows is to effectively capture this degeneracy as the free parameter $\boldsymbol{S}$. It gives us a set of solutions (given by different $\boldsymbol{S}$’s), whereas previous methods give a single admissible solution that may depend on initialization and other factors (Figure~\ref{supp-figure-3}).

\begin{figure}[h]
\vskip 0.2in
\begin{center} 
\centerline{\includegraphics[width=3.2in]{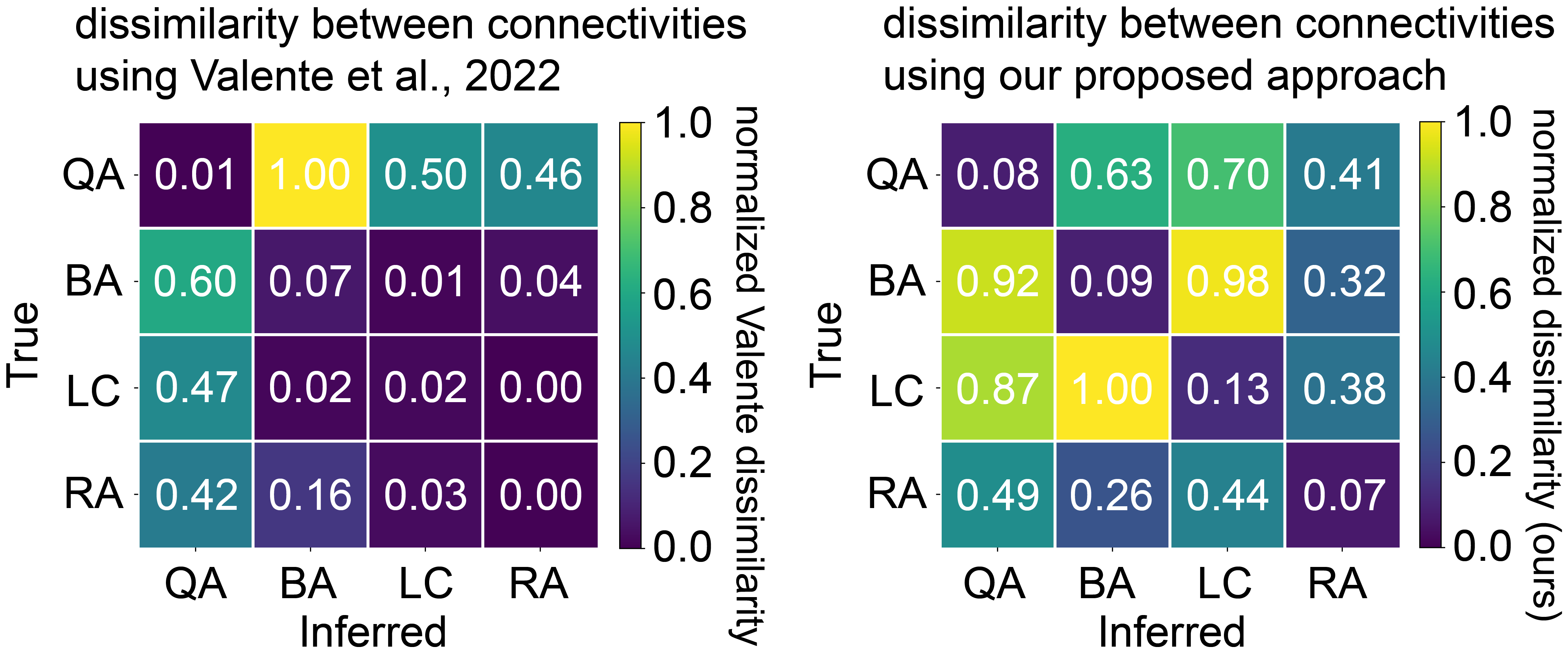}}
\caption{Connectivity dissimilarity $D(\textrm{True},\textrm{Inferred})$ computed using the measure in \citet{valente22} and our measure in Equation~(\ref{eq:our_distance_measure}). QA: Quadstable Attractors (Figure~\ref{figure-2}), BA: Bistable Attractors (Figure~\ref{supp-figure-6}A--F), LC: Limit Cycles (Figure~\ref{supp-figure-6}G--J), RA: Ring Attractors (Figure~\ref{supp-figure-6}K--N).
}
\label{figure-2-part2}
\end{center}
\vskip -0.2in
\end{figure}

Finally, we compared connectivity dissimilarity using the measure based on \citet{valente22} and ours (Section~\ref{main_distance}). Our measure correctly assigned the lowest dissimilarity when inferred and ground-truth dynamics matched, whereas the \citet{valente22} measure did not (see diagonals of dissimilarity matrices in Figure~\ref{figure-2-part2}). We obtained similar results when we replaced the $1 - \textrm{corr}(\cdot, \cdot)$ with mean-squared error in the \citet{valente22} measure. This suggests that comparisons of effective connectivity can be improved by incorporating the sources of degeneracy in Section~\ref{main_identify} and incorporating cases where the two networks being compared do not have one-to-one correspondence between neurons.

\subsection{Context-Dependent Decision-Making RNNs} \label{cddm_rnns}

We next analyzed neural activity from an lrRNN trained on a context-dependent decision-making task with external inputs (Figure~\ref{figure-3}A; \citet{dubreuil22, valente22}). LINT recovered latent trajectories and rates matching the ground truth (Figure~\ref{figure-3}B). However,  as in Section~\ref{generalized_hopfield}, the inferred loadings $\boldsymbol{N}$ did not match the ground truth, reflecting degeneracy from source {\bf (3)} in Section~\ref{main_identify} (Figure~\ref{figure-3}C). Applying Connector, we obtained a set of solutions for $p(\boldsymbol{n}|\boldsymbol{m},\boldsymbol{b})$; among these, $\boldsymbol{S} = \boldsymbol{I}_R$ approximately matched the ground truth connectivity (Figure~\ref{figure-3}D).

For this dataset, we also experimented with whether we can infer connectivity distribution based on the latents and loadings of LVMs more general than lrRNNs, as our results in Section~\ref{main_identify} showed that LVM models in general suffer more from non-identifiability compared to lrRNNs, especially when there are external inputs (Appendix~\ref{identify}). We found that, indeed, this is empirically the case (Appendix~\ref{appendix_cddm}, Figures~\ref{supp-figure-18}--\ref{supp-figure-19}). This suggests that identifying interpretable connectivity structures with expressive LVMs, such as ones based on Transformers (e.g., \citet{Ye_2021}), could be more difficult (consistent with, e.g., \citet{vafa2025what}).

\begin{figure*}[ht]
\vskip 0.2in
\begin{center} 
\centerline{\includegraphics[width=5.88in]{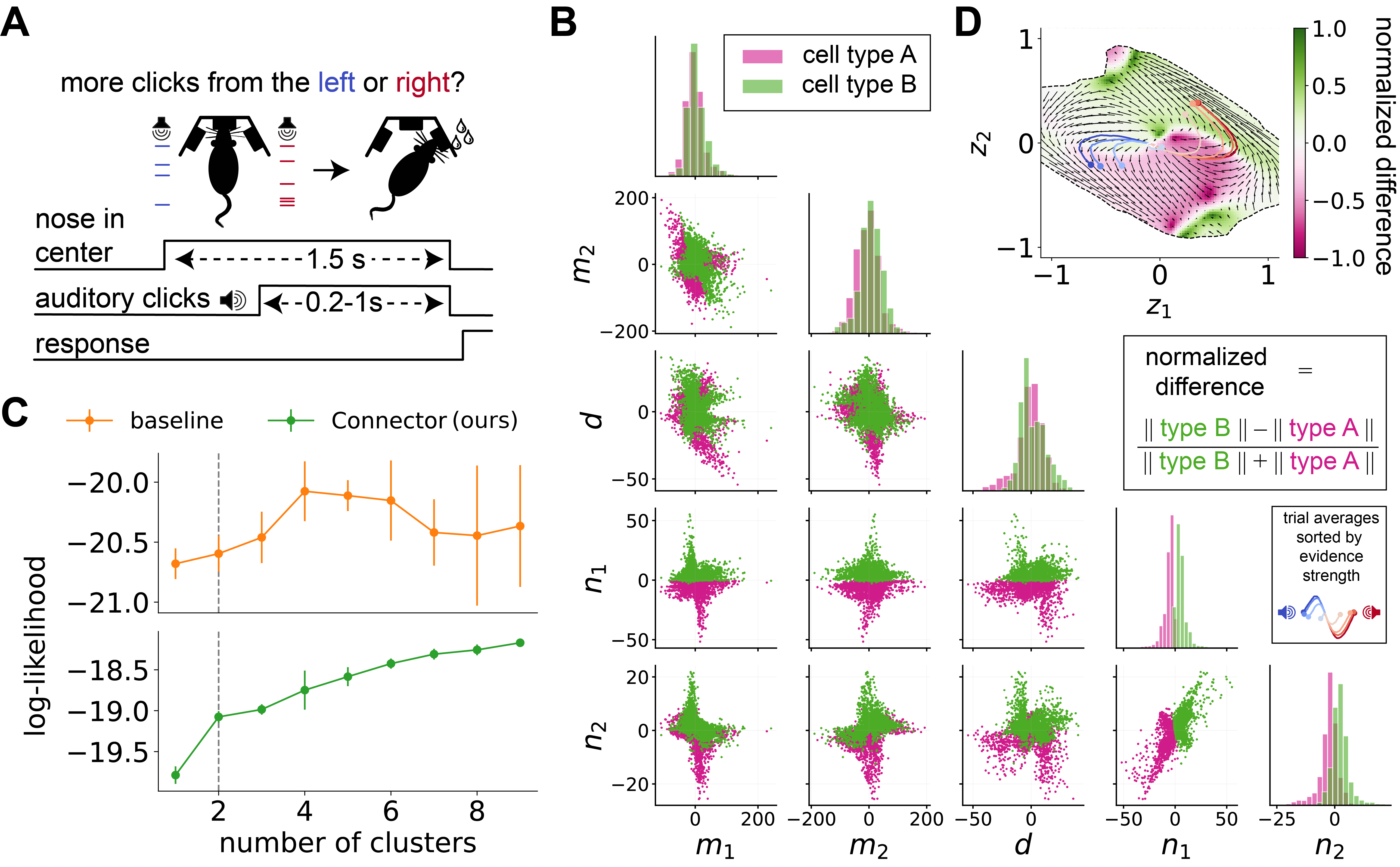}}
\caption{Connector-inferred connectivity distribution reveals computational cell types and their roles in neural dynamics. ({\bf A}) Dataset from \citet{LuoKim2025}. The rat listened to a stream of clicks from the left and right speakers and oriented to the side that had more clicks. Neurons from frontal cortex were recorded during this task ($K_{obs}=240$). Figure adapted from \citet{kim2025findr}.
({\bf B}) Connectivity distribution inferred with Connector. Each dot denotes a neuron sampled from the distribution ($K=$ 5,000).
({\bf C}) We clustered the neurons based on the connectivity inferred in {\bf B} using GMM. The 5-fold cross-validated log-likelihood (mean $\pm$ std) was computed to identify the ``elbow''. Connector-inferred connectivity suggests at least 2 clusters,  which we plot in pink and green colors in {\bf B}. ({\bf D}) Flow field (in quiver plot, showing only inside the dotted line---the part traversed by the single-trial latent trajectories---to be consistent with \citet{LuoKim2025}) with normalized difference showing relative contributions of cell type A and B to the dynamics. The colored trajectories are trial-averaged latent trajectories grouped by the number of clicks (darker red: more right clicks; darker blue: more left clicks).
}
\label{figure-4}
\end{center}
\vskip -0.2in
\end{figure*}

\subsection{Computational Cell Types in Frontal Cortex of Rats During Decision-Making} \label{luo_section}

A key advantage of mechanistic models such as lrRNN is that, unlike low-dimensional state space models such as recurrent switching linear dynamical system (rSLDS; \citet{linderman17}) or FINDR \cite{kim2025findr}, they allow us to quantify how the activity of each neuron contributes to the low-dimensional dynamics. This enables {\it in silico} manipulations, such as selectively silencing (Figure~\ref{supp-figure-5}, Appendix~\ref{perturb}) or isolating subsets of neurons (Appendix~\ref{cell-type-specific-dynamics}), to assess their roles in network dynamics.

To see whether this idea holds for real neural activity, we looked into dataset published in \citet{LuoKim2025}. Because training lrRNNs directly on the single-trial spiking activity of this dataset is challenging (as reported by \citet{LuoKim2025}), we used a knowledge distillation approach where the lrRNN was trained to match the latent dynamics from FINDR while ensuring that the activity of the lrRNN units matched the task-relevant firing rates (Appendix~\ref{details_luo_section}). The resulting lrRNN reproduced flow field similar to FINDR (Figure~\ref{figure-4}D, Figure~\ref{supp-figure-15}A--B). This distillation was needed because connectivity distributions are not directly recoverable from most state space models, including FINDR. In lrRNNs, the latent update is a weighted sum of neural activity (Equation~(\ref{w_update})), enabling us to quantify each neuron's contribution to the latent, but Equation~(\ref{w_update}) may not always hold if we use general state space models (Appendix~\ref{details_luo_section}).

Using the distilled lrRNN, we inferred the connectivity distribution with Connector. Large networks sampled from this distribution generated flow fields nearly identical to the original flow field (Figure~\ref{supp-figure-15}C). The inferred distribution was non-Gaussian, with heavy tails, multimodality, and skew (Figure~\ref{figure-4}B). We found that at least 2 or more clusters (or cell types) may be present, as shown by cross-validated log-likelihoods (Figure~\ref{figure-4}C, Connector) and silhouette scores (Figure~\ref{supp-figure-15}D) from GMM and $k$-means clustering, respectively. This was less clear if we use the connectivity matrix directly from the lrRNN for clustering (Figure~\ref{figure-4}C, baseline).

We labeled the two clusters as cell types A and B (Figure~\ref{figure-4}B) and quantified their relative contributions to the dynamics by generating flow fields using neurons sampled from each type separately (Figure~\ref{supp-figure-15}E--F), and using what we call the normalized difference index. Intuitively, this index is a measure of {\it difference in speed} between cell-type-A-specific dynamics and cell-type-B-specific dynamics, normalized so that it lies between $[-1, 1]$, and is similar to the one developed in \citet{LuoKim2025} but for cell types. This index revealed that the two cell types contribute differently across the state space (Figure~\ref{figure-4}D; see Appendix~\ref{NDI} for precise definition of the index). Using alternative definitions of the index gave similar results (Figure~\ref{supp-figure-15}I; Appendix~\ref{NDI}). Notably, changes in the relative contributions of the two cell types coincided with turning points in trial-averaged latent trajectories, which have been associated with decision commitment in this task \cite{LuoKim2025}. If we directly use the connectivity matrix learned from the lrRNN and perform the same clustering, we do not get similarly interpretable cell types (Figure~\ref{supp-figure-15}H). Results were robust across clustering methods ($k$-means and GMMs; Figure~\ref{supp-figure-15}J) and across a range of conditional covariance assumptions for $\boldsymbol{n}$ (Figure~\ref{supp-figure-15}K); Figure~\ref{figure-4}B shows the case $\boldsymbol{S} = \boldsymbol{0}$.

We emphasize that we do not interpret the identified cell types as distinct decision-related neuron classes (e.g., ``accumulation'' versus ``commitment'' neurons). Rather, our goal is to demonstrate that Connector enables unsupervised discovery of {\it computational cell types} whose roles vary across state space, and may generate testable hypotheses for future experiments.

\section{Discussion}

We identified three sources of degeneracy in inferring connectivity from neural dynamics in low-rank recurrent neural networks (lrRNNs) and derived conditions under which a unique, maximum entropy connectivity distribution can be identified. We then introduced Connector, a framework using continuous normalizing flows (CNFs) to infer such distributions from data.

We showed that this degeneracy is empirically observed in connectivity weights of lrRNNs trained on neural data, and that, depending on initialization or training, learned connectivities can differ substantially, despite producing nearly identical dynamics. While all such connectivities may be admissible, they are not equally simple. Connector finds the simplest (maximum entropy) distribution consistent with data, avoiding spurious structures that arise from arbitrary optimization choices. Applied to recordings from rat frontal cortex during decision-making, Connector identified interpretable computational cell types that were less apparent with the standard approach, demonstrating the benefit of the maximum entropy criterion in isolating minimal necessary structure.

Several assumptions bound the scope of this work. For example, the connectivity inferred in this work is not constrained to obey Dale's law, and is assumed throughout to be time-invariant. A natural next direction is to explore how much of our analyses carry over to networks with excitation-inhibition balance \citep{van_Vreeswijk}, and to networks in which the connectivity changes over time \citep{pellegrino, lu2025netformer}.

More broadly, Connector bridges mean-field theory and modern data-driven modeling by learning flexible, structured connectivity distributions directly from neural data while retaining mechanistic interpretability. By distinguishing necessary from arbitrary connectivity features, our framework shifts circuit inference toward principled assessment of computational necessity, enabling sharper hypotheses, more targeted perturbations, and better-grounded mechanistic theories of neural computation.

\section*{Software and Data}
Our code is available as a GitHub repository: \url{https://github.com/AllenInstitute/connector}.

\section*{Acknowledgements}
This research was supported by the Allen Institute, founded by Jody Allen---chair and co-founder of Allen Family Philanthropies, and the late Paul G. Allen---investor, philanthropist, and co-founder of Microsoft. We gratefully acknowledge their vision and generosity, which make this work possible. We are also grateful to Bing Brunton, Michael Buice, Mia Cameron, Christof Koch, and Karel Svoboda for their feedback on this work. TDK and YW were supported by the Shanahan Family Foundation Fellowship at the Interface of Data and Neuroscience at the Allen Institute and the University of Washington, supported in part by the Allen Institute.

\section*{Impact Statement}

This paper presents work whose goal is to advance the field of Machine Learning and Computational Neuroscience. There are many potential societal consequences of our work, none which we feel must be specifically highlighted here.

\bibliography{icml_rnfn}
\bibliographystyle{icml2026}


\newpage
\renewcommand{\thefigure}{S\arabic{figure}}
\renewcommand{\thetable}{S\arabic{table}}
\setcounter{figure}{0}
\setcounter{table}{0}
\setcounter{section}{1}

\appendix

\onecolumn
\section{Low-Rank Recurrent Neural Networks} \label{background_lrrnn}
Here we present a brief note on low-rank recurrent neural networks (lrRNNs) introduced in \citet{MASTROGIUSEPPE18}. We start with a form of recurrent neural networks (RNNs) often considered in neuroscience, physics, and cognitive science (e.g., \citet{sompolinsky88}): \begin{equation} \label{eq:sompolinsky}
\tau \dot{\boldsymbol{h}} = -\boldsymbol{h} + \boldsymbol{J}\phi(\boldsymbol{h}) + \boldsymbol{B}\boldsymbol{u} + \boldsymbol{d}.
\end{equation} Here, the strength of interaction between the units $\boldsymbol{h} \in \mathbb{R}^{K}$ are represented by the connectivity matrix $\boldsymbol{J} \in \mathbb{R}^{K \times K}$. Each unit $\boldsymbol{h}_i$ may be interpreted as the membrane potential of neuron $i$, and $\boldsymbol{r}_i = \phi(\boldsymbol{h}_i)$ may be interpreted as the output firing rate of neuron $i$. The activation function $\phi$, which in this work is assumed to be strictly monotonically increasing, is applied element-wise to the elements of the vector $\boldsymbol{h}$. In this work, $\phi$ is set to be $\tanh$ unless stated otherwise. The external input $\boldsymbol{u} \in \mathbb{R}^{K_{in}}$ projects to the network via $\boldsymbol{B} \in \mathbb{R}^{K \times K_{in}}$. The baseline internal voltages of the neurons are represented by $\boldsymbol{d} \in \mathbb{R}^K$. The time constant $\tau$ dictates the timescale of the network activity. To simulate the network activity $\boldsymbol{h}$ over time, one simple way is to discretize Equation~(\ref{eq:sompolinsky}) with the forward Euler method: \begin{equation} \label{eq:disc_sompolinsky}
\tau\frac{\boldsymbol{h}_{t} - \boldsymbol{h}_{t-1}}{\Delta t} = -\boldsymbol{h}_{t-1} + \boldsymbol{J} \phi (\boldsymbol{h}_{t-1}) + \boldsymbol{B}\boldsymbol{u}_t + \boldsymbol{d},
\end{equation} and update the internal voltages of neurons on the current time step $t$, $\boldsymbol{h}_t$, based on the internal voltages of neurons on the previous time step $t-1$, $\boldsymbol{h}_{t-1}$. The dependence of the current network state $\boldsymbol{h}_t$ on the network state on the previous time step $\boldsymbol{h}_{t-1}$ makes it clear why this network is recurrent.

Low-rank RNNs further assume that the connectivity matrix $\boldsymbol{J}$ has rank $R < K$. More specifically, $\boldsymbol{J} = \frac{1}{K}\boldsymbol{M}\boldsymbol{N}^\top$, where $\boldsymbol{M} \in \mathbb{R}^{K \times R}$, $\boldsymbol{N} \in \mathbb{R}^{K \times R}$. Then, Equation~(\ref{eq:disc_sompolinsky}) becomes \begin{equation} \label{eq:lr-RNN}
\tau\frac{\boldsymbol{h}_{t} - \boldsymbol{h}_{t-1}}{\Delta t} = -\boldsymbol{h}_{t-1} + \frac{1}{K} \boldsymbol{M}\boldsymbol{N}^\top \phi (\boldsymbol{h}_{t-1}) + \boldsymbol{B}\boldsymbol{u}_t + \boldsymbol{d}.
\end{equation} The network state $\boldsymbol{h}_t$ at any time point $t$ can be expressed as: \begin{equation} \label{eq:what_is_h}
\boldsymbol{h}_t = \boldsymbol{M} \boldsymbol{z}_t + \boldsymbol{B} \boldsymbol{v}_t + \boldsymbol{d},
\end{equation} where $\boldsymbol{z}_t \in \mathbb{R}^R$ and $\boldsymbol{v}_t \in \mathbb{R}^{K_{in}}$. Here, $\boldsymbol{z}_t$ are a set of latent variables representing the collective low-dimensional dynamics of the network activity $\boldsymbol{h}_t$, and $\boldsymbol{v}_t$ is a low-pass filtering of the input $\boldsymbol{u}_t$ \citep{valente22}: \begin{equation} \label{eq:low-pass_filter}
\tau \frac{\boldsymbol{v}_{t} - \boldsymbol{v}_{t-1}}{\Delta t} = - \boldsymbol{v}_{t-1} + \boldsymbol{u}_t.
\end{equation} If we plug in Equation~(\ref{eq:what_is_h}) into Equation~(\ref{eq:lr-RNN}), we have \begin{equation} \label{eq:20}
\tau\frac{\boldsymbol{M} (\boldsymbol{z}_{t} - \boldsymbol{z}_{t-1})}{\Delta t} = -\boldsymbol{M} \boldsymbol{z}_{t-1} + \frac{1}{K} \boldsymbol{M}\boldsymbol{N}^\top \phi (\boldsymbol{M} \boldsymbol{z}_{t-1} + \boldsymbol{B} \boldsymbol{v}_{t-1} + \boldsymbol{d}).
\end{equation} Left-multiplying both sides of Equation~(\ref{eq:20}) with the pseudo-inverse of $\boldsymbol{M}$ (i.e., $\boldsymbol{M}^\dag = (\boldsymbol{M}^\top \boldsymbol{M})^{-1} \boldsymbol{M}^\top$), we get \begin{equation} \label{eq:endeq}
\tau\frac{\boldsymbol{z}_{t} - \boldsymbol{z}_{t-1}}{\Delta t} = -\boldsymbol{z}_{t-1} + \frac{1}{K} \boldsymbol{N}^\top \phi (\boldsymbol{M} \boldsymbol{z}_{t-1} + \boldsymbol{B} \boldsymbol{v}_{t-1} + \boldsymbol{d}).
\end{equation} Therefore, we have shown that we can express the $K$-dimensional network dynamics in Equation~(\ref{eq:lr-RNN}) in terms of the dynamics of the latent variable $\boldsymbol{z}$ in Equation~(\ref{eq:endeq}). Note that $\boldsymbol{N}^\top \phi (\boldsymbol{M} \boldsymbol{z}_{t-1} + \boldsymbol{B} \boldsymbol{v}_{t-1})$ is a multilayer perceptron (MLP) with a single hidden layer. We can view Equation~(\ref{eq:endeq}) as a special instance of a neural ODE, discretized with the forward Euler method. Let us define \begin{equation} \boldsymbol{M} = \begin{bmatrix}
    \longdash \boldsymbol{m}^\top_1 \longdash\\
    ...\\
    \longdash \boldsymbol{m}^\top_K \longdash
\end{bmatrix}, \quad \boldsymbol{N} = \begin{bmatrix}
    \longdash \boldsymbol{n}^\top_1 \longdash\\
    ...\\
    \longdash \boldsymbol{n}^\top_K \longdash
\end{bmatrix}, \quad \boldsymbol{B} = \begin{bmatrix}
    \longdash \boldsymbol{b}^\top_1 \longdash\\
    ...\\
    \longdash \boldsymbol{b}^\top_K \longdash
\end{bmatrix}. \end{equation} Then, Equation~(\ref{eq:endeq}) can be rewritten as \begin{equation} \label{eq:lrRNNeq}
\tau\frac{\boldsymbol{z}_{t} - \boldsymbol{z}_{t-1}}{\Delta t} = -\boldsymbol{z}_{t-1} + \frac{1}{K} \sum_{i=1}^K \boldsymbol{n}_i \phi (\boldsymbol{m}_i^\top \boldsymbol{z}_{t-1} + \boldsymbol{b}_i^\top \boldsymbol{v}_{t-1} + \boldsymbol{d}_i).
\end{equation} where $\boldsymbol{m}_i, \boldsymbol{n}_i \in \mathbb{R}^R$, $\boldsymbol{b}_i \in \mathbb{R}^{K_{in}}$, and $\boldsymbol{d}_i \in \mathbb{R}$. If $\boldsymbol{m}_i, \boldsymbol{n}_i, \boldsymbol{b}_i, \boldsymbol{d}_i \overset{\text{iid}}{\sim} p(\boldsymbol{m}, \boldsymbol{n}, \boldsymbol{b}, \boldsymbol{d})$, then as $K \rightarrow \infty$, we get \begin{equation} \label{eq:mf-lr-RNN}
\tau\frac{\boldsymbol{z}_{t} - \boldsymbol{z}_{t-1}}{\Delta t} = -\boldsymbol{z}_{t-1} + \mathbb{E}[\boldsymbol{n} \phi (\boldsymbol{m}^\top \boldsymbol{z}_{t-1} + \boldsymbol{b}^\top \boldsymbol{v}_{t-1} + \boldsymbol{d})],
\end{equation} by the law of large numbers, giving us the mean-field dynamics of low-rank RNNs. We refer to $p(\boldsymbol{m}, \boldsymbol{n}, \boldsymbol{b}, \boldsymbol{d})$ as the {\it connectivity distribution}. By the law of iterated expectations, Equation~(\ref{eq:mf-lr-RNN}) is equivalent to \begin{equation} \label{eq:mf-lr-RNN-final}
\tau\frac{\boldsymbol{z}_{t} - \boldsymbol{z}_{t-1}}{\Delta t} = -\boldsymbol{z}_{t-1} + \mathbb{E}[\mathbb{E}[\boldsymbol{n}|\boldsymbol{m}, \boldsymbol{b}, \boldsymbol{d}] \phi (\boldsymbol{m}^\top \boldsymbol{z}_{t-1} + \boldsymbol{b}^\top \boldsymbol{v}_{t-1} + \boldsymbol{d})].
\end{equation}

\subsection{When the connectivity distribution is Gaussian} \label{beiran_proof}

When the connectivity distribution $p(\boldsymbol{m}, \boldsymbol{n}, \boldsymbol{b}, \boldsymbol{d})$ is jointly Gaussian, we can simplify Equation~(\ref{eq:mf-lr-RNN}) further \citep{beiran21}. Let \begin{equation}
\boldsymbol{z}^{rec}_{t-1} = \mathbb{E}[\mathbb{E}[\boldsymbol{n}|\boldsymbol{m}, \boldsymbol{b}, \boldsymbol{d}] \phi (\boldsymbol{m}^\top \boldsymbol{z}_{t-1} + \boldsymbol{b}^\top \boldsymbol{v}_{t-1} + \boldsymbol{d})].
\end{equation} If the joint distribution $p(\boldsymbol{m}, \boldsymbol{n}, \boldsymbol{b}, \boldsymbol{d})$ is \begin{equation}
\boldsymbol{n}, \boldsymbol{m}, \boldsymbol{b}, \boldsymbol{d} \sim \mathcal{N}\left( \begin{bmatrix}
    \boldsymbol{a}_{n}\\
    \boldsymbol{a}_{m}\\
    \boldsymbol{a}_{b}\\
    \boldsymbol{a}_{d}
\end{bmatrix}, \begin{bmatrix}
    \boldsymbol{\Sigma}_{nn} & \boldsymbol{\Sigma}_{nm} & \boldsymbol{\Sigma}_{nb} & \boldsymbol{\Sigma}_{nd}\\
    \boldsymbol{\Sigma}_{mn} & \boldsymbol{\Sigma}_{mm} & \boldsymbol{\Sigma}_{mb} & \boldsymbol{\Sigma}_{md}\\
    \boldsymbol{\Sigma}_{bn} & \boldsymbol{\Sigma}_{bm} & \boldsymbol{\Sigma}_{bb} & \boldsymbol{\Sigma}_{bd}\\
    \boldsymbol{\Sigma}_{dn} & \boldsymbol{\Sigma}_{dm} & \boldsymbol{\Sigma}_{db} & \boldsymbol{\Sigma}_{dd}
\end{bmatrix} \right),
\end{equation} with \begin{equation} \begin{split}
\boldsymbol{y}_{t-1} &= \begin{bmatrix}
    \boldsymbol{z}_{t-1} \\ \boldsymbol{v}_{t-1} \\ 1
\end{bmatrix}, \\
\boldsymbol{x} &= \begin{bmatrix}
    \boldsymbol{m} \\ \boldsymbol{b} \\ \boldsymbol{d}
\end{bmatrix}, \\
\boldsymbol{a}_x &= \begin{bmatrix}
    \boldsymbol{a}_m \\ \boldsymbol{a}_b \\ \boldsymbol{a}_d
\end{bmatrix}, \\
\boldsymbol{\Sigma}_c &= \begin{bmatrix}
    \boldsymbol{\Sigma}_{nm} & \boldsymbol{\Sigma}_{nb} & \boldsymbol{\Sigma}_{nd}
\end{bmatrix}, \\
\boldsymbol{\Sigma}_{x} &= \begin{bmatrix}
\boldsymbol{\Sigma}_{mm} & \boldsymbol{\Sigma}_{mb} & \boldsymbol{\Sigma}_{md}\\
    \boldsymbol{\Sigma}_{bm} & \boldsymbol{\Sigma}_{bb} & \boldsymbol{\Sigma}_{bd}\\
    \boldsymbol{\Sigma}_{dm} & \boldsymbol{\Sigma}_{db} & \boldsymbol{\Sigma}_{dd}
\end{bmatrix},
\end{split}
\end{equation} then $p(\boldsymbol{n} | \boldsymbol{m}, \boldsymbol{b}, \boldsymbol{d})$ can be expressed as \begin{equation}
\boldsymbol{n} \sim \mathcal{N}(\boldsymbol{\mu}, \boldsymbol{\Sigma}),
\end{equation} where \begin{equation}
\begin{split}
\boldsymbol{\mu} &= \boldsymbol{a}_n + \boldsymbol{\Sigma}_c \boldsymbol{\Sigma}^{-1}_x \left(
    \boldsymbol{x} - \boldsymbol{a}_x \right),\\
\boldsymbol{\Sigma} &= \boldsymbol{\Sigma}_{nn} - \boldsymbol{\Sigma}_c\boldsymbol{\Sigma}^{-1}_{x} \boldsymbol{\Sigma}^\top_c.
\end{split}
\end{equation} Thus, $\mathbb{E}[\boldsymbol{n}|\boldsymbol{m}, \boldsymbol{b}, \boldsymbol{d}] = \boldsymbol{\mu}$, and this means \begin{equation}
\begin{split}
\boldsymbol{z}^{rec}_{t-1} &= \mathbb{E}[\mathbb{E}[\boldsymbol{n}|\boldsymbol{m}, \boldsymbol{b}, \boldsymbol{d}] \phi (\boldsymbol{m}^\top \boldsymbol{z}_{t-1} + \boldsymbol{b}^\top \boldsymbol{v}_{t-1} + \boldsymbol{d})] \\
&= \mathbb{E}[ (\boldsymbol{a}_n + \boldsymbol{\Sigma}_c \boldsymbol{\Sigma}^{-1}_x \left(
    \boldsymbol{x} - \boldsymbol{a}_x \right)) \phi (\boldsymbol{x}^\top \boldsymbol{y}_{t-1})] \\
&= \boldsymbol{a}_n\mathbb{E}[\phi (\boldsymbol{x}^\top \boldsymbol{y}_{t-1})] + \boldsymbol{\Sigma}_c \boldsymbol{\Sigma}^{-1}_x \mathbb{E}[ \left(
    \boldsymbol{x} - \boldsymbol{a}_x \right) \phi (\boldsymbol{x}^\top \boldsymbol{y}_{t-1})].
\end{split}
\end{equation} By Stein's lemma, \begin{equation}
\mathbb{E}[\left(
    \boldsymbol{x} - \boldsymbol{a}_x \right) \phi (\boldsymbol{x}^\top \boldsymbol{y}_{t-1})] = \boldsymbol{\Sigma}_x \boldsymbol{y}_{t-1} \mathbb{E}[
    \phi' (\boldsymbol{x}^\top \boldsymbol{y}_{t-1})].
\end{equation} Therefore, \begin{equation} \begin{split}
\boldsymbol{z}^{rec}_{t-1} &= \boldsymbol{a}_n\mathbb{E}[\phi (\boldsymbol{x}^\top \boldsymbol{y}_{t-1})] + \boldsymbol{\Sigma}_c \boldsymbol{\Sigma}^{-1}_x \boldsymbol{\Sigma}_x \boldsymbol{y}_{t-1}  \mathbb{E}[
    \phi' (\boldsymbol{x}^\top \boldsymbol{y}_{t-1})], \\ 
    &=  \boldsymbol{a}_n\mathbb{E}[\phi (\boldsymbol{x}^\top \boldsymbol{y}_{t-1})] + \boldsymbol{\Sigma}_c \boldsymbol{y}_{t-1}  \mathbb{E}[
    \phi' (\boldsymbol{x}^\top \boldsymbol{y}_{t-1})],\\
    &=  \boldsymbol{a}_n\int \phi (s) \mathcal{N}(s|\boldsymbol{a}_x^\top \boldsymbol{y}_{t-1}, \boldsymbol{y}_{t-1}^\top \boldsymbol{\Sigma}_x \boldsymbol{y}_{t-1})) ds \\
    &\quad+ \boldsymbol{\Sigma}_c \boldsymbol{y}_{t-1}  \int \phi' (s) \mathcal{N}(s|\boldsymbol{a}_x^\top \boldsymbol{y}_{t-1}, \boldsymbol{y}_{t-1}^\top \boldsymbol{\Sigma}_x \boldsymbol{y}_{t-1})) ds.
\end{split}
\end{equation} This suggests that $\boldsymbol{n}$ affects $\boldsymbol{z}^{rec}_{t-1}$ only through its mean $\boldsymbol{a}_n$, and its covariance $\boldsymbol{\Sigma}_c$ between variables $\boldsymbol{m}$, $\boldsymbol{b}$, and $\boldsymbol{d}$. $\boldsymbol{\Sigma}_{nn}$ has no direct influence on $\boldsymbol{z}^{rec}_{t-1}$. The first term $\bar{r}_{t-1} = \int \phi (s) \mathcal{N}(s|\boldsymbol{a}_x^\top \boldsymbol{y}_{t-1}, \boldsymbol{y}_{t-1}^\top \boldsymbol{\Sigma}_x \boldsymbol{y}_{t-1})) ds$ can be interpreted as the average activity of neurons. The second term $\bar{g}_{t-1} = \int \phi' (s) \mathcal{N}(s|\boldsymbol{a}_x^\top \boldsymbol{y}_{t-1}, \boldsymbol{y}_{t-1}^\top \boldsymbol{\Sigma}_x \boldsymbol{y}_{t-1})) ds$ can be interpreted as the average gain of neurons. Then, \begin{equation}
\boldsymbol{z}^{rec}_{t-1} = \bar{r}_{t-1}\boldsymbol{a}_n + \bar{g}_{t-1}\boldsymbol{\Sigma}_c \boldsymbol{y}_{t-1}
\end{equation} where we interpret $\bar{r}_{t-1}\boldsymbol{a}_n$ as the effective input to $\boldsymbol{z}^{rec}_{t-1}$, and $\bar{g}_{t-1}\boldsymbol{\Sigma}_c$ as the effective connectivity. Finally, \begin{equation} \label{eq:final_beiran}
\tau\frac{\boldsymbol{z}_{t} - \boldsymbol{z}_{t-1}}{\Delta t} = -\boldsymbol{z}_{t-1} + \bar{r}_{t-1}\boldsymbol{a}_n + \bar{g}_{t-1}[\boldsymbol{\Sigma}_{nm}\boldsymbol{z}_{t-1} + \boldsymbol{\Sigma}_{nb}\boldsymbol{v}_{t-1} + \boldsymbol{\Sigma}_{nd}].
\end{equation} Here, $\boldsymbol{\Sigma}_{nd}$ is a vector, not a matrix. The average activity $\bar{r}_{t}$ and gain $\bar{g}_t$ of neurons can change over time, but given a time point $t$, interactions between variables $\boldsymbol{z}_{t}$ and input $\boldsymbol{b}_{t}$ are linear. When $\phi$ is the identity function, $\bar{r}_{t-1} = \boldsymbol{a}_x^\top\boldsymbol{y}_{t-1}$, and $\bar{g}_{t-1} = 1$. An alternative derivation to the one presented here can be found in \citet{beiran21}.

\subsection{When the connectivity distribution is a mixture of Gaussians} \label{mog_proof}

We can easily extend Appendix~\ref{beiran_proof} to a mixture of Gaussians \cite{beiran21}: \begin{equation} \label{eq:mog}
\tau\frac{\boldsymbol{z}_{t} - \boldsymbol{z}_{t-1}}{\Delta t} = -\boldsymbol{z}_{t-1} + \sum_{p=1}^P \alpha_p \left[\bar{r}^{(p)}_{t-1}\boldsymbol{a}^{(p)}_n + g^{(p)}_{t-1}[\boldsymbol{\Sigma}^{(p)}_{nm}\boldsymbol{z}_{t-1} + \boldsymbol{\Sigma}^{(p)}_{nb}\boldsymbol{v}_{t-1} + \boldsymbol{\Sigma}^{(p)}_{nd}] \right],
\end{equation} where $P$ is the number of Gaussians and $\sum \alpha_p = 1$.

\subsection{When the connectivity distribution is Student-$t$} \label{student-t}

When the connectivity distribution $p(\boldsymbol{m}, \boldsymbol{n}, \boldsymbol{b}, \boldsymbol{d})$ is jointly Student-$t$, it can be expressed as an infinite mixture of Gaussians where the weights of the Gaussians are distributed as a Gamma distribution. That is, \begin{equation}
\int \mathcal{N}(s|\bar{a}, \bar{\sigma}^2/\lambda) P(\lambda) d\lambda = t_\nu(s|\bar{a}, \bar{\sigma}^2)
\end{equation} where $P(\lambda)$ is $\Gamma(\nu/2, \nu/2)$ \citep{andrews1974scale}. Let $\bar{a}_{t-1} = \boldsymbol{a}_x^\top \boldsymbol{y}_{t-1}$ and $\bar{\sigma}^2_{t-1} = \boldsymbol{y}_{t-1}^\top \boldsymbol{\Sigma}_x \boldsymbol{y}_{t-1}$. Also, let $\bar{r}_{t-1}(\lambda) = \int \phi (s) \mathcal{N}(s|\bar{a}_{t-1}, \bar{\sigma}^2_{t-1}/\lambda) ds$, and $\bar{g}_{t-1}(\lambda) = \int \phi' (s) \mathcal{N}(s|\bar{a}_{t-1}, \bar{\sigma}_{t-1}^2/\lambda) ds$. Then, \begin{equation}
\begin{split}
\boldsymbol{z}^{rec}_{t-1} &=\left[ \int \bar{r}_{t-1}(\lambda) \boldsymbol{a}_n P(\lambda) d\lambda \right] + \left[ \int \bar{g}_{t-1}(\lambda) (\boldsymbol{\Sigma}_c/\lambda)P(\lambda) d\lambda \right] \boldsymbol{y}_{t-1} \\
&=\left[ \int \bar{r}_{t-1}(\lambda) P(\lambda) d\lambda \right] \boldsymbol{a}_n + \left[ \int \frac{\bar{g}_{t-1}(\lambda)}{\lambda} P(\lambda) d\lambda \right] \boldsymbol{\Sigma}_c \boldsymbol{y}_{t-1} \\
&=\tilde{r}_{t-1} \boldsymbol{a}_n + \tilde{g}_{t-1}\boldsymbol{\Sigma}_c \boldsymbol{y}_{t-1}.
\end{split}
\end{equation} Note that by Fubini's Theorem, \begin{equation}
\tilde{r}_{t-1} = \int \bar{r}_{t-1}(\lambda) P(\lambda) d\lambda = \int \phi(s) t_{\nu}(s | \bar{a}_{t-1}, \bar{\sigma}_{t-1}^2)ds.
\end{equation}

\noindent Also, \begin{equation} \begin{split}
\tilde{g}_{t-1} &= \int_0^{\infty} \frac{\bar{g}_{t-1}(\lambda)}{\lambda} P(\lambda) d\lambda \\
&= \int_0^{\infty} \left[ \int \frac{1}{\lambda} \phi'(s) \mathcal{N}(s | \bar{a}_{t-1}, \bar{\sigma}_{t-1}^2/\lambda)ds \right] d\lambda \\
&= \int_0^{\infty} \left[ \int \frac{1}{\lambda} \frac{(s - \bar{a}_{t-1})}{\bar{\sigma}_{t-1}^2/\lambda}\phi(s) \mathcal{N}(s | \bar{a}_{t-1}, \bar{\sigma}_{t-1}^2/\lambda)ds \right] d\lambda \\
&= \int_0^{\infty} \left[ \int \frac{(s - \bar{a}_{t-1})}{\bar{\sigma}_{t-1}^2}\phi(s) \mathcal{N}(s | \bar{a}_{t-1}, \bar{\sigma}_{t-1}^2/\lambda)ds \right] d\lambda \\
&= \int \left[ \int_0^{\infty} \frac{(s - \bar{a}_{t-1})}{\bar{\sigma}_{t-1}^2}\phi(s) \mathcal{N}(s | \bar{a}_{t-1}, \bar{\sigma}_{t-1}^2/\lambda)d\lambda \right] ds \\
&= \int \frac{(s - \bar{a}_{t-1})}{\bar{\sigma}_{t-1}^2}\phi(s) t_\nu(s | \bar{a}_{t-1}, \bar{\sigma}_{t-1}^2) ds \\
&= \mathbb{E} \left[ \frac{(s - \bar{a}_{t-1})}{\bar{\sigma}_{t-1}^2}\phi(s)\right]
\end{split}
\end{equation} where we have used Stein's Lemma in the third row, and Fubini's Theorem in the fifth row. Note that \begin{equation}
t_\nu(s | \bar{a}, \bar{\sigma}^2) = \frac{\Gamma\left(\frac{\nu + 1}{2} \right)}{\sqrt{\pi \nu \bar{\sigma}^2} \Gamma\left(\frac{\nu}{2} \right)}\left( 1 + \frac{(s - \bar{a})^2}{\nu \bar{\sigma}^2} \right)^{-\frac{\nu + 1}{2}}
\end{equation} where we have suppressed the subscript $_{t-1}$ for $\bar{a}$ and $\bar{\sigma}^2$ for brevity. Then, \begin{equation}
\begin{split}
t'_\nu(s | \bar{a}, \bar{\sigma}^2) &= \frac{\Gamma\left(\frac{\nu + 1}{2} \right)}{\sqrt{\pi \nu \bar{\sigma}^2} \Gamma\left(\frac{\nu}{2} \right)}\left( 1 + \frac{(s - \bar{a})^2}{\nu \bar{\sigma}^2} \right)^{-\frac{\nu + 1}{2} - 1} \left(-\frac{\nu + 1}{2} \right) \left( \frac{2(s - \bar{a})}{\nu \bar{\sigma}^2} \right) \\
&= \frac{\Gamma\left(\frac{\nu + 1}{2} \right)}{\sqrt{\pi \nu \bar{\sigma}^2} \Gamma\left(\frac{\nu}{2} \right)}\left( 1 + \frac{(s - \bar{a})^2}{\nu \bar{\sigma}^2} \right)^{-\frac{\nu + 1}{2} - 1} \left(-\frac{\nu + 1}{\nu} \right) \left( \frac{s - \bar{a}}{\bar{\sigma}^2} \right)\\
&= t_\nu(s | \bar{a}, \bar{\sigma}^2) \left( 1 + \frac{(s - \bar{a})^2}{\nu \bar{\sigma}^2} \right)^{-1} \left(-\frac{\nu + 1}{\nu} \right) \left( \frac{s - \bar{a}}{\bar{\sigma}^2} \right)
\end{split}
\end{equation}

\noindent Therefore,
\begin{equation}
\begin{split}
\left(-\frac{\nu}{\nu + 1} \right) \left( 1 + \frac{(s - \bar{a})^2}{\nu \bar{\sigma}^2} \right) t'_\nu(s | \bar{a}, \bar{\sigma}^2) &= t_\nu(s | \bar{a}, \bar{\sigma}^2) \left( \frac{s - \bar{a}}{\bar{\sigma}^2} \right) \\
\left(-\frac{\nu\bar{\sigma}^2 + (s - \bar{a})^2}{(\nu + 1) \bar{\sigma}^2} \right) t'_\nu(s | \bar{a}, \bar{\sigma}^2) &= t_\nu(s | \bar{a}, \bar{\sigma}^2) \left( \frac{s - \bar{a}}{\bar{\sigma}^2} \right)
\end{split}
\end{equation}

\noindent Now, 
\begin{equation}
\begin{split}
\mathbb{E} \left[ \frac{(s - \bar{a})}{\bar{\sigma}^2}\phi(s)\right] & = \int \frac{(s - \bar{a})}{\bar{\sigma}^2}\phi(s) t_\nu(s | \bar{a}, \bar{\sigma}^2) ds \\
&= \int \left(-\frac{\nu\bar{\sigma}^2 + (s - \bar{a})^2}{(\nu + 1) \bar{\sigma}^2} \right) \phi(s) t'_\nu(s | \bar{a}, \bar{\sigma}^2) ds \\
&= \left(-\frac{\nu\bar{\sigma}^2 + (s - \bar{a})^2}{(\nu + 1) \bar{\sigma}^2} \right) \phi(s) t_\nu(s | \bar{a}, \bar{\sigma}^2) \bigg|^{\infty}_{-\infty} - \int \left(\left(-\frac{\nu\bar{\sigma}^2 + (s - \bar{a})^2}{(\nu + 1) \bar{\sigma}^2} \right) \phi(s)\right)' t_\nu(s | \bar{a}, \bar{\sigma}^2) ds \\
&= \int \left(\left(\frac{\nu\bar{\sigma}^2 + (s - \bar{a})^2}{(\nu + 1) \bar{\sigma}^2} \right) \phi(s)\right)' t_\nu(s | \bar{a}, \bar{\sigma}^2) ds
\end{split}
\end{equation} where we have integrated by parts in the third row. We can re-express
\begin{equation}
\begin{split}
\left(\left(\frac{\nu\bar{\sigma}^2 + (s - \bar{a})^2}{(\nu + 1) \bar{\sigma}^2} \right) \phi(s)\right)' &= \left(\frac{\nu\bar{\sigma}^2 + (s - \bar{a})^2}{(\nu + 1) \bar{\sigma}^2} \right)' \phi(s) + \left(\frac{\nu\bar{\sigma}^2 + (s - \bar{a})^2}{(\nu + 1) \bar{\sigma}^2} \right) \phi'(s) \\
&= \left(\frac{2(s - \bar{a})}{(\nu + 1) \bar{\sigma}^2} \right) \phi(s) + \left(\frac{\nu\bar{\sigma}^2 + (s - \bar{a})^2}{(\nu + 1) \bar{\sigma}^2} \right) \phi'(s).
\end{split}
\end{equation}

\noindent Therefore, \begin{equation} \begin{split}
\mathbb{E} \left[ \frac{(s - \bar{a})}{\bar{\sigma}^2}\phi(s)\right] & = \mathbb{E} \left[ \frac{2(s - \bar{a})}{(\nu + 1) \bar{\sigma}^2} \phi(s)\right] + \mathbb{E}\left[\frac{\nu\bar{\sigma}^2 + (s - \bar{a})^2}{(\nu + 1) \bar{\sigma}^2} \phi'(s)\right] \\
\mathbb{E} \left[ \frac{(s - \bar{a})}{\bar{\sigma}^2}\phi(s) - \frac{2(s - \bar{a})}{(\nu + 1) \bar{\sigma}^2} \phi(s)\right] & = \mathbb{E}\left[\frac{\nu\bar{\sigma}^2 + (s - \bar{a})^2}{(\nu + 1) \bar{\sigma}^2} \phi'(s)\right] \\
\mathbb{E} \left[ \frac{(s - \bar{a})}{\bar{\sigma}^2}\phi(s) \left( 1 - \frac{2}{(\nu + 1)}\right) \right] & = \mathbb{E}\left[\frac{\nu\bar{\sigma}^2 + (s - \bar{a})^2}{(\nu + 1) \bar{\sigma}^2} \phi'(s)\right] \\
\frac{\nu - 1}{\nu + 1} \mathbb{E} \left[ \frac{(s - \bar{a})}{\bar{\sigma}^2}\phi(s) \right] & = \mathbb{E}\left[\frac{\nu\bar{\sigma}^2 + (s - \bar{a})^2}{(\nu + 1) \bar{\sigma}^2} \phi'(s)\right]\\
\mathbb{E} \left[ \frac{(s - \bar{a})}{\bar{\sigma}^2}\phi(s) \right] & = \mathbb{E}\left[\frac{\nu\bar{\sigma}^2 + (s - \bar{a})^2}{(\nu - 1) \bar{\sigma}^2} \phi'(s)\right]
\end{split}
\end{equation}

\noindent This gives us the final expression for \begin{equation} \begin{split}
\tilde{g}_{t-1} &= \int \frac{\nu\bar{\sigma}^2_{t-1} + (s-\bar{a}_{t-1})^2}{(\nu - 1)\bar{\sigma}^2_{t-1}} \phi'(s) t_{\nu}(s | \bar{a}_{t-1}, \bar{\sigma}_{t-1}^2)ds \\
&= \frac{\nu}{\nu - 1} \int \phi'(s) t_{\nu}(s | \bar{a}_{t-1}, \bar{\sigma}_{t-1}^2)ds + \frac{1}{\nu - 1} \int \frac{(s-\bar{a}_{t-1})^2}{\bar{\sigma}^2_{t-1}} \phi'(s) t_{\nu}(s | \bar{a}_{t-1}, \bar{\sigma}_{t-1}^2)ds.
\end{split}
\end{equation} Therefore, $\tilde{r}_{t-1}$ is the average activity of neurons under the Student-$t$, while $\tilde{g}_{t-1}$ is the average gain of neurons under the Student-$t$ plus the average of the gain multiplied by the squared $z$-score of the neurons under the Student-$t$ (which we might call the average of the ``variance-weighted gain''). This term suggests that the gain of neurons has a stronger effect on the dynamics when the neuron is a ``rare'' neuron at the tail of the distribution. As $\nu \rightarrow \infty$, $\bar{r}_{t-1} = \tilde{r}_{t-1}$ and $\bar{g}_{t-1} = \tilde{g}_{t-1}$. Note that when $\nu = 1$, this distribution is Cauchy and $\tilde{g}_{t-1}$ is undefined, consistent with the first and second moments of Cauchy being undefined. $\nu >1$ for the above to be valid.

\section{Latent Variable Models and State Space Models}

Many latent variable models (LVMs) in neuroscience have the form
\begin{equation} \label{eq:main_lvm}
\begin{split}
\boldsymbol{r}_t &= \phi(\boldsymbol{M}\boldsymbol{z}_t + \boldsymbol{B}\boldsymbol{v}_t + \boldsymbol{d}), \\
\hat{\boldsymbol{r}}_t &\sim p(\boldsymbol{r}^{data}_t | \boldsymbol{r}_t).
\end{split}
\end{equation} Given some neural population activity $\boldsymbol{r}^{data}_{1:T} \in \mathbb{R}^{K_{obs} \times T}$ and external input $\boldsymbol{v}_{1:T} \in \mathbb{R}^{K_{in} \times T}$, LVMs recover the underlying $\boldsymbol{z}_{1:T} \in \mathbb{R}^{R \times T}$, assuming that the LVM-reconstructed observation $\boldsymbol{r}_t$ at time $t$ depends only on the current $\boldsymbol{z}_t$ and $\boldsymbol{v}_t$, and not on previous time steps. Typically, parameters $\boldsymbol{M} \in \mathbb{R}^{K_{obs} \times R}$, $\boldsymbol{B} \in \mathbb{R}^{K_{obs} \times K_{in}}$ and $\boldsymbol{d} \in \mathbb{R}^{K_{obs}}$ are called loadings, and are directly learned in these models, where $K_{obs}$ is the number of neurons observed in the data. Models such as factor analysis \cite{bartholomew2011latent, yu_gpfa_2008}, and variational latent Gaussian Process \cite{zhao2017variational} are some prominent examples. In addition, if we have a model of how the latent variable $\boldsymbol{z}_t$ evolves over time, \begin{equation} \label{eq:main_ssm}
\tau\frac{\boldsymbol{z}_{t} - \boldsymbol{z}_{t-1}}{\Delta t} = -\boldsymbol{z}_{t-1} + f(\boldsymbol{z}_{t-1}, \boldsymbol{v}_{t-1}) + \boldsymbol{\eta}_{t-1},
\end{equation} then we call them state space models (SSMs). Here, we defined some generic function $f$ that determines the dynamics of the neural population, evolving with some noise $\boldsymbol{\eta} \sim \mathcal{N}(\boldsymbol{0}, \boldsymbol{Q})$. The positive semi-definite $\boldsymbol{Q}$ may or may not be learned depending on the model. The function $f$ is learned in this model along with other parameters. Poisson linear dynamical system (PLDS; \citet{macke11}) is a special case of this form, where $\boldsymbol{B} = \boldsymbol{0}$, and $f$ is linear; recurrent switching linear dynamical system (rSLDS; \citet{linderman17}) instead assumes that $f$ is switching linear (by introducing an additional discrete latent variable that depends on $\boldsymbol{z}$); flow-field inference from neural data using deep recurrent networks (FINDR; \citet{kim2025findr}) assumes that $f$ is a gated multilayer perceptron (MLP). If the observed neural activity $\boldsymbol{r}^{data}_t$ represents spike counts, this is typically modeled with $P(\boldsymbol{r}^{data}_t|\boldsymbol{r}_t) = \textrm{Poisson}(\boldsymbol{r}^{data}_t | \lambda = \Delta t \boldsymbol{r}_t)$. If $\boldsymbol{r}_t$ represents averaged neural activity or principal components of a larger population, or calcium signals, the emission probability distribution $P(\boldsymbol{r}^{data}_t|\boldsymbol{r}_t)$ can be changed accordingly.

Low-rank RNN models that are trained on neural data, such as \citet{valente22} and \citet{pals24} can also be cast into this form. In these models, $\boldsymbol{B} \neq \boldsymbol{0}$ in $\boldsymbol{r}_t = \phi(\boldsymbol{M}\boldsymbol{z}_t + \boldsymbol{B}\boldsymbol{v}_t + \boldsymbol{d})$. They further assume that the parameters $\boldsymbol{M}$, $\boldsymbol{B}$ and $\boldsymbol{d}$ are also the parameters that determine $f$, such that $f(\boldsymbol{z}_{t}, \boldsymbol{v}_{t}) =  \frac{1}{K_{obs}} \boldsymbol{N}^{\top}\phi(\boldsymbol{M}\boldsymbol{z}_{t} + \boldsymbol{B}\boldsymbol{v}_{t} + \boldsymbol{d})$. In addition to $\boldsymbol{M}$, $\boldsymbol{B}$ and $\boldsymbol{d}$, the parameters $\boldsymbol{N}$ are learned.

\section{Identifiability of Latent Variables and Connectivity Distributions} \label{identify}

There are three possible sources that can influence the identifiability of the connectivity distribution $p(\boldsymbol{m}, \boldsymbol{n}, \boldsymbol{b}, \boldsymbol{d}) = p(\boldsymbol{m}, \boldsymbol{b}, \boldsymbol{d})p(\boldsymbol{n} | \boldsymbol{m}, \boldsymbol{b}, \boldsymbol{d})$: {\bf (1)} the identifiability of the latent variable $\boldsymbol{z}$, {\bf (2)} $p(\boldsymbol{m}, \boldsymbol{b}, \boldsymbol{d})$, and {\bf (3)} $p(\boldsymbol{n} | \boldsymbol{m}, \boldsymbol{b}, \boldsymbol{d})$. Here we assume that there is a unique  $\boldsymbol{h}^*_{1:T}$ that minimizes some loss $L(\boldsymbol{r}^{data}_{1:T}, \phi(\boldsymbol{h}_{1:T}))$, and that our training gets us to this $\boldsymbol{h}^*_{1:T}$.

{\bf (1)} Identifiability of $\boldsymbol{z}$: The first source comes from the identifiability of the latent variable $\boldsymbol{z}$. The latent variable $\boldsymbol{z}$ is identifiable only up to affine transformations (noted in e.g., \citet{zhao2017variational}, but generally true for LVMs of the form in Equation~(\ref{eq:main_lvm})). For LVMs of the form in Equation~(\ref{eq:main_lvm}), whether the latent variable $\boldsymbol{z}$ is identifiable is equivalent to asking, 
\begin{equation}
\begin{split}
    \Big( \boldsymbol{h}_{1:T} = \boldsymbol{M}\boldsymbol{z}_{1:T} + \boldsymbol{B}\boldsymbol{v}_{1:T} + \boldsymbol{d}\boldsymbol{1}^\top_T &= \boldsymbol{M}'\boldsymbol{z}'_{1:T} + \boldsymbol{B}'\boldsymbol{v}_{1:T} + \boldsymbol{d}'\boldsymbol{1}^\top_T \Big)\\
    &\Rightarrow \\
    \Big( \boldsymbol{z}_{1:T} &= \boldsymbol{z}'_{1:T} \Big)?
    \end{split}
\end{equation}

Let $\boldsymbol{z}'_t = \boldsymbol{A}\boldsymbol{z}_t + \boldsymbol{C}\boldsymbol{v}_t + \boldsymbol{k}$, where $\boldsymbol{A} \in \mathbb{R}^{R\times R}$ is any invertible matrix, $\boldsymbol{C} \in \mathbb{R}^{R\times R}$ is any matrix, and $\boldsymbol{k} \in \mathbb{R}^{R}$ is any vector. Because we can set $\boldsymbol{M}' = \boldsymbol{M}\boldsymbol{A}^{-1}$ and $\boldsymbol{B}'=\boldsymbol{B} - \boldsymbol{M}\boldsymbol{A}^{-1}\boldsymbol{C}$, and $\boldsymbol{d}' = \boldsymbol{d} - \boldsymbol{M}\boldsymbol{A}^{-1}\boldsymbol{k}$, it is not necessarily that $\boldsymbol{M} = \boldsymbol{M}'$, $\boldsymbol{z}_{1:T} = \boldsymbol{z}'_{1:T}$, and $\boldsymbol{d} = \boldsymbol{d}'$ when $\boldsymbol{M}\boldsymbol{z}_{1:T} + \boldsymbol{B}\boldsymbol{v}_{1:T} + \boldsymbol{d}\boldsymbol{1}^\top_T = \boldsymbol{M}'\boldsymbol{z}'_{1:T} + \boldsymbol{B}'\boldsymbol{v}_{1:T} + \boldsymbol{d}'\boldsymbol{1}^\top_T$.

Therefore, $\boldsymbol{z}_t$ is identifiable only up to affine transformations of the form $\boldsymbol{z}'_t = \boldsymbol{A}\boldsymbol{z}_t + \boldsymbol{C}\boldsymbol{v}_t + \boldsymbol{k}$. Importantly, this implies that $\boldsymbol{M}$ and $\boldsymbol{B}$ are not identifiable in general LVMs. However, in Appendix~\ref{latent_identify}, we show that when we further assume that the latent variable $\boldsymbol{z}_t$ evolves over time with the dynamics function $f$ being an lrRNN, as long as $\boldsymbol{1}^\top_{T} \notin \textrm{rowspan}(\boldsymbol{u}_{1:T})$, and as long as $\boldsymbol{u}_{1:T} \neq \boldsymbol{0}$, $\boldsymbol{B}$ and $\boldsymbol{d}$ are identifiable. Even in this case, $\boldsymbol{z}$ is identifiable only up to the linear transformation $\boldsymbol{z}'_t = \boldsymbol{A}\boldsymbol{z}_t$. For a given linear transformation $\boldsymbol{z}'_t = \boldsymbol{A}\boldsymbol{z}_t$, there is a corresponding transformation for $p(\boldsymbol{m}, \boldsymbol{n}, \boldsymbol{b}, \boldsymbol{d})$. In Appendix~\ref{id_connectivity}, we derive this transformation. This makes $p(\boldsymbol{m}, \boldsymbol{n}, \boldsymbol{b}, \boldsymbol{d})$ identifiable only up to the transformation shown in Appendix~\ref{id_connectivity}.

{\bf (2)} $p(\boldsymbol{m}, \boldsymbol{b}, \boldsymbol{d})$: Suppose that $\boldsymbol{z}_{1:T}$ is already determined. Below, we show that even when $\boldsymbol{z}_{1:T}$ is known, $\boldsymbol{M}$, $\boldsymbol{B}$, and $\boldsymbol{d}$ are identifiable only under certain conditions. Let's rewrite $\boldsymbol{M}\boldsymbol{z}_{1:T} + \boldsymbol{B}\boldsymbol{v}_{1:T} + \boldsymbol{d}\boldsymbol{1}^\top_T$ as \begin{equation}
\begin{split}
\boldsymbol{M}\boldsymbol{z}_{1:T} + \boldsymbol{B}\boldsymbol{v}_{1:T} + \boldsymbol{d}\boldsymbol{1}^\top_T &= \tilde{\boldsymbol{M}}\tilde{\boldsymbol{z}}_{1:T} + \boldsymbol{d}\boldsymbol{1}^\top_T, \\ \tilde{\boldsymbol{M}} &= \begin{bmatrix}\boldsymbol{M} & \boldsymbol{B}\end{bmatrix}, \\ \tilde{\boldsymbol{z}}_{1:T} &= \begin{bmatrix}
    \boldsymbol{z}_{1:T} \\ \boldsymbol{v}_{1:T}
\end{bmatrix}.
\end{split}
\end{equation}

Is it true that \begin{equation}
\left(\tilde{\boldsymbol{M}}\tilde{\boldsymbol{z}}_{1:T} + \boldsymbol{d}\boldsymbol{1}^\top_T = \tilde{\boldsymbol{M}}'\tilde{\boldsymbol{z}}_{1:T} + \boldsymbol{d}'\boldsymbol{1}^\top_T \right) \Rightarrow
\left(\tilde{\boldsymbol{M}} = \tilde{\boldsymbol{M}}', \boldsymbol{d} = \boldsymbol{d}' \right) ?
\end{equation} This is equivalent to asking, \begin{equation}
\left( \tilde{\boldsymbol{M}}\tilde{\boldsymbol{z}}_{1:T} + \boldsymbol{d}\boldsymbol{1}^\top_T = \boldsymbol{0} \right) \Rightarrow 
    \left( \tilde{\boldsymbol{M}} = \boldsymbol{0}, \boldsymbol{d} = \boldsymbol{0} \right) ?
\end{equation} If the left-hand side is true, then\begin{equation}
\tilde{\boldsymbol{M}}\tilde{\boldsymbol{z}}_{1:T} = - \boldsymbol{d}\boldsymbol{1}^\top_T,
\end{equation} which means that the $i$-th row is \begin{equation}
\tilde{\boldsymbol{m}}_i\tilde{\boldsymbol{z}}_{1:T} = - \boldsymbol{d}_i\boldsymbol{1}^\top_T.
\end{equation} What this implies is that when $\boldsymbol{1}^\top_T \in \textrm{rowspan}(\tilde{\boldsymbol{z}}_{1:T})$, it is not necessarily that $\tilde{\boldsymbol{M}} = \boldsymbol{0}$ and $\boldsymbol{d} = \boldsymbol{0}$ when $\tilde{\boldsymbol{M}}\tilde{\boldsymbol{z}}_{1:T} + \boldsymbol{d}\boldsymbol{1}^\top_T = \boldsymbol{0}$, thus $\boldsymbol{M}$, $\boldsymbol{B}$, and $\boldsymbol{d}$ are not uniquely identifiable. If $\boldsymbol{1}^\top_T \notin \textrm{rowspan}(\tilde{\boldsymbol{z}}_{1:T})$, necessarily, $\boldsymbol{d} =\boldsymbol{0}$. In such a case, for $\tilde{\boldsymbol{M}}\tilde{\boldsymbol{z}}_{1:T} = \boldsymbol{0} \Rightarrow \tilde{\boldsymbol{M}} = \boldsymbol{0}$, the right inverse $\tilde{\boldsymbol{z}}_{1:T}(\tilde{\boldsymbol{z}}_{1:T}\tilde{\boldsymbol{z}}^\top_{1:T})^{-1}$ should exist. The condition for existence is that $\textrm{rank}(\tilde{\boldsymbol{z}}_{1:T}) = R + K_{in}$. Therefore, given $\boldsymbol{z}_{1:T}$, $\boldsymbol{M}$, $\boldsymbol{B}$, and $\boldsymbol{d}$ are identifiable only when $\textrm{rank}(\tilde{\boldsymbol{z}}_{1:T}) = R + K_{in}$ and when $\boldsymbol{1}^\top_T \notin \textrm{rowspan}(\tilde{\boldsymbol{z}}_{1:T})$.

If $\boldsymbol{m}_i$, $\boldsymbol{b}_i$ and $\boldsymbol{d}_i$ are the $i$-th rows of $\boldsymbol{M}$, $\boldsymbol{B}$, and $\boldsymbol{d}$, and if they are taken to be i.i.d. samples from the probability distribution $p(\boldsymbol{m}, \boldsymbol{b}, \boldsymbol{d})$, then $p(\boldsymbol{m}, \boldsymbol{b}, \boldsymbol{d})$ is identifiable (up to Equation~(\ref{eq:p_transformation})) in the limit of infinite data (i.e., $K_{obs} \rightarrow \infty$ and $T \geq R + K_{in}$) as long as $\textrm{rank}(\tilde{\boldsymbol{z}}_{1:T}) = R + K_{in}$ and $\boldsymbol{1}^\top_T \notin \textrm{rowspan}(\tilde{\boldsymbol{z}}_{1:T})$. In practice, we can check whether $\textrm{rank}(\tilde{\boldsymbol{z}}_{1:T}) = R + K_{in}$ and $\boldsymbol{1}^\top_T \notin \textrm{rowspan}(\tilde{\boldsymbol{z}}_{1:T})$ after training LVM.

Even when $\boldsymbol{m}_i$, $\boldsymbol{b}_i$ and $\boldsymbol{d}_i$ are not taken to be i.i.d. samples from $p(\boldsymbol{m}, \boldsymbol{b}, \boldsymbol{d})$ and $K_obs$ is finite, the loadings $\boldsymbol{B}$ and $\boldsymbol{d}$ are uniquely identifiable, and $\boldsymbol{M}$ is identifiable up to linear transformations. As we show in Appendix~\ref{latent_identify}--\ref{general_identify}, lrRNN parameters are more identifiable than general SSMs/LVMs.

\subsection{Identifiability of the latent variable $\boldsymbol{z}$ when the LVM is an lrRNN} \label{latent_identify}

We start with Equation~(\ref{eq:main_zlrrnn}). Let $\boldsymbol{z}'_t = \boldsymbol{A}\boldsymbol{z}_t + \boldsymbol{C}\boldsymbol{v}_t + \boldsymbol{k}$. Then, $\boldsymbol{z}_t = \boldsymbol{A}^{-1}(\boldsymbol{z}'_t - \boldsymbol{C}\boldsymbol{v}_t - \boldsymbol{k})$. Plugging this into Equation~(\ref{eq:main_zlrrnn}), we get \begin{equation}\begin{split}
\tau\frac{\boldsymbol{A}^{-1}(\boldsymbol{z}'_t - \boldsymbol{C}\boldsymbol{v}_t) - \boldsymbol{A}^{-1}(\boldsymbol{z}'_{t-1} - \boldsymbol{C}\boldsymbol{v}_{t-1})}{\Delta t} = &-\boldsymbol{A}^{-1}(\boldsymbol{z}'_{t-1} - \boldsymbol{C}\boldsymbol{v}_{t-1} - \boldsymbol{k}) \\
&+\frac{1}{K} \boldsymbol{N}\phi(\boldsymbol{M}\boldsymbol{A}^{-1}(\boldsymbol{z}'_{t-1} - \boldsymbol{C}\boldsymbol{v}_{t-1} - \boldsymbol{k}) + \boldsymbol{B}\boldsymbol{v}_{t-1} + \boldsymbol{d}).
\end{split}
\end{equation} We can further simplify this into
\begin{equation}\begin{split}
\tau\frac{\boldsymbol{z}'_t - \boldsymbol{z}'_{t-1}}{\Delta t} = &-\boldsymbol{z}'_{t-1} + \boldsymbol{C}\boldsymbol{v}_{t-1} + \boldsymbol{k} + \tau\frac{\boldsymbol{C}\boldsymbol{v}_t - \boldsymbol{C}\boldsymbol{v}_{t-1}}{\Delta t} \\
&+\frac{1}{K} \boldsymbol{A}\boldsymbol{N}\phi(\boldsymbol{M}\boldsymbol{A}^{-1}\boldsymbol{z}'_{t-1} + (\boldsymbol{B} - \boldsymbol{M}\boldsymbol{A}^{-1}\boldsymbol{C})\boldsymbol{v}_{t-1} + (\boldsymbol{d} - \boldsymbol{M}\boldsymbol{A}^{-1}\boldsymbol{k})).
\end{split}
\end{equation} Due to Equation~(\ref{eq:low-pass_filter}), this is equivalent to \begin{equation}\begin{split}
\tau\frac{\boldsymbol{z}'_t - \boldsymbol{z}'_{t-1}}{\Delta t} = &-\boldsymbol{z}'_{t-1} + \boldsymbol{C}\boldsymbol{u}_{t} + \boldsymbol{k} + \frac{1}{K} \boldsymbol{A}\boldsymbol{N}\phi(\boldsymbol{M}\boldsymbol{A}^{-1}\boldsymbol{z}'_{t-1} + (\boldsymbol{B} - \boldsymbol{M}\boldsymbol{A}^{-1}\boldsymbol{C})\boldsymbol{v}_{t-1} + (\boldsymbol{d} - \boldsymbol{M}\boldsymbol{A}^{-1}\boldsymbol{k})).
\end{split}
\end{equation} If we set $\boldsymbol{N}' = \boldsymbol{A}\boldsymbol{N}$, $\boldsymbol{M}' = \boldsymbol{M}\boldsymbol{A}^{-1}$, $\boldsymbol{B}' = \boldsymbol{B} - \boldsymbol{M}\boldsymbol{A}^{-1}\boldsymbol{C}$, and $\boldsymbol{d}' = \boldsymbol{d} - \boldsymbol{M}\boldsymbol{A}^{-1}\boldsymbol{k}$, then we have \begin{equation}\begin{split}
\tau\frac{\boldsymbol{z}'_t - \boldsymbol{z}'_{t-1}}{\Delta t} = &-\boldsymbol{z}'_{t-1} + \boldsymbol{C}\boldsymbol{u}_{t} + \boldsymbol{k} + \frac{1}{K} \boldsymbol{N}'\phi(\boldsymbol{M}'\boldsymbol{z}'_{t-1} + \boldsymbol{B}'\boldsymbol{v}_{t-1} + \boldsymbol{d}').
\end{split}
\end{equation} For this Equation to be of the form in Equation~(\ref{eq:main_zlrrnn}), we should have $\boldsymbol{C}\boldsymbol{u}_{t} + \boldsymbol{k} = \boldsymbol{0}$. Similar to the reasoning in {\bf (2)} above, this implies that we should have $\boldsymbol{1}^\top_{T} \notin \textrm{rowspan}(\boldsymbol{u}_{1:T})$ and $\boldsymbol{u}_{1:T} \neq \boldsymbol{0}$ in order for $\boldsymbol{C} = \boldsymbol{k} = \boldsymbol{0}$, and therefore for $\boldsymbol{B}$ and $\boldsymbol{d}$ to be identifiable. One corollary of this result is that if we have constant input $\boldsymbol{u}$, then $\boldsymbol{M}$, $\boldsymbol{B}$, and $\boldsymbol{d}$ trade off with each other and are therefore not identifiable.

\subsection{Identifiability of the latent variable $\boldsymbol{z}$ when the LVM is an SSM} \label{general_identify}

Let $f$ in Equation~(\ref{eq:main_ssm}) be linear, switching-linear or an MLP. Similar to Appendix~\ref{latent_identify}, we plug in $\boldsymbol{z}_t = \boldsymbol{A}^{-1}(\boldsymbol{z}'_t - \boldsymbol{C}\boldsymbol{v}_t - \boldsymbol{k})$ to $\tau \frac{\boldsymbol{z}_t - \boldsymbol{z}_{t-1}}{\Delta t} = -\boldsymbol{z}_{t-1} + f(\boldsymbol{z}_{t-1}, \boldsymbol{v}_{t-1})$: \begin{equation}
\tau\frac{\boldsymbol{A}^{-1}(\boldsymbol{z}'_t - \boldsymbol{C}\boldsymbol{v}_t) - \boldsymbol{A}^{-1}(\boldsymbol{z}'_{t-1} - \boldsymbol{C}\boldsymbol{v}_{t-1})}{\Delta t} = - \boldsymbol{A}^{-1}(\boldsymbol{z}'_{t-1} - \boldsymbol{C}\boldsymbol{v}_{t-1} - \boldsymbol{k}) + f(\boldsymbol{A}^{-1}(\boldsymbol{z}'_{t-1} - \boldsymbol{C}\boldsymbol{v}_{t-1} - \boldsymbol{k}), \boldsymbol{v}_{t-1}).
\end{equation} This simplifies to \begin{equation} \begin{split}
\tau\frac{\boldsymbol{z}'_t - \boldsymbol{z}'_{t-1}}{\Delta t} &= - \boldsymbol{z}'_{t-1} + \boldsymbol{C}\boldsymbol{v}_{t-1} + \tau\frac{\boldsymbol{C}\boldsymbol{v}_t - \boldsymbol{C}\boldsymbol{v}_{t-1}}{\Delta t} +\boldsymbol{k} + \boldsymbol{A}f(\boldsymbol{A}^{-1}(\boldsymbol{z}'_{t-1} - \boldsymbol{C}\boldsymbol{v}_{t-1} - \boldsymbol{k}), \boldsymbol{v}_{t-1}) \\
& = - \boldsymbol{z}'_{t-1} + \boldsymbol{C}\boldsymbol{u}_{t} + \boldsymbol{k} + f'(\boldsymbol{z}'_{t-1} - \boldsymbol{C}\boldsymbol{v}_{t-1} - \boldsymbol{k}, \boldsymbol{v}_{t-1}).
\end{split}
\end{equation} If $f$ is linear, $f'$ is also linear. If $f$ is switching-linear or MLP, $f'$ is also switching-linear or MLP, respectively. However, in many cases, linear, switching-linear, or MLP models have bias terms which may trade off with $\boldsymbol{k}$. Therefore, the latent $\boldsymbol{z}_t$ is identifiable up to $\boldsymbol{z}'_t = \boldsymbol{A}\boldsymbol{z}_t + \boldsymbol{k}$ as long as $\boldsymbol{1}^\top_{T} \notin \textrm{rowspan}(\boldsymbol{u}_{1:T})$. When $\boldsymbol{1}^\top_{T} \notin \textrm{rowspan}(\boldsymbol{u}_{1:T})$ and $\boldsymbol{u}_{1:T} \neq \boldsymbol{0}$, $\boldsymbol{C}$ must be $\boldsymbol{0}$.

\subsection{Identifiability of the connectivity distribution} \label{id_connectivity}

Due to Appendix~\ref{latent_identify}, the latent variable $\boldsymbol{z}$ is identifiable up to invertible linear transformations: $\boldsymbol{z}'_t = \boldsymbol{A}\boldsymbol{z}_t$. Then, plugging in $\boldsymbol{z}_t = \boldsymbol{A}^{-1}\boldsymbol{z}'_t$ into Equation~(\ref{eq:lrRNNeq}), \begin{equation}
\tau\frac{\boldsymbol{A}^{-1}\boldsymbol{z}'_t - \boldsymbol{A}^{-1}\boldsymbol{z}'_{t-1}}{\Delta t} = -\boldsymbol{A}^{-1}\boldsymbol{z}'_{t-1} + \frac{1}{K} \sum_{i=1}^K \boldsymbol{n}_i \phi (\boldsymbol{m}_i^\top \boldsymbol{A}^{-1}\boldsymbol{z}'_{t-1} + \boldsymbol{b}_i^\top \boldsymbol{v}_{t-1} + \boldsymbol{d}_i).
\end{equation} Multiplying both sides by $\boldsymbol{A}$, the dynamics in the transformed space of $\boldsymbol{z}'_t$ is 
\begin{equation}
\begin{split}
\tau \frac{\boldsymbol{z}'_t - \boldsymbol{z}'_{t-1}}{\Delta t} &= -\boldsymbol{z}'_{t-1} + \frac{1}{K}\sum_{i=1}^K \boldsymbol{A}\boldsymbol{n}_i \phi(\boldsymbol{m}_i^\top\boldsymbol{A}^{-1}\boldsymbol{z}'_{t-1} + \boldsymbol{b}^\top_i \boldsymbol{v}_{t-1} + \boldsymbol{d}_i) \\
&= -\boldsymbol{z}'_{t-1} + \frac{1}{K}\sum_{i=1}^K \boldsymbol{n}'_i\phi(\boldsymbol{m}'^\top_i\boldsymbol{z}'_{t-1} + \boldsymbol{b}'^\top_i \boldsymbol{v}_{t-1} + \boldsymbol{d}'_i),
\end{split}
\end{equation} where $\boldsymbol{n}' = \boldsymbol{A}\boldsymbol{n}$, $\boldsymbol{m}' = \boldsymbol{A}^{-1\top}\boldsymbol{m}$, $\boldsymbol{b}' = \boldsymbol{b}$ and $\boldsymbol{d}' = \boldsymbol{d}$. To find the transformation that the connectivity distribution $p(\boldsymbol{m}, \boldsymbol{n}, \boldsymbol{b}, \boldsymbol{d})=p(\boldsymbol{n}| \boldsymbol{m}, \boldsymbol{b}, \boldsymbol{d})p(\boldsymbol{m}, \boldsymbol{b}, \boldsymbol{d})$ is identifiable up to, we need to solve the following problem: how should $p(\boldsymbol{m}, \boldsymbol{n}, \boldsymbol{b}, \boldsymbol{d})$ be transformed when the latent variable $\boldsymbol{z}_t$ is transformed to $\boldsymbol{z}'_t$? \\

\noindent Let \begin{equation}
    \Theta = \begin{bmatrix}
    \boldsymbol{m}\\
    \boldsymbol{n}\\
    \boldsymbol{b}\\
    \boldsymbol{d}
\end{bmatrix}, \quad \Theta' = \begin{bmatrix}
    \boldsymbol{m}'\\
    \boldsymbol{n}'\\
    \boldsymbol{b}'\\
    \boldsymbol{d}'
\end{bmatrix} = \begin{bmatrix}
\boldsymbol{A}^{-1\top}\boldsymbol{m}\\
\boldsymbol{A}\boldsymbol{n}\\
    \boldsymbol{b}\\
    \boldsymbol{d}
\end{bmatrix} = \begin{bmatrix}
\boldsymbol{A}^{-1\top} & \boldsymbol{0} & \boldsymbol{0} & \boldsymbol{0}\\
\boldsymbol{0} & \boldsymbol{A} & \boldsymbol{0} & \boldsymbol{0}\\
\boldsymbol{0} & \boldsymbol{0} & \boldsymbol{I} & \boldsymbol{0}\\
\boldsymbol{0}& \boldsymbol{0}&\boldsymbol{0}&\boldsymbol{I}
\end{bmatrix} \begin{bmatrix}
\boldsymbol{m}\\
\boldsymbol{n}\\
    \boldsymbol{b}\\
    \boldsymbol{d}
\end{bmatrix}.
\end{equation} This means that \begin{equation}
    \Theta = \begin{bmatrix}
    \boldsymbol{m}\\
    \boldsymbol{n}\\
    \boldsymbol{b}\\
    \boldsymbol{d}
\end{bmatrix} = \begin{bmatrix}
\boldsymbol{A}^\top \boldsymbol{m}'\\
\boldsymbol{A}^{-1}\boldsymbol{n}'\\
\boldsymbol{b}'\\
\boldsymbol{d}'
\end{bmatrix} = \begin{bmatrix}
\boldsymbol{A}^\top & \boldsymbol{0} & \boldsymbol{0} & \boldsymbol{0}\\
\boldsymbol{0} & \boldsymbol{A}^{-1} & \boldsymbol{0} & \boldsymbol{0}\\
\boldsymbol{0} & \boldsymbol{0} & \boldsymbol{I} & \boldsymbol{0}\\
\boldsymbol{0} & \boldsymbol{0}&\boldsymbol{0}&\boldsymbol{I}
\end{bmatrix} \begin{bmatrix}
\boldsymbol{m}'\\
\boldsymbol{n}'\\
    \boldsymbol{b}'\\
    \boldsymbol{d}'
\end{bmatrix} = \boldsymbol{J}\Theta'.
\end{equation} By change of variables \begin{equation}
p'(\Theta') = p(\boldsymbol{J}\Theta')|\det \boldsymbol{J}|,
\end{equation} which gives us \begin{equation} \label{eq:P_identify} \begin{split}
p'(\boldsymbol{m}',\boldsymbol{n}',\boldsymbol{b}',\boldsymbol{d}') &= p(\boldsymbol{A}^\top\boldsymbol{m}',\boldsymbol{A}^{-1}\boldsymbol{n}',\boldsymbol{b}',\boldsymbol{d}')|\det \boldsymbol{A}^\top||\det \boldsymbol{A}^{-1}|\\
 &= p(\boldsymbol{A}^\top\boldsymbol{m}',\boldsymbol{A}^{-1}\boldsymbol{n}',\boldsymbol{b}',\boldsymbol{d}')
\end{split}
\end{equation} We can also decompose \begin{equation}
\begin{split}
p'(\boldsymbol{m}',\boldsymbol{n}',\boldsymbol{b}',\boldsymbol{d}') &= p'(\boldsymbol{n}'| \boldsymbol{m}',\boldsymbol{b}',\boldsymbol{d}')P'(\boldsymbol{m}',\boldsymbol{b}',\boldsymbol{d}')\\
&=p(\boldsymbol{A}^{-1}\boldsymbol{n}'|\boldsymbol{A}^\top\boldsymbol{m}',\boldsymbol{b}',\boldsymbol{d}')p(\boldsymbol{A}^\top\boldsymbol{m}',\boldsymbol{b}',\boldsymbol{d}')
\end{split}
\end{equation} Therefore, when the latent variable $\boldsymbol{z}_t$ is identifiable up to linear transformations, the connectivity distribution is identifiable up to the transformation in Equation~(\ref{eq:P_identify}). If we require $\boldsymbol{M}$ to be semi-orthogonal, then $\boldsymbol{A}$ must be an orthogonal transformation. Thus in such a case, the connectivity distribution $p(\boldsymbol{m}, \boldsymbol{n}, \boldsymbol{b}, \boldsymbol{d})$ is identifiable up to rotations and reflections of $\boldsymbol{m}$ and $\boldsymbol{n}$.

\subsection{Effective connectivity in Valente et al., 2022} \label{remark}

In Appendix A.4 of \citet{valente22}, the authors mention that all components of $\boldsymbol{n}$ orthogonal to the subspace spanned by $\boldsymbol{m}$, $\boldsymbol{b}$ and $\boldsymbol{d}$ are irrelevant for the dynamics, which is true when $\phi$ is a linear function. To see why, note that for some finite $K$, \begin{equation}
\begin{split}
\boldsymbol{z}^{rec}_{t-1} &= \frac{1}{K} \boldsymbol{N}^\top \phi (\boldsymbol{M} \boldsymbol{z}_{t-1} + \boldsymbol{B} \boldsymbol{v}_{t-1} + \boldsymbol{d}) \\
&= \frac{1}{K}\left[ \boldsymbol{N}_{\perp}^\top \phi (\boldsymbol{M} \boldsymbol{z}_{t-1} + \boldsymbol{B} \boldsymbol{v}_{t-1} + \boldsymbol{d}) + \boldsymbol{N}_{\parallel}^\top \phi (\boldsymbol{M} \boldsymbol{z}_{t-1} + \boldsymbol{B} \boldsymbol{v}_{t-1} + \boldsymbol{d})\right],
\end{split}
\end{equation} where $\boldsymbol{N}_{\parallel}$ is the component of $\boldsymbol{N}$ projected to the span of $\boldsymbol{m}$, $\boldsymbol{b}$ and $\boldsymbol{d}$, and $\boldsymbol{N}_{\perp}$ is the component of $\boldsymbol{N}$ orthogonal to the span of $\boldsymbol{m}$, $\boldsymbol{b}$ and $\boldsymbol{d}$. If $\phi(x)$ is linear (e.g., $\phi(x) = \gamma x$ for some scalar $\gamma$), then, by definition, $\gamma\boldsymbol{N}^\top_\perp \boldsymbol{M} = \boldsymbol{0}$, $\gamma\boldsymbol{N}^\top_\perp \boldsymbol{B} = \boldsymbol{0}$, $\gamma\boldsymbol{N}^\top_\perp \boldsymbol{d} = \boldsymbol{0}$, and thus $\boldsymbol{N}$ really does affect the dynamics $\boldsymbol{z}_{rec}$ only with its component that spans $\boldsymbol{m}$, $\boldsymbol{b}$ and $\boldsymbol{d}$: \begin{equation}
\begin{split}
\boldsymbol{z}_{t-1}^{rec} &= \frac{1}{K} \gamma\boldsymbol{N}^\top (\boldsymbol{M} \boldsymbol{z}_{t-1} + \boldsymbol{B} \boldsymbol{v}_{t-1} + \boldsymbol{d}) \\
&= \frac{1}{K} \gamma \boldsymbol{N}_{\parallel}^\top (\boldsymbol{M} \boldsymbol{z}_{t-1} + \boldsymbol{B} \boldsymbol{v}_{t-1} + \boldsymbol{d}).
\end{split}
\end{equation} However, in the more general case where $\phi$ is {\it not} linear, then $\boldsymbol{N}_{\perp}^\top \phi (\boldsymbol{M} \boldsymbol{z} + \boldsymbol{B} \boldsymbol{v} + \boldsymbol{d}) \neq \boldsymbol{0}$. Note that if $\boldsymbol{m}$, $\boldsymbol{n}$ are distributed as jointly Gaussian, and $K$ approaches infinity, \begin{equation}
\begin{split}
\lim_{K\to\infty} \frac{1}{K} \boldsymbol{N}_{\perp}^\top \phi (\boldsymbol{M} \boldsymbol{z}_{t-1} + \boldsymbol{B} \boldsymbol{v}_{t-1} + \boldsymbol{d}) &= \mathbb{E}[\boldsymbol{n}_\perp] \mathbb{E}[\phi (\boldsymbol{m}^\top \boldsymbol{z}_{t-1} + \boldsymbol{b}^\top \boldsymbol{v}_{t-1} + \boldsymbol{d})] = \bar{r}_{t-1}\boldsymbol{a}_{n_\perp}, \\
\lim_{K\to\infty} \frac{1}{K} \boldsymbol{N}_{\parallel}^\top \phi (\boldsymbol{M} \boldsymbol{z}_{t-1} + \boldsymbol{B} \boldsymbol{v}_{t-1} + \boldsymbol{d}) &= \bar{r}_{t-1}\boldsymbol{a}_{n_\parallel} + g_{t-1}[\boldsymbol{\Sigma}_{nm}\boldsymbol{z}_{t-1} + \boldsymbol{\Sigma}_{nb}\boldsymbol{v}_{t-1} + \boldsymbol{\Sigma}_{nd}],
\end{split}
\end{equation} by following the derivation in \ref{beiran_proof}. Since $\bar{r}_{t-1}\boldsymbol{a}_{n} = \bar{r}_{t-1}\boldsymbol{a}_{n_\perp} + \bar{r}_{t-1}\boldsymbol{a}_{n_\parallel}$, we arrive at Equation~(\ref{eq:final_beiran}).

Therefore, $\boldsymbol{N}_\perp$ may be relevant to the dynamics, when $\phi$ is nonlinear. 

\newpage
\section{Connector: \underline{Connect}ivity distributions of l\underline{o}w-rank \underline{R}NNs from neural population dynamics}

\subsection{Additional details for Section~\ref{main_fm}} \label{fm}

In Section~\ref{main_fm}, we primarily used continuous normalizing flows (CNFs) to model $p(\boldsymbol{m}, \boldsymbol{b}, \boldsymbol{d})$. When the number of observed neurons is low, estimating $p(\boldsymbol{m}, \boldsymbol{b}, \boldsymbol{d})$ using CNFs can lead to overfitting. In such a case, having a strong prior on the parametric form of $p(\boldsymbol{m}, \boldsymbol{b}, \boldsymbol{d})$ can be helpful. One possible choice is the Gaussian Mixture Model (GMM). Note that maximum entropy $p(\boldsymbol{n}|\boldsymbol{m},\boldsymbol{b},\boldsymbol{d})$ given $\boldsymbol{S}$ is $\mathcal{N}(\boldsymbol{\mu}(\boldsymbol{m},\boldsymbol{b},\boldsymbol{d}), \boldsymbol{S})$. Thus, if $p(\boldsymbol{m}, \boldsymbol{b}, \boldsymbol{d})$ is a mixture of Gaussians, $p(\boldsymbol{m},\boldsymbol{n},\boldsymbol{b},\boldsymbol{d}) = p(\boldsymbol{m}, \boldsymbol{b}, \boldsymbol{d})p(\boldsymbol{n}|\boldsymbol{m},\boldsymbol{b},\boldsymbol{d})$ is also a mixture of Gaussians. When $p(\boldsymbol{m},\boldsymbol{n},\boldsymbol{b},\boldsymbol{d})$ is a mixture of Gaussians, it can be shown that the low-dimensional mean-field dynamics of lrRNN in Equation~(\ref{eq:main_mf-lrrnn}) can be further simplified and expressed in terms of GMM parameters (\citet{beiran21}; see Appendix~\ref{beiran_proof} for our version of the derivation). This expression allows us to interpret the low-dimensional mean-field dynamics equation of the lrRNN as an effective neural circuit model consisting of effective input and effective coupling of the latent variables, where the effective input is modulated by the average activity of neurons, and effective coupling is modulated by the average gain of neurons (Appendix~\ref{mog_proof}, Equation~(\ref{eq:mog})). 

\subsection{Additional details for Section~\ref{main_inference}} \label{inference}

Suppose the neural activity data $\boldsymbol{r}^{data}_{1:T} \in \mathbb{R}^{K_{obs} \times T}$, external input $\boldsymbol{v}_{1:T} \in \mathbb{R}^{K_{in} \times T}$, and the loadings $\boldsymbol{M}$, $\boldsymbol{B}$ and $\boldsymbol{d}$ are given. Then the posterior $p(\boldsymbol{N}|\boldsymbol{M}, \boldsymbol{B}, \boldsymbol{d}, \boldsymbol{z}_{1:T}, \boldsymbol{v}_{1:T})$ can be expressed as 
\begin{equation}
\begin{split}
p(\boldsymbol{N}|\boldsymbol{M}, \boldsymbol{B}, \boldsymbol{d}, \boldsymbol{z}_{1:T}, \boldsymbol{v}_{1:T}) = &\ p(\boldsymbol{z}_{1:T} | \boldsymbol{N}, \boldsymbol{M}, \boldsymbol{B}, \boldsymbol{d}, \boldsymbol{v}_{1:T}) p(\boldsymbol{N}|\boldsymbol{M}, \boldsymbol{B}, \boldsymbol{d}, \boldsymbol{v}_{1:T}) / p(\boldsymbol{z}_{1:T} | \boldsymbol{M}, \boldsymbol{B}, \boldsymbol{d}, \boldsymbol{v}_{1:T}) \\
\propto &\ p(\boldsymbol{z}_{1:T} | \boldsymbol{N}, \boldsymbol{M}, \boldsymbol{B}, \boldsymbol{d}, \boldsymbol{v}_{1:T}) p(\boldsymbol{N}|\boldsymbol{M}, \boldsymbol{B}, \boldsymbol{d}, \boldsymbol{v}_{1:T})\\
= &\ p(\boldsymbol{z}_{1} |\boldsymbol{N}, \boldsymbol{M}, \boldsymbol{B}, \boldsymbol{d}) \prod_{t=1}^{T-1} p(\boldsymbol{z}_{t+1} |\boldsymbol{z}_{t}, \boldsymbol{N}, \boldsymbol{M}, \boldsymbol{B}, \boldsymbol{d}, \boldsymbol{v}_{t}) \prod_{i=1}^K p(\boldsymbol{n}_i|\boldsymbol{m}_i, \boldsymbol{b}_i, \boldsymbol{d}_i, \boldsymbol{v}_{1:T})\\
=&\ p(\boldsymbol{z}_{1} |\boldsymbol{N}, \boldsymbol{M}, \boldsymbol{B}, \boldsymbol{d})\\
&\prod_{t=1}^{T-1} \mathcal{N}(\boldsymbol{z}_{t+1} |\boldsymbol{z}_{t} + \alpha(-\boldsymbol{z}_{t} + \frac{1}{K} \boldsymbol{N}^\top \phi(\boldsymbol{M}\boldsymbol{z}_{t} +\boldsymbol{B}\boldsymbol{v}_{t} + \boldsymbol{d})), \boldsymbol{Q}) \\
&\prod_{i=1}^K \mathcal{N}(\boldsymbol{n}_i|\boldsymbol{\mu}_0, \boldsymbol{Q}_0),
\end{split}
\end{equation} 
where $\boldsymbol{Q}$ is either given by the latent variable model or assumed, and $\alpha = \Delta t / \tau$. Also, the parameters $\boldsymbol{\mu}_0 \in \mathbb{R}^R$ and $\boldsymbol{Q}_0 \in \mathbb{R}^{R \times R}$ for the prior $p(\boldsymbol{n}_i|\boldsymbol{m}_i, \boldsymbol{b}_i, \boldsymbol{d}_i, \boldsymbol{v}_{1:T})$ are assumed to be given. Consistent with the formulation in Equation~(\ref{eq:mf-lr-RNN}), we have assumed that the rows of the matrix $\boldsymbol{N}$ are independent. Taking logs, \begin{equation}
\begin{split}    -\log p(\boldsymbol{N}|\boldsymbol{M}, \boldsymbol{B}, \boldsymbol{d}, \boldsymbol{z}_{1:T}, \boldsymbol{v}_{1:T})
= &\ -\log p(\boldsymbol{z}_{1} |\boldsymbol{N}, \boldsymbol{M}, \boldsymbol{B}, \boldsymbol{d})+\\
&\sum_{t=1}^{T-1} \Bigg[ -\log \mathcal{N}(\boldsymbol{z}_{t+1} |\boldsymbol{z}_{t} + \alpha(-\boldsymbol{z}_{t} + \frac{1}{K} \boldsymbol{N}^\top \phi(\boldsymbol{M}\boldsymbol{z}_{t}+\boldsymbol{B}\boldsymbol{v}_t + \boldsymbol{d})), \boldsymbol{Q}) \Bigg] + \\
&\sum_{i=1}^K \Bigg[ -\log \mathcal{N}(\boldsymbol{n}_i|\boldsymbol{\mu}_0, \boldsymbol{Q}_0) \Bigg] + \textrm{constant}\\
= &\sum_{t=1}^{T-1} \Bigg[ \frac{1}{2}(\boldsymbol{w}_t - \frac{\alpha}{K}\boldsymbol{N}^\top \boldsymbol{r}_{t})^\top \boldsymbol{Q}^{-1} (\boldsymbol{w}_t - \frac{\alpha}{K}\boldsymbol{N}^\top \boldsymbol{r}_{t}) \Bigg] +\\
&\sum_{i=1}^K \Bigg[ \frac{1}{2}(\boldsymbol{n}_i -\boldsymbol{\mu}_0)^\top \boldsymbol{Q}_0^{-1} (\boldsymbol{n}_i -\boldsymbol{\mu}_0)\Bigg] + \textrm{constant} \\
= &\ \frac{1}{2} \textrm{tr}\Bigg[ (\boldsymbol{w}_{1:T} - \frac{\alpha}{K}\boldsymbol{N}^\top \boldsymbol{r}_{1:T})^\top \boldsymbol{Q}^{-1} (\boldsymbol{w}_{1:T} - \frac{\alpha}{K}\boldsymbol{N}^\top \boldsymbol{r}_{1:T}) \Bigg]+ \\
&\ \frac{1}{2} \textrm{tr} \Bigg[ (\boldsymbol{N} -{\bf M}_0) \boldsymbol{Q}_0^{-1} (\boldsymbol{N} -{\bf M}_0)^\top \Bigg] + \textrm{constant}
\end{split}
\end{equation} 

where $\boldsymbol{w}_t = \boldsymbol{z}_{t+1} + (\alpha-1)\boldsymbol{z}_{t}$, and $\boldsymbol{r}_t = \phi(\boldsymbol{M}\boldsymbol{z}_{t}+\boldsymbol{B}\boldsymbol{v}_t+\boldsymbol{d})$, and $\boldsymbol{w}_{1:(T-1)} \in \mathbb{R}^{R \times (T-1)}$, $\boldsymbol{r}_{1:(T-1)} \in \mathbb{R}^{K \times (T-1)}$, and each row of ${\bf M}_0 \in \mathbb{R}^{K \times R}$ is $\boldsymbol{\mu}_0$. Let $\boldsymbol{R} = \sum_{t=1}^{T-1} \boldsymbol{r}_t \boldsymbol{r}^\top_t = (\boldsymbol{r}_{1:(T-1)}) (\boldsymbol{r}_{1:(T-1)})^\top$, and $\boldsymbol{W} = \sum_{t=1}^{T-1} \boldsymbol{r}_t \boldsymbol{w}^\top_t = (\boldsymbol{r}_{1:(T-1)})(\boldsymbol{w}_{1:(T-1)})^\top$. Taking the derivative with respect to $\boldsymbol{N}$, \begin{equation} \label{eq:first_derivative}
\begin{split}    -\frac{\partial \log p(\boldsymbol{N}|\boldsymbol{M},\boldsymbol{B},\boldsymbol{d}, \hat{\boldsymbol{z}}_{1:T}, \boldsymbol{v}_{1:T})}{\partial \boldsymbol{N}}
&= -\frac{\alpha}{K} \sum_{t=1}^{T-1} \boldsymbol{r}_t (\boldsymbol{w}_t - \frac{\alpha}{K}\boldsymbol{N}^\top \boldsymbol{r}_t)^\top \boldsymbol{Q}^{-1} + (\boldsymbol{N} - {\bf M}_0) \boldsymbol{Q}_0^{-1} \\
&=-\frac{\alpha}{K}\boldsymbol{W}\boldsymbol{Q}^{-1} + \frac{\alpha^2}{K^2}\boldsymbol{R}\boldsymbol{N} \boldsymbol{Q}^{-1} + (\boldsymbol{N} - {\bf M}_0) \boldsymbol{Q}_0^{-1}.
\end{split}
\end{equation}

Setting this derivative to $\boldsymbol{0}$ and re-arranging,
\begin{equation} \label{eq:sylvester}
    \frac{\alpha^2}{K^2}\boldsymbol{R}\boldsymbol{N} \boldsymbol{Q}^{-1}  + \boldsymbol{N}\boldsymbol{Q}_0^{-1} =
    \frac{\alpha}{K}\boldsymbol{W}\boldsymbol{Q}^{-1}
    + {\bf M}_0 \boldsymbol{Q}_0^{-1}.
\end{equation} Right multiplying the equation by $\boldsymbol{Q}$ gives \begin{equation}
    \frac{\alpha^2}{K^2}\boldsymbol{R}\boldsymbol{N}  + \boldsymbol{N}(\boldsymbol{Q}_0^{-1} \boldsymbol{Q}) =
    \left(\frac{\alpha}{K}\boldsymbol{W}\boldsymbol{Q}^{-1}
    + {\bf M}_0 \boldsymbol{Q}_0^{-1}\right)\boldsymbol{Q}.
\end{equation} Note that this is a Sylvester equation with $A = \frac{\alpha^2}{K^2}\boldsymbol{R}$, $B = \boldsymbol{Q}_0^{-1} \boldsymbol{Q}$ and $C = \left(\frac{\alpha}{K}\boldsymbol{W}\boldsymbol{Q}^{-1}
    + {\bf M}_0 \boldsymbol{Q}_0^{-1}\right)\boldsymbol{Q}$. That is, \begin{equation}
    A\boldsymbol{N}+\boldsymbol{N}B = C.
\end{equation} We can re-write Equation~(\ref{eq:sylvester}) by vectorizing and using the Kronecker product notation: \begin{equation}
    \left[\frac{\alpha^2}{K^2}(\boldsymbol{Q}^{-1} \otimes \boldsymbol{R})  + (\boldsymbol{Q}_0^{-1} \otimes \boldsymbol{I}_K)\right]\textrm{vec}(\boldsymbol{N}) =
    \textrm{vec}\left(\frac{\alpha}{K}\boldsymbol{W}\boldsymbol{Q}^{-1}
    + {\bf M}_0 \boldsymbol{Q}_0^{-1}\right),
\end{equation} where $\textrm{vec}: \mathbb{R} ^{K\times R} \rightarrow \mathbb{R}^{KR}$. Therefore, \begin{equation}
    \textrm{vec}(\boldsymbol{N}) =
    \left[\frac{\alpha^2}{K^2}(\boldsymbol{Q}^{-1} \otimes \boldsymbol{R})  + (\boldsymbol{Q}_0^{-1} \otimes \boldsymbol{I}_K)\right]^{-1}\textrm{vec}\left(\frac{\alpha}{K}\boldsymbol{W}\boldsymbol{Q}^{-1}
    + {\bf M}_0 \boldsymbol{Q}_0^{-1}\right),
\end{equation} which can be un-vectorized (i.e., by taking $\textrm{vec}^{-1}: \mathbb{R} ^{KR} \rightarrow \mathbb{R}^{K \times R}$) to obtain the solution $\boldsymbol{N}$ that is the maximum of $p(\boldsymbol{N}|\boldsymbol{M},\boldsymbol{B},\boldsymbol{d}, \hat{\boldsymbol{z}}_{1:T},\boldsymbol{v}_{1:T})$. Since $p(\boldsymbol{N}|\boldsymbol{M},\boldsymbol{B},\boldsymbol{d}, \hat{\boldsymbol{z}}_{1:T},\boldsymbol{v}_{1:T})$ is a Gaussian, this $\boldsymbol{N}$ is the mean of this Gaussian, which we denote as $\mathbb{E}[\boldsymbol{N}|\boldsymbol{M}, \boldsymbol{B}, \boldsymbol{d}, \hat{\boldsymbol{z}}_{1:T}, \boldsymbol{v}_{1:T}]$. More precisely, \begin{equation}
    \mathbb{E}[\textrm{vec}(\boldsymbol{N})|\boldsymbol{M}, \boldsymbol{B}, \boldsymbol{d}, \hat{\boldsymbol{z}}_{1:T}, \boldsymbol{v}_{1:T}] =
    \left[\frac{\alpha^2}{K^2}(\boldsymbol{Q}^{-1} \otimes \boldsymbol{R})  + (\boldsymbol{Q}_0^{-1} \otimes \boldsymbol{I}_K)\right]^{-1}\textrm{vec}\left(\frac{\alpha}{K}\boldsymbol{W}\boldsymbol{Q}^{-1}
    + {\bf M}_0 \boldsymbol{Q}_0^{-1}\right).
\end{equation} Taking the derivative of Equation~(\ref{eq:first_derivative}) with respect to $\boldsymbol{N}$, we get \begin{equation} \label{eq:var}
    \mathbb{V}[\textrm{vec}(\boldsymbol{N})|\boldsymbol{M}, \boldsymbol{B}, \boldsymbol{d}, \hat{\boldsymbol{z}}_{1:T}, \boldsymbol{v}_{1:T}] = \left[ -\frac{\partial^2 \log P(\boldsymbol{N}|\boldsymbol{M}, \boldsymbol{B}, \boldsymbol{d}, \hat{\boldsymbol{z}}_{1:T}, \boldsymbol{v}_{1:T})}{\partial \boldsymbol{N}^2}\right]^{-1} = 
    \left[\frac{\alpha^2}{K^2}\boldsymbol{Q}^{-1}\otimes\boldsymbol{R} + \boldsymbol{Q}_0^{-1}\otimes\boldsymbol{I}_K \right]^{-1}.
\end{equation} In the simple case when $\boldsymbol{Q} = \lambda \boldsymbol{I}_R$, $\boldsymbol{Q}_0 = \lambda_0 \boldsymbol{I}_R$, and ${\bf M}_0 = \boldsymbol{0}$, we have \begin{equation} \label{eq:E[n|m]}
    \mathbb{E}[\boldsymbol{N}|\boldsymbol{M}, \boldsymbol{B},\boldsymbol{d}, \hat{\boldsymbol{z}}_{1:T}, \boldsymbol{v}_{1:T}] =
    \frac{\lambda_0 \alpha}{\lambda K}\left(\frac{\lambda_0\alpha^2}{\lambda K^2}\boldsymbol{R} + \boldsymbol{I}_K\right)^{-1}\boldsymbol{W} = \left(\frac{\alpha}{K}\boldsymbol{R} + \frac{\lambda K}{\lambda_0 \alpha} \boldsymbol{I}_K \right)^{-1} \boldsymbol{W},
\end{equation} 

\begin{equation}
    \mathbb{V}[\textrm{vec}(\boldsymbol{N})|\boldsymbol{M}, \boldsymbol{B},\boldsymbol{d}, \hat{\boldsymbol{z}}_{1:T}, \boldsymbol{v}_{1:T}] = \lambda_0\left(\frac{\lambda_0\alpha^2}{\lambda K^2}\boldsymbol{I}_R\otimes\boldsymbol{R} + \boldsymbol{I}_{KR} \right)^{-1}.
\end{equation} In this case, the posterior $p(\boldsymbol{N}|\boldsymbol{M}, \boldsymbol{B},\boldsymbol{d}, \hat{\boldsymbol{z}}_{1:T}, \boldsymbol{v}_{1:T})$ is proportional to a matrix Gaussian distribution of the form \begin{equation}
    \boldsymbol{N} \sim \mathcal{MN}\left(\frac{\lambda_0 \alpha}{\lambda K}\left(\frac{\lambda_0\alpha^2}{\lambda K^2}\boldsymbol{R} + \boldsymbol{I}_K\right)^{-1}\boldsymbol{W},  \left(\frac{\lambda_0\alpha^2}{\lambda K^2}\boldsymbol{R} + \boldsymbol{I}_{K} \right)^{-1}, \lambda_0\boldsymbol{I}_R \right).
\end{equation} As $K \rightarrow \infty$, the covariance becomes isotropic, scaled by $\lambda_0$.

In our experiments, we let $\lambda = 1$, and let the ridge coefficient $c = \frac{1}{\lambda_0}$. Then, \begin{equation} \label{eq:n_hat_distrib}
    \boldsymbol{N} \sim \mathcal{MN}\left(\frac{K}{\alpha}\left(\boldsymbol{R} + \frac{cK^2}{\alpha^2} \boldsymbol{I}_K\right)^{-1}\boldsymbol{W},  \left(\frac{\alpha^2}{cK^2}\boldsymbol{R} + \boldsymbol{I}_{K} \right)^{-1}, c^{-1}\boldsymbol{I}_R \right),
\end{equation} and we compute $\hat{\boldsymbol{N}} = \frac{K}{\alpha}\left(\boldsymbol{R} + \frac{cK^2}{\alpha^2} \boldsymbol{I}_K\right)^{-1}\boldsymbol{W}$. For sufficiently large $K$, notice that the $i$-th row of $\hat{\boldsymbol{N}}$ quickly becomes more dependent only on the $i$-th row of $\boldsymbol{M}$, $\boldsymbol{B}$, and $\boldsymbol{d}$, and not other rows (and of course, $\hat{\boldsymbol{N}}$ depends on $\boldsymbol{z}_{1:T}$ and $\boldsymbol{v}_{1:T}$). Thus for sufficiently large $K$ and some non-negligible $c > 0$, the $i$-th row of $\hat{\boldsymbol{N}}$ can be interpreted as an approximation of $\boldsymbol{\mu}(\boldsymbol{m}_i, \boldsymbol{b}_i, \boldsymbol{d}_i) = \mathbb{E}[\boldsymbol{n}|\boldsymbol{m}_i, \boldsymbol{b}_i, \boldsymbol{d}_i]$ in Equation~(\ref{eq:maxent}). We numerically validated that this is the case in our experiments. 

\subsection{Summary}

Steps in Sections~\ref{main_fm}--\ref{last_step} can be summarized into Algorithm~\ref{alg:connector} below. 
\begin{algorithm}[h]
  \caption{Connectivity distribution inference using Connector}
\label{alg:connector}
\begin{algorithmic}[1]
\STATE {\bfseries Input:} Observed neural activity data $\boldsymbol{r}^{data}_{1:T}$, external input data $\boldsymbol{v}_{1:T}$ (optional), hyperparameters $\alpha$, $c$, $K$, covariance $\boldsymbol{S}$
\STATE
\STATE \textbf{Step 1: Train lrRNN}

\STATE Train lrRNN to learn:
\STATE \hspace{1em} - Latent trajectories $\boldsymbol{z}_{1:T}$
\STATE \hspace{1em} - Observation loadings $\boldsymbol{M}$, $\boldsymbol{B}$, $\boldsymbol{d}$

\STATE
\STATE \textbf{Step 2: Learn $p(\boldsymbol{m}, \boldsymbol{b}, \boldsymbol{d})$}
\STATE Extract rows $\boldsymbol{m}_i$, $\boldsymbol{n}_i$, $\boldsymbol{d}_i$ from $\boldsymbol{M}$, $\boldsymbol{B}$, $\boldsymbol{d}$ as data samples
\STATE Train neural network using flow matching objective to infer $p(\boldsymbol{m}, \boldsymbol{b}, \boldsymbol{d})$ from the data samples

\STATE
\STATE \textbf{Step 3: Learn $p(\boldsymbol{n}|\boldsymbol{m}, \boldsymbol{b}, \boldsymbol{d})$}
\FOR{each epoch}
    \STATE Sample $\boldsymbol{m}_i, \boldsymbol{b}_i, \boldsymbol{d}_i \sim p(\boldsymbol{m}, \boldsymbol{b}, \boldsymbol{d})$ for $i = 1, \ldots, K$
    \STATE Construct matrices $\boldsymbol{M}$, $\boldsymbol{B}$, and vector $\boldsymbol{d}$
    \STATE Compute $\hat{\boldsymbol{N}} = \frac{K}{\alpha}\left(\boldsymbol{R} + \frac{cK^2}{\alpha^2} \boldsymbol{I}_K\right)^{-1}\boldsymbol{W}$
    \STATE Extract rows $\hat{\boldsymbol{n}}_i$ from $\hat{\boldsymbol{N}}$
    \STATE Generate training samples $\hat{\boldsymbol{n}}_i + \boldsymbol{\xi}_i$ where $\boldsymbol{\xi}_i \sim \mathcal{N}(\boldsymbol{0}, \boldsymbol{S})$
    \STATE Train neural network via conditional flow matching with inputs $\boldsymbol{m}_i, \boldsymbol{b}_i, \boldsymbol{d}_i$
\ENDFOR

\STATE

{\bf return} $p(\boldsymbol{m}, \boldsymbol{n}, \boldsymbol{b}, \boldsymbol{d}) = p(\boldsymbol{m}, \boldsymbol{b}, \boldsymbol{d}) p(\boldsymbol{n}|\boldsymbol{m}, \boldsymbol{b}, \boldsymbol{d})$
\end{algorithmic}
\end{algorithm}

\newpage
\section{Low-Rank RNNs Under Experimental Perturbations} \label{perturb}

If there are experimental perturbations (e.g., if we are optogenetically manipulating single cells), then we assume that the RNN dynamics follow \begin{equation}
\tau\frac{\boldsymbol{h}_{i, t} - \boldsymbol{h}_{i, t-1}}{\Delta t} = -\boldsymbol{h}_{i, t-1} + \frac{1}{K} \boldsymbol{m}^\top_i \left[ \sum_{j \notin \mathcal{P}}  \boldsymbol{n}_j \phi (\boldsymbol{h}_{j, t-1}) + \sum_{l \in \mathcal{P}} \boldsymbol{n}_l \phi (\boldsymbol{q}_{l, t-1}) \right] + \boldsymbol{d}_i
\end{equation} if $i \notin \mathcal{P}$, and otherwise $\boldsymbol{h}_{i,t} = \boldsymbol{q}_{i,t}$. In other words, for any neuron $l$ that is perturbed (i.e., $l$ is in set $\mathcal{P}$), its activity $\boldsymbol{h}_{l, t-1}$ is clamped to $\boldsymbol{q}_{l, t-1}$. If $\boldsymbol{h}_{j, t} = \boldsymbol{m}_j^\top \boldsymbol{z}_t + \boldsymbol{d}_j$ where $j \notin \mathcal{P}$, we can rewrite this equation into \begin{equation} \label{eq:ilrRNNz}
\tau\frac{\boldsymbol{z}_{t} - \boldsymbol{z}_{t-1}}{\Delta t} = -\boldsymbol{z}_{t-1} + \frac{1}{K} \left[ \sum_{j \notin \mathcal{P}}  \boldsymbol{n}_j \phi (\boldsymbol{m}^\top_j \boldsymbol{z}_{t-1} + \boldsymbol{d}_j) + \sum_{l \in \mathcal{P}} \boldsymbol{n}_l \phi (\boldsymbol{q}_{l, t-1}) \right].
\end{equation} This represents the collective dynamics of the unperturbed neurons. The dynamics of the perturbed neurons are given by $\boldsymbol{q}_t$. We define $\boldsymbol{q}_t = \boldsymbol{0}$ to mean silencing in our experiments. The neurons were silenced throughout the entire $t$ in our experiments. More generally, other perturbations where $\boldsymbol{q}_t$ is not $\boldsymbol{0}$ are possible. It is straightforward to extend this framework to a setup with external inputs as well.

\section{Cell-Type-Specific Dynamics of Low-Rank RNNs} \label{cell-type-specific-dynamics}

Let us suppose that the connectivity distribution $p(\boldsymbol{m}, \boldsymbol{n}, \boldsymbol{b},\boldsymbol{d})$ is composed of $P$ different cell types. That is, $p(\boldsymbol{m}, \boldsymbol{n}, \boldsymbol{b},\boldsymbol{d}) = \sum_{p=1}^P \alpha_p p_p(\boldsymbol{m}, \boldsymbol{n}, \boldsymbol{b},\boldsymbol{d})$. Then, the mean-field dynamics of this network is \begin{equation} \label{eq:ctsd1}
\tau\frac{\boldsymbol{z}_{t} - \boldsymbol{z}_{t-1}}{\Delta t} = -\boldsymbol{z}_{t-1} + \sum_{p=1}^P\alpha_p \mathbb{E}_{p_p}[\boldsymbol{n} \phi (\boldsymbol{m}^\top \boldsymbol{z}_{t-1} + \boldsymbol{b}^\top \boldsymbol{v}_{t-1} + \boldsymbol{d})],
\end{equation} due to Equation~(\ref{eq:mf-lr-RNN}). This is a more general form of Equation~(\ref{eq:mog}). As discussed in Section~\ref{luo_section}, we can generate dynamics from only a specific group of neurons, group $p$, with \begin{equation} \label{eq:ctsd2}
\tau\frac{\boldsymbol{z}_{t} - \boldsymbol{z}_{t-1}}{\Delta t} = -\boldsymbol{z}_{t-1} + \mathbb{E}_{p_p}[\boldsymbol{n} \phi (\boldsymbol{m}^\top \boldsymbol{z}_{t-1} + \boldsymbol{b}^\top \boldsymbol{v}_{t-1} + \boldsymbol{d})].
\end{equation}

\noindent Practically, in our experiments, we sample from our inferred connectivity distribution $p(\boldsymbol{m}, \boldsymbol{n}, \boldsymbol{b},\boldsymbol{d})$ multiple times ($K$ times, with some large $K$), and cluster the samples into $P$ groups. Thus, the dynamics of this network is \begin{equation}
\begin{split}
\tau\frac{\boldsymbol{z}_{t} - \boldsymbol{z}_{t-1}}{\Delta t} = -\boldsymbol{z}_{t-1} + \frac{1}{K} \Bigg( &\sum_{i=1}^{K_1} \boldsymbol{n}^{(1)}_i \phi (\boldsymbol{m}^{(1)\top}_i \boldsymbol{z}_{t-1} + \boldsymbol{b}^{(1)\top}_i \boldsymbol{v}_{t-1} + \boldsymbol{d}^{(1)}_i) +\\ &\sum_{i=1}^{K_2}\boldsymbol{n}^{(2)}_i \phi (\boldsymbol{m}^{(2)\top}_i \boldsymbol{z}_{t-1} + \boldsymbol{b}^{(2)\top}_i \boldsymbol{v}_{t-1} + \boldsymbol{d}^{(2)}_i) +\\
&\quad\quad\quad\quad\quad\quad\quad\quad\quad...\\
&\sum_{i=1}^{K_P}\boldsymbol{n}^{(P)}_i \phi (\boldsymbol{m}^{(P)\top}_i \boldsymbol{z}_{t-1} + \boldsymbol{b}^{(P)\top}_i \boldsymbol{v}_{t-1} + \boldsymbol{d}^{(P)}_i)
\Bigg),
\end{split}
\end{equation} where $\boldsymbol{m}^{(p)}_i, \boldsymbol{n}^{(p)}_i, \boldsymbol{b}^{(p)}_i, \boldsymbol{d}^{(p)}_i$ indicate the $i$-th neuron in group $p$, with $K=K_1+K_2+...+K_P$. This becomes equivalent to Equation~(\ref{eq:ctsd1}), as $K \rightarrow \infty$, with $\alpha_p = K_p/K$, and $\boldsymbol{m}^{(p)}_i, \boldsymbol{n}^{(p)}_i, \boldsymbol{b}^{(p)}_i, \boldsymbol{d}^{(p)}_i \overset{\text{iid}}{\sim} p_p(\boldsymbol{m}, \boldsymbol{n}, \boldsymbol{b}, \boldsymbol{d})$, similar to Equation~(\ref{eq:lrRNNeq}). We can generate dynamics from only group $p$ by \begin{equation} \label{eq:lrRNNeq_p}
\tau\frac{\boldsymbol{z}_{t} - \boldsymbol{z}_{t-1}}{\Delta t} = -\boldsymbol{z}_{t-1} + \frac{1}{K_p} \sum_{i=1}^{K_p} \boldsymbol{n}_i \phi (\boldsymbol{m}_i^\top \boldsymbol{z}_{t-1} + \boldsymbol{b}_i^\top \boldsymbol{v}_{t-1} + \boldsymbol{d}_i).
\end{equation} This becomes equivalent to Equation~(\ref{eq:ctsd2}) as $K_p \rightarrow \infty$.

\subsection{Normalized Difference Index for Cell Types} \label{NDI}

In Figure~\ref{figure-4}D, we quantified the relative contributions of cell types A and B to the mean-field dynamics of the network at any given point $\boldsymbol{z}$ in the latent state space using \begin{equation} \label{eq:NDI}
    \textrm{normalized difference index }(\boldsymbol{z}) = \frac{\| -\boldsymbol{z} + \mathbb{E}_{p_B}[\boldsymbol{n} \phi (\boldsymbol{m}^\top \boldsymbol{z} + \boldsymbol{d})] \| - \| -\boldsymbol{z} + \mathbb{E}_{p_A}[\boldsymbol{n} \phi (\boldsymbol{m}^\top \boldsymbol{z} + \boldsymbol{d})] \|}{\| -\boldsymbol{z} + \mathbb{E}_{p_B}[\boldsymbol{n} \phi (\boldsymbol{m}^\top \boldsymbol{z} + \boldsymbol{d})] \| + \| -\boldsymbol{z} + \mathbb{E}_{p_A}[\boldsymbol{n} \phi (\boldsymbol{m}^\top \boldsymbol{z} + \boldsymbol{d})] \|},
\end{equation} where the expectations were approximated via sampling, with $K=K_A+K_B=$ 5,000. Intuitively, this index is a measure of {\it difference in speed} between cell-type-A-specific dynamics and cell-type-B-specific dynamics, normalized so that it lies between $[-1, 1]$. This definition is similar to the definition of normalized difference index in Fig. 3C of \citet{LuoKim2025}, where they have used this index to quantify whether autonomous dynamics are more dominant compared to input-driven dynamics at a given point $\boldsymbol{z}$ in the state space. Here, the difference is calculated not between autonomous and input dynamics, but between cell type A and cell type B. 

We also considered the following alternative definitions for measuring relative contributions of cell types A and B to dynamics \begin{equation} \label{eq:NDI2}
    \textrm{full-to-type-A difference index }(\boldsymbol{z}) = \frac{\| -\boldsymbol{z} + \mathbb{E}_{p}[\boldsymbol{n} \phi (\boldsymbol{m}^\top \boldsymbol{z} + \boldsymbol{d})] \| - \| -\boldsymbol{z} + \mathbb{E}_{p_A}[\boldsymbol{n} \phi (\boldsymbol{m}^\top \boldsymbol{z} + \boldsymbol{d})] \|}{\textrm{max}(\textrm{abs}(\| -\boldsymbol{z} + \mathbb{E}_{p}[\boldsymbol{n} \phi (\boldsymbol{m}^\top \boldsymbol{z} + \boldsymbol{d})] \| - \| -\boldsymbol{z} + \mathbb{E}_{p_A}[\boldsymbol{n} \phi (\boldsymbol{m}^\top \boldsymbol{z} + \boldsymbol{d})] \|))},
\end{equation} and \begin{equation} \label{eq:NDI3}
    \textrm{full-to-type-B difference index }(\boldsymbol{z}) = \frac{\| -\boldsymbol{z} + \mathbb{E}_{p}[\boldsymbol{n} \phi (\boldsymbol{m}^\top \boldsymbol{z} + \boldsymbol{d})] \| - \| -\boldsymbol{z} + \mathbb{E}_{p_B}[\boldsymbol{n} \phi (\boldsymbol{m}^\top \boldsymbol{z} + \boldsymbol{d})] \|}{\textrm{max}(\textrm{abs}(\| -\boldsymbol{z} + \mathbb{E}_{p}[\boldsymbol{n} \phi (\boldsymbol{m}^\top \boldsymbol{z} + \boldsymbol{d})] \| - \| -\boldsymbol{z} + \mathbb{E}_{p_B}[\boldsymbol{n} \phi (\boldsymbol{m}^\top \boldsymbol{z} + \boldsymbol{d})] \|))},
\end{equation} where the expectations here were also approximated via sampling, with $K=K_A+K_B=$ 5,000. The $\textrm{max}$ operation was over $\boldsymbol{z}$, where $\boldsymbol{z}$ is inside the part of the state space traversed by single-trial latent trajectories (i.e., inside the dotted line in Figure~\ref{figure-4}D). This normalization ensured that the maximum value of this index applied to this state space is 1. Our results were robust to which normalization we use, either the one in Equation~(\ref{eq:NDI}) or the one in Equations~(\ref{eq:NDI2}--\ref{eq:NDI3}).

\newpage
\section{Experiments}

\subsection{Numerical validation of result in Section~\ref{main_inference}} \label{num_val}

\begin{figure*}[h]
\vskip 0.2in
\begin{center} 
\centerline{\includegraphics[width=6in]{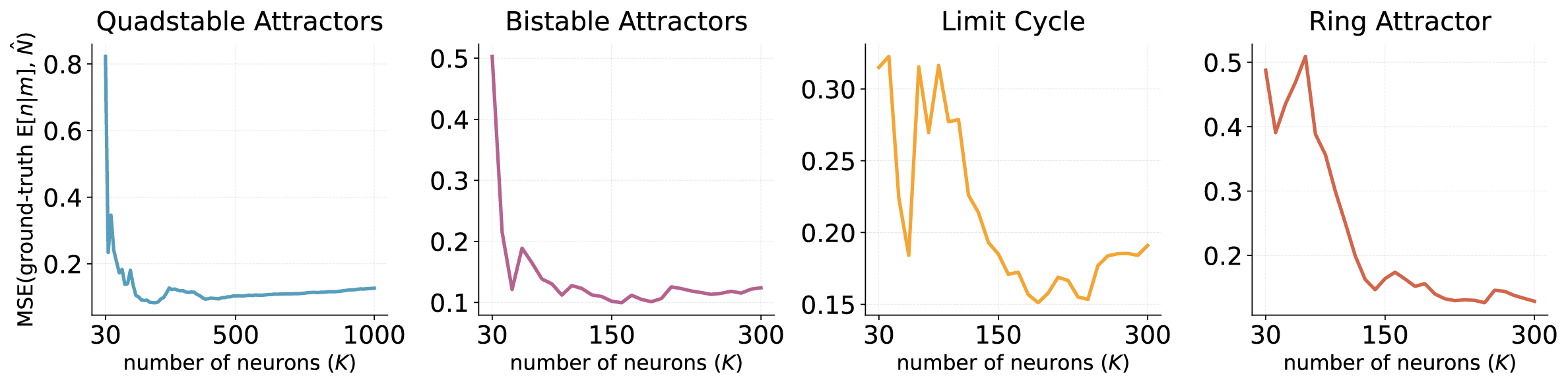}}
\caption{
MSE between the ground-truth $\mathbb{E}[\boldsymbol{n}|\boldsymbol{m}]$ and our estimate $\hat{\boldsymbol{N}}$. The regularization coefficient $c=10^{-2}$ for this experiment. For this value of $c$, our estimate does not converge to the ground-truth $\mathbb{E}[\boldsymbol{n}|\boldsymbol{m}]$ as $K$ becomes large.
}
\label{num-val-figure}
\end{center}
\vskip -0.2in
\end{figure*}

Because we know the ground-truth $\mathbb{E}[\boldsymbol{n}|\boldsymbol{m}]$ for the synthetic datasets used in Section~\ref{generalized_hopfield}, we can compare this ground truth against our estimate $\hat{\boldsymbol{N}}$ in Equation~(\ref{eq:main_nhat}). In all four synthetic datasets used in Section~\ref{generalized_hopfield}, we found that the mean-squared error (MSE) between our estimate $\hat{\boldsymbol{N}}$ and the ground-truth $\mathbb{E}[\boldsymbol{n}|\boldsymbol{m}]$ approaches 0 as $K$ becomes large, given that the regularization coefficient $c$ is sufficiently small (Figure~\ref{num-val-figure}--\ref{num-val-figure-1e-6}). This suggests that if we sample from the connectivity distribution many times (large $K$), $\hat{\boldsymbol{N}}$ should closely approximate $\mathbb{E}[\boldsymbol{n}|\boldsymbol{m}]$.

\begin{figure*}[h]
\vskip 0.2in
\begin{center} 
\centerline{\includegraphics[width=6in]{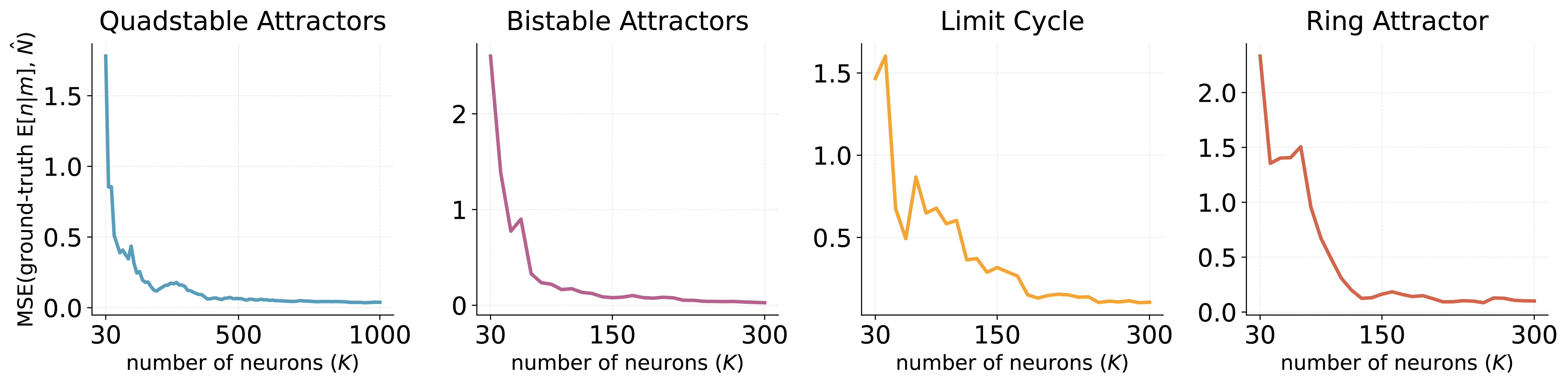}}
\caption{
MSE between the ground-truth $\mathbb{E}[\boldsymbol{n}|\boldsymbol{m}]$ and our estimate $\hat{\boldsymbol{N}}$. The regularization coefficient $c=10^{-4}$ for this experiment.
}
\end{center}
\vskip -0.2in
\end{figure*}

\begin{figure*}[h]
\vskip 0.2in
\begin{center} 
\centerline{\includegraphics[width=6in]{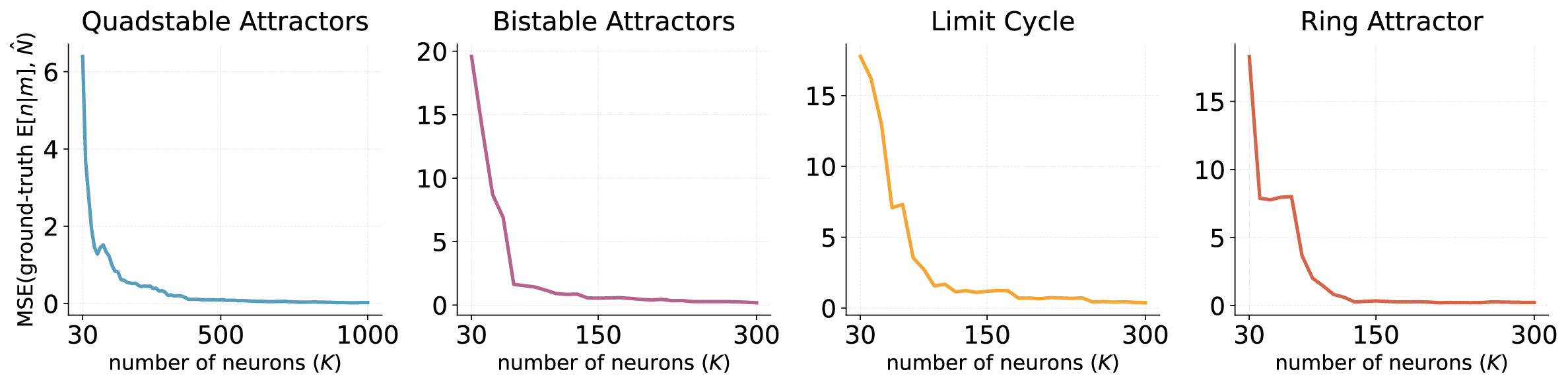}}
\caption{
MSE between the ground-truth $\mathbb{E}[\boldsymbol{n}|\boldsymbol{m}]$ and our estimate $\hat{\boldsymbol{N}}$. The regularization coefficient $c=10^{-6}$ for this experiment.
}
\label{num-val-figure-1e-6}
\end{center}
\vskip -0.2in
\end{figure*}

\subsection{Additional details for Section~\ref{generalized_hopfield}}

\subsubsection{Multiple connectivity distributions can generate quadstable attractor dynamics} \label{cnf_section}

\begin{wrapfigure}{r}{0.45\textwidth}
\begin{center} 
\centerline{\includegraphics[width=3.1in]{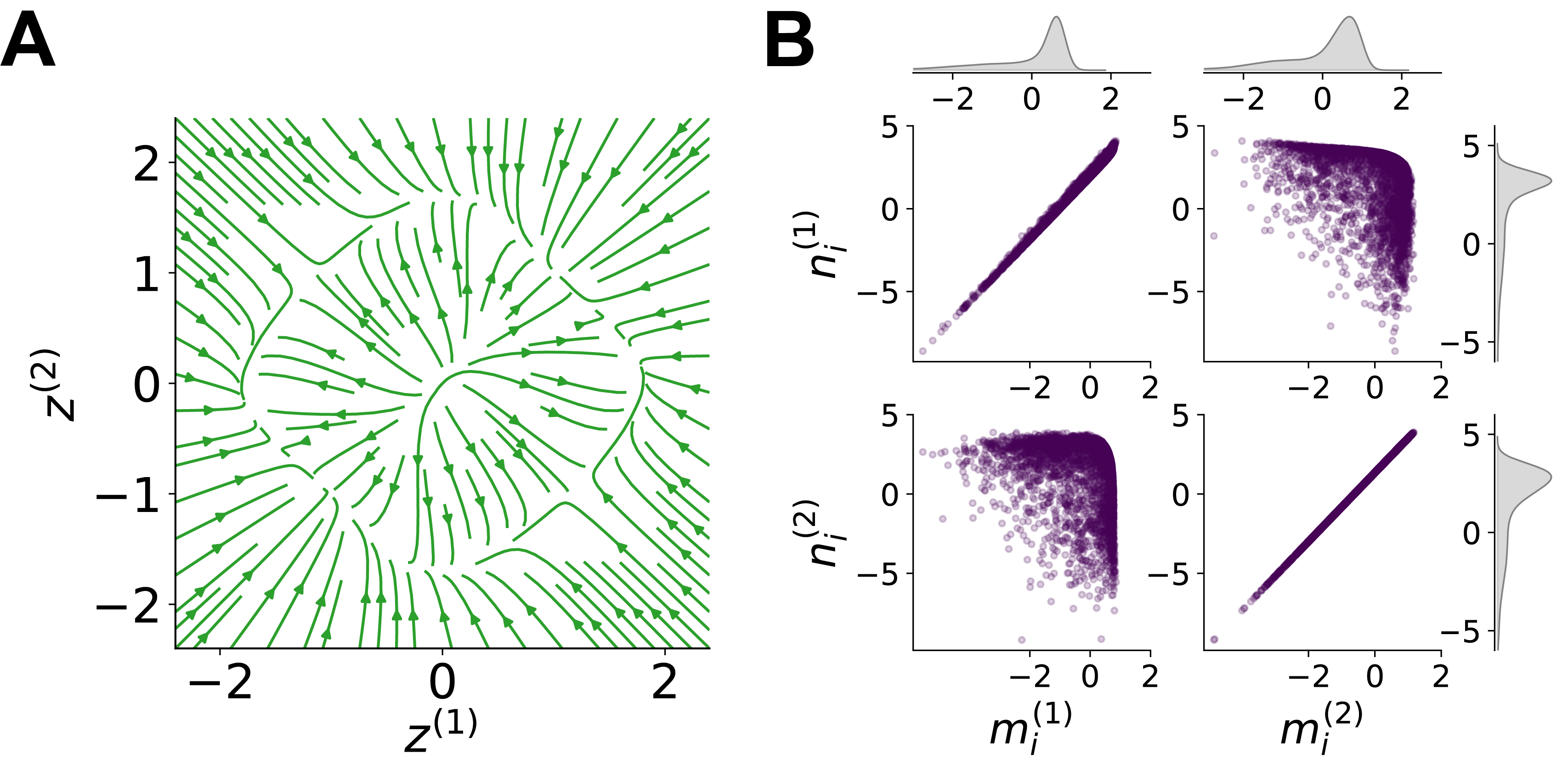}}
\caption{There are many $p(\boldsymbol{m}, \boldsymbol{n})$'s that give rise to quadstable attractors, not just the mixture of four Gaussians in Figure~\ref{figure-2}B. Notice that $p(\boldsymbol{m})$ learned by this network is unimodal. ({\bf A}) Quadstable attractor dynamics generated from $p(\boldsymbol{m}, \boldsymbol{n})$ in {\bf B}.
({\bf B}) Connectivity distribution of the quadstable attractors in {\bf A}.
}
\label{supp-figure-2}
\end{center}
\end{wrapfigure}

We found that there can be multiple non-unique $p(\boldsymbol{m}, \boldsymbol{n})$'s that can generate the low-dimensional quadstable attractor-like structure in Figure~\ref{figure-2}A. This suggests that we need to know how the latents map onto the neural population activity in order to identify the correct $p(\boldsymbol{m}, \boldsymbol{n})$. Figure~\ref{supp-figure-2} shows one example. For this example, we trained a CNF network $p_\theta(\boldsymbol{m},\boldsymbol{n})$ via backpropagation such that the following loss is minimized: \begin{equation} \label{eq:cnf_loss}
\mathcal{L}_{CNF} = \sum_{t=2}^T \left\Vert \tau \frac{\boldsymbol{z}_t - \boldsymbol{z}_{t-1}}{\Delta t} + \boldsymbol{z}_{t-1} - \frac{1}{K} \boldsymbol{N} \phi(\boldsymbol{M}\boldsymbol{z}_{t-1}) \right\Vert^2.
\end{equation} 

Here, $\{ \boldsymbol{z}_t \}_{t=1}^T$ were generated from the ground-truth quadstable attractor network and were used as data for this training procedure. The matrices $\boldsymbol{M}$ and $\boldsymbol{N}$ were generated from the CNF network $p_\theta(\boldsymbol{m}, \boldsymbol{n})$ by sampling from the distribution that this network is modeling. We sampled $K=$ 1,000 times from this distribution each epoch of the training to get $\{ \boldsymbol{m}_i, \boldsymbol{n}_i \}_{i=1}^K$, which we can stack to produce $\boldsymbol{M}$ and $\boldsymbol{N}$ (of size $K \times R$). We used ADAM \cite{kingma2017adammethodstochasticoptimization} to minimize $\mathcal{L}_{CNF}$ with respect to $\theta$, the parameters of the CNF network. Note that unlike other networks in the main text, flow matching was not used in training this network for generating Figure~\ref{supp-figure-2}. We also put a regularizing loss term in addition to $\mathcal{L}_{CNF}$ to encourage the network to learn a distribution with zero mean. Depending on this regularization, the learned $p(\boldsymbol{m}, \boldsymbol{n})$ varied, despite producing quadstable attractor-like structure similar to Figure~\ref{supp-figure-2}A.

\subsubsection{Dissimilarity between connectivities} \label{diss_supp}

To quantify dissimilarity between two connectivity distributions, we developed a measure in Equation~(\ref{eq:our_distance_measure}) based on the identifiability results in Section~\ref{main_identify}. In Equation~(\ref{eq:our_distance_measure}), $W$ in the first term denotes the 2-Wasserstein distance computed via Sinkhorn divergence with blur parameter 0.05, and $\mathbb{E}_{\boldsymbol{n} \sim p^{(1)}} [\boldsymbol{n}|\boldsymbol{x}]$ in the second term represents the conditional mean value of $\boldsymbol{n}$ given $\boldsymbol{x} \sim p^{(2)}(\boldsymbol{m}, \boldsymbol{b}, \boldsymbol{d})$. The first term captures the distance between $p^{(1)}(\boldsymbol{m}, \boldsymbol{b}, \boldsymbol{d})$ and $p^{(2)}(\boldsymbol{m}, \boldsymbol{b}, \boldsymbol{d})$, while the second term measures the mismatch in the conditional means. We do not quantify mismatch between $p^{(1)}(\boldsymbol{n}|\boldsymbol{m}, \boldsymbol{b}, \boldsymbol{d})$ and $p^{(2)}(\boldsymbol{n}|\boldsymbol{m}, \boldsymbol{b}, \boldsymbol{d})$ because of the degeneracy discussed in {\bf (3)} of Section~\ref{main_identify}, and rather quantify the mismatch in their means.

We implemented the Wasserstein distance using the GeomLoss library \cite{feydy2018interpolatingoptimaltransportmmd} with 1000 samples per distribution. These samples were also used to compute an estimate of the second term.

\subsubsection{Difference between the ground-truth and inferred $p(\boldsymbol{n})$}

To quantify histogram overlap in Figure~\ref{figure-2}G, we computed the optimal transport distance $W$, the same one used in Equation~(\ref{eq:our_distance_measure}). For details on $W$, see Appendix~\ref{diss_supp}. The distance $W$ between the ground truth and LINT (in Figure~\ref{figure-2}G) was $6.980$, and the distance $W$ between the ground truth and Connector (with $\boldsymbol{S}=\boldsymbol{I}_R$, and using equal number of neurons as the ground truth and LINT) was $0.056$.

\subsubsection{Similarity between flow fields under perturbation in Figure~\ref{supp-figure-5}} \label{flow_diff}

We quantified the similarity between perturbed flow fields by taking $30 \times 30=900$ velocity vectors from $30$-by-$30$ grid points from Figure~\ref{supp-figure-5}A and similarly taking $900$ vectors from the same grid points from Figure~\ref{supp-figure-5}B and taking the MSE between these vectors ($=0.0894$). The MSE for Figure~\ref{supp-figure-5}A and Figure~\ref{supp-figure-5}C was $0.0097$. The number of attractive slow points (speed $<$ 3e-3) for each flow field in Figure~\ref{supp-figure-5} was $2$. The basin geometries in Figure~\ref{supp-figure-5} were points (no continuous attractors), based on our numerical identification of slow points. We also computed the similarity between latent trajectories by starting the latent trajectories from one of the $30$-by-$30$ grid points, and running them for $100$ timesteps, flattening the latent components $\boldsymbol{z}_1$ and $\boldsymbol{z}_2$ into a single vector, then computing the $R^2$ (between the vector in Figure~\ref{supp-figure-5}A and Figure~\ref{supp-figure-5}B, and between Figure~\ref{supp-figure-5}A and Figure~\ref{supp-figure-5}C). We show a summary in Table~\ref{tab:comparison}. 

\begin{table}[h]
\centering
\begin{tabular}{lccc}
\hline
 & \begin{tabular}[c]{@{}c@{}}flow-field similarity to\\ ground truth (MSE)\end{tabular}
 & \begin{tabular}[c]{@{}c@{}}\# of approximate point\\ attractors\end{tabular}
 & \begin{tabular}[c]{@{}c@{}}latent trajectory similarity to\\ ground truth ($R^2$)\end{tabular} \\
\hline
LINT (Figure~\ref{supp-figure-5}B) & 0.0894 & 2 & 0.695 \\
Connector (Figure~\ref{supp-figure-5}C) & 0.0097 & 2 & 0.929 \\
\hline
\end{tabular}
\caption{Similarity between inferred flow fields and ground-truth flow field under perturbation.}
\label{tab:comparison}
\end{table}

\subsection{Additional details for Section~\ref{cddm_rnns}} \label{appendix_cddm}
 
For this Section, we used the rates of an lrRNN trained to perform a context-dependent decision-making (CDDM) task as neural activity data \cite{dubreuil22, valente22}. The rank-1 lrRNN receives a context input, and noisy sensory stimuli from two separate channels. Its objective is to make a binary choice by integrating the sensory input and correctly indicating the sign of the integrated value. The context input instructs
the lrRNN which channel to attend to.
The lrRNN was trained to encode its decision on each trial in its one-dimensional latent variable $\boldsymbol{z}$ (Figure~\ref{figure-3}A; see \citet{dubreuil22} for details of the task and training of this lrRNN). This example does not have the bias term $\boldsymbol{d}$. We found that, similar to our experiments in Section~\ref{generalized_hopfield}, degeneracy due to source {\bf (3)} in Section~\ref{main_identify} induces a mismatch between the ground-truth and inferred connectivities. Whereas LINT-inferred connectivity gives a single admissible solution that generates the latent dynamics nearly identical to the ground truth, Connector gives a set of solutions consistent with the ground truth dynamics. See Figure~\ref{figure-3} for details.

In this Section, we also experimented with whether we can apply Connector to networks other than lrRNNs (e.g., a general SSM or an LVM based on transformers, like NDT \citep{Ye_2021}). When we trained transformer models similar to NDT on neural activity from lrRNNs trained on this task, even though we could capture neural activity well ($R^2=0.99$), we could not recover the ground-truth connectivity due to the loadings in these models being less identifiable compared to lrRNNs or SSMs, validating our analyses in Appendix~\ref{identify}. See Figures~\ref{supp-figure-18}--~\ref{supp-figure-19} for details.

\subsection{Additional details for Section~\ref{luo_section}} \label{details_luo_section}

\subsubsection{Training lrRNNs via knowledge distillation from FINDR} \label{kd_lrrnn_findr}

Here we trained an lrRNN on the dataset published in \citet{LuoKim2025}. We found that training it directly on the single-trial spiking activity is challenging (potentially for reasons discussed in \citet{kim2025findr}), so we instead used a knowledge distillation approach where the lrRNN was trained to match the latent dynamics from FINDR \cite{kim2025findr} while making sure that the activity of the lrRNN units matched the task-relevant firing rates.

More specifically, we trained FINDR on a representative session from \citet{LuoKim2025}, and, using the trained gated MLP network $F$ from FINDR, generated multiple trajectories of $\boldsymbol{z}$ for one second, by evolving $\dot{\boldsymbol{z}} = F(\boldsymbol{z}, \boldsymbol{v})$ over time. (The duration of the auditory clicks stimuli on each trial was a maximum of one second \cite{LuoKim2025}.) In this work, we focused on the autonomous dynamics, so the external input $\boldsymbol{v}$ was fixed at zero throughout the evolution of the trajectory. The trajectories were generated from multiple initial conditions, which formed a 30-by-30 grid over the state space.

In FINDR, firing rate at time $t$ is given by \begin{equation} \label{eq:findr_r} \begin{split}
\boldsymbol{r}_t &= \textrm{softplus}(\boldsymbol{h}_t),\\
\boldsymbol{h}_t &= \boldsymbol{M}\boldsymbol{z}_t + \boldsymbol{\omega}_t,
\end{split}
\end{equation} where $\boldsymbol{M}\boldsymbol{z}_t$ represents the task-relevant component of the neural activity and $\boldsymbol{\omega}_t$ represents the time-varying task-irrelevant component of the neural activity \cite{kim2025findr}. Thus, we took the loading $\boldsymbol{M}$ from FINDR and used it for training the rest of the parameters, $\boldsymbol{N}$ and $\boldsymbol{d}$ of our lrRNN. We used the following loss to train our lrRNN: \begin{equation}
\mathcal{L}_{\textrm{distill}} = \sum_{t=2}^T \left\Vert \tau \frac{\boldsymbol{z}_t - \boldsymbol{z}_{t-1}}{\Delta t} + \boldsymbol{z}_{t-1} - \frac{1}{K} \boldsymbol{N} \phi(\boldsymbol{M}\boldsymbol{z}_{t-1} + \boldsymbol{d}) \right\Vert^2,
\end{equation} where, similar to the setting in Equation~(\ref{eq:cnf_loss}), $\{ \boldsymbol{z}_t \}_{t=1}^T$ generated from FINDR were used as data for this training procedure. We call this loss and $\mathcal{L}_{CNF}$ in Equation~(\ref{eq:cnf_loss}) the ``velocity matching'' objective. This objective is conceptually similar to the KL divergence term in FINDR \cite{kim2025findr}, and avoids backpropagating through time. Here, $\Delta t/\tau = 0.1$, same as what was used during the training of FINDR. Similar to Appendix~\ref{cnf_section}, ADAM \cite{kingma2017adammethodstochasticoptimization} was used to minimize $\mathcal{L}_{\textrm{distill}}$, but with respect to only $\boldsymbol{N}$ and $\boldsymbol{d}$, and not $\boldsymbol{M}$, which was fixed. If we do not train $\boldsymbol{d}$, we found that we are not able to get the lrRNN to reproduce the flow field inferred by FINDR. Without $\boldsymbol{d}$, the class of functions that can be represented by training only $\boldsymbol{N}$ is limited \cite{arora2025efficient}. 

After distillation, we applied Connector to the latents and loadings of this lrRNN. The results presented in the main text are from CNFs trained via flow matching \cite{lipman2023flow}. We found results similar to Figure~\ref{figure-4} when flow matching was not used, and CNFs were trained via methods in \citet{chen18, grathwohl2018scalable}.

\subsubsection{Connectivity inference from general SSMs} \label{id_ssm_findr}

Here, we justify why distillation was needed and why we did not directly apply Connector to the latents and loadings learned by FINDR. More broadly, this discussion highlights why caution is needed when applying Connector directly to general SSMs, not just FINDR.

As discussed in Appendix~\ref{general_identify}, in SSMs, the latent variable $\boldsymbol{z}_t$ is identifiable up to affine transformations (i.e., $\boldsymbol{z}'_t = \boldsymbol{A}\boldsymbol{z}_t + \boldsymbol{k}$), unlike lrRNNs where $\boldsymbol{z}_t$ is identifiable up to linear transformations. As a result, the following is equivalent to Equation~(\ref{eq:findr_r}): \begin{equation} \begin{split}
\boldsymbol{h}_t &=\boldsymbol{M}\boldsymbol{A}^{-1}(\boldsymbol{z}'_t - \boldsymbol{k}) + \boldsymbol{\omega}_t, \\
&=\boldsymbol{M}\boldsymbol{A}^{-1}\boldsymbol{z}'_t - \boldsymbol{M}\boldsymbol{A}^{-1}\boldsymbol{k} + \boldsymbol{\omega}_t,\\
&= \boldsymbol{M}'\boldsymbol{z}'_t + \boldsymbol{\omega}'_t,
\end{split}
\end{equation} where $\boldsymbol{M}' = \boldsymbol{M}\boldsymbol{A}^{-1}$ and $\boldsymbol{\omega}'_t = - \boldsymbol{M}\boldsymbol{A}^{-1}\boldsymbol{k} + \boldsymbol{\omega}_t$. Thus, as in lrRNNs, the loading matrix $\boldsymbol{M}$ is identifiable up to linear invertible transformations $\boldsymbol{A}$, and in our distillation procedure, we therefore fixed $\boldsymbol{M}$. However, unlike in lrRNNs, the affine offset $\boldsymbol{k}$ need not be $\boldsymbol{0}$ in general SSMs. In FINDR specifically, this offset is further complicated by the separation of task-relevant and -irrelevant components of neural activity. We could have introduced any $\boldsymbol{d}$ in \begin{equation}
\boldsymbol{h}_t = \boldsymbol{M}\boldsymbol{z}_t + \boldsymbol{d} - \boldsymbol{d} + \boldsymbol{\omega}_t,
\end{equation} and defined the task-relevant component of neural activity as $\boldsymbol{M}\boldsymbol{z}_t + \boldsymbol{d}$ and the task-irrelevant component of neural activity as $-\boldsymbol{d} + \boldsymbol{\omega}_t$. Due to this degeneracy from affine freedom, we had to learn $\boldsymbol{d}$, along with $\boldsymbol{N}$ in Appendix~\ref{kd_lrrnn_findr}, to determine what they should be in an lrRNN model, which must obey Equation~(\ref{w_update}).

More generally, Equation~(\ref{w_update}) does not necessarily hold in SSMs. If the dynamics learned by the SSM are sufficiently complex that they cannot be approximated by a finite-width lrRNN (constrained by $K_{obs}$ and also constrained by the loadings $\boldsymbol{M}$, $\boldsymbol{B}$ and $\boldsymbol{d}$), then no choice of $\boldsymbol{N}$ will satisfy Equation~(\ref{w_update}), unless the SSM has Equation~(\ref{w_update}) built into it as one of the constraints. This makes applying Connector directly to the loadings learned by SSMs difficult, motivating the distillation step used here. 

\subsubsection{Flow fields from FINDR, distilled lrRNN, and Connector-sampled lrRNN}

In Figure~\ref{supp-figure-15}A-C, the differences between the three flow fields (flow fields from FINDR, distilled lrRNN, and Connector-sampled lrRNN) were small: between Figure~\ref{supp-figure-15}A and Figure~\ref{supp-figure-15}B (MSE=$0.005$; computed in the same way as above in Appendix~\ref{flow_diff}), and Figure~\ref{supp-figure-15}A and Figure~\ref{supp-figure-15}C (=$0.072$). The MSEs here are of a similar order of magnitude to MSEs between the ground-truth flow field and flow fields learned by LINT in Figure~\ref{supp-figure-3}.

\subsubsection{Sampling new neurons from inferred connectivity distribution}

One interesting aspect of our model is that it provides a generative model of connectivity---if we randomly sample from the connectivity distribution in Figure~\ref{figure-4}, then we can sample new neurons not observed in the data. Although the activity of these sampled neurons does not correspond in a one-to-one manner with the activity of individual neurons observed in the data, we can analyze the overall population-level statistics from sampled neurons compared to observed neurons.

In Figure~\ref{figure-4}, sampling neurons is not straightforward because FINDR decomposes neural activity into task-relevant and -irrelevant components, and the flow field represents the only task-relevant component. We have trained our lrRNN to match the FINDR flow field (i.e., the task-relevant component), so even if we sample from the lrRNN connectivity from Connector and run the dynamics forward, this would only generate the task-relevant component of the new neurons not observed in the data, but not the task-irrelevant component, which we did not model in this work. Nevertheless, the distribution of the task-relevant neural activity from FINDR and the distribution of task-relevant neural activity from Connector empirically match (Figure~\ref{supp-figure-16}).

\subsection{Independent and identically distributed (iid) assumption in Equation~(\ref{eq:main_mf-lrrnn})} \label{iid_justification}
\begin{figure*}[ht]
\vskip 0.2in
\begin{center} 
\centerline{\includegraphics[width=6in]{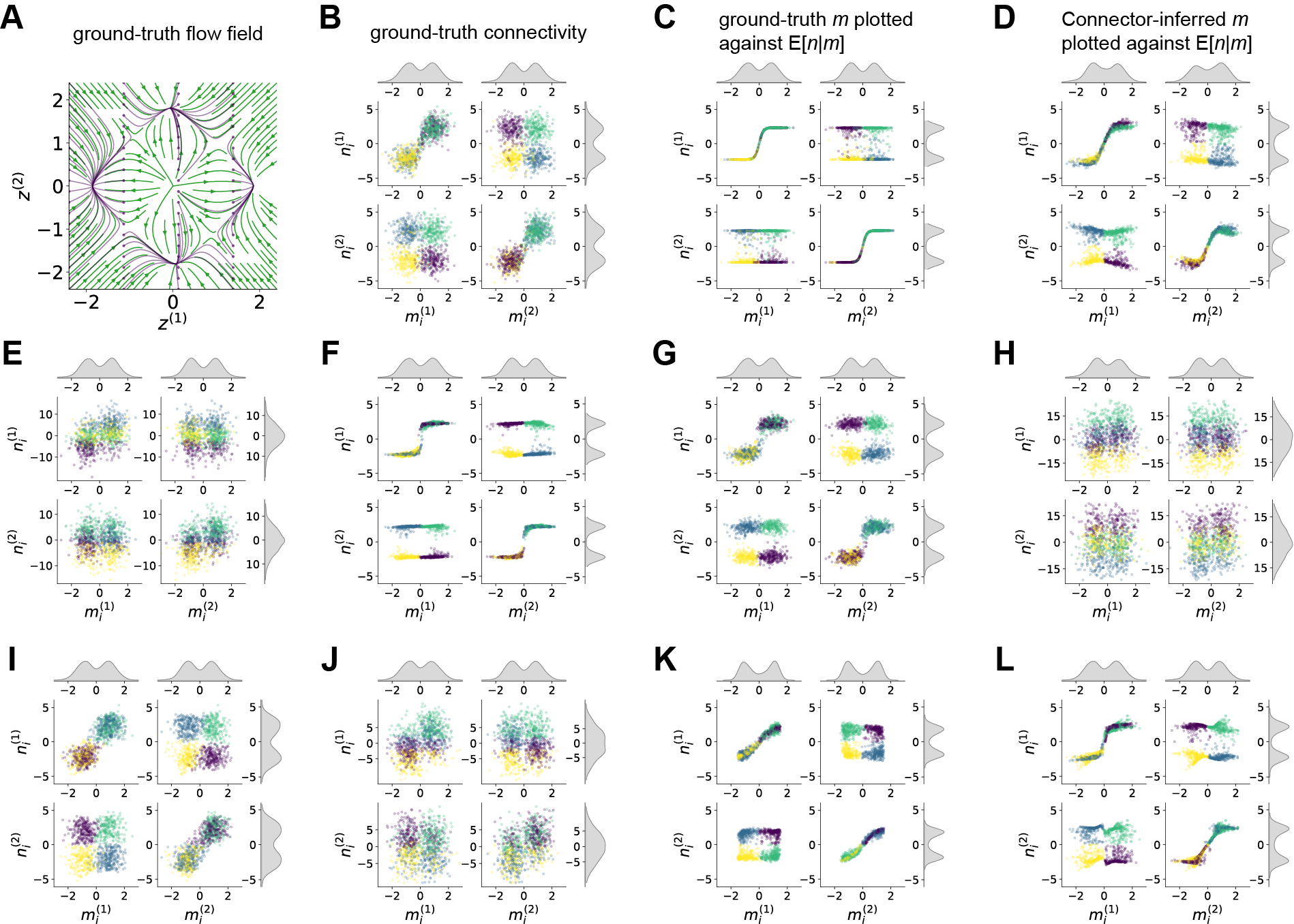}}
\caption{The learned $p(\boldsymbol{n}|\boldsymbol{m})$ by LINT depends on its hyperparameters (e.g., learning rate, regularization) and initialization, in contrast to Connector, which learns consistent $p(\boldsymbol{n}|\boldsymbol{m})$ and associated $\mathbb{E}[\boldsymbol{n}|\boldsymbol{m}]$. ({\bf A}) Quadstable attractor dynamics generated from a generalized Hopfield network \cite{beiran21}. (Same as Figure~\ref{figure-2}A.)
({\bf B}) Connectivity distribution of the generalized Hopfield network is a mixture of four Gaussians. The conditional covariance of $p(\boldsymbol{n}|\boldsymbol{m})$ is set to be $\boldsymbol{S} = \boldsymbol{I}_R$.(Same as Figure~\ref{figure-2}B.)
({\bf C}) Ground-truth $\boldsymbol{m}_i$'s plotted against $\mathbb{E}[\boldsymbol{n}|\boldsymbol{m}_i]$'s. (Same as Figure~\ref{figure-2}C.)
({\bf D}) Connector-inferred $\boldsymbol{m}_i$'s plotted against Connector-inferred $\mathbb{E}[\boldsymbol{n}|\boldsymbol{m}_i]$'s. We used the learned $\mathbb{E}[\boldsymbol{n}|\boldsymbol{m}_i]$ here, and set $\boldsymbol{S} = \boldsymbol{I}_R$ in the connectivity distribution inferred by Connector in Figure~\ref{figure-2}F.
({\bf E}) Connectivity inferred from LINT \cite{valente22}, where network parameters were initialized from $\mathcal{N}(\mu = 0, \sigma=2)$. Learning rate was $0.001$. No regularization was applied. Color code based on $4$-means clustering. (Same as Figure~\ref{figure-2}E.)
({\bf F}) Connectivity inferred from LINT \cite{valente22}, where orthogonal initialization \cite{saxe2013exact} with gain $=1$ was used. Learning rate was $0.001$. No regularization was applied.
({\bf G}) Connectivity inferred from LINT \cite{valente22}, where orthogonal initialization with gain $=10$ was used. Learning rate was $0.001$. No regularization was applied.
({\bf H}) Connectivity inferred from LINT \cite{valente22}, where network parameters were initialized from $\mathcal{U}[-5,5]$. Learning rate was $0.005$. No regularization was applied.
({\bf I}) Connectivity inferred from LINT \cite{valente22}, where network parameters were initialized from $\mathcal{U}[-1,1]$. Learning rate was $0.005$. No regularization was applied.
({\bf J}) Connectivity inferred from LINT \cite{valente22}, where network parameters were initialized from $\mathcal{U}[-3,3]$. Learning rate was $0.003$. Regularization with coefficient $=\textrm{1e-5}$ was applied.
({\bf K}) Connectivity inferred from LINT \cite{valente22}, where network parameters were initialized from $\mathcal{U}[-3,3]$. Learning rate was $0.001$. Regularization with coefficient $=\textrm{5e-4}$ was applied.
({\bf L}) Connectivity inferred from LINT \cite{valente22}, where network was initialized with Xavier Uniform \cite{glorot10} with gain $=\textrm{1e-3}$. Learning rate was $0.001$. Unlike the models above, we didn't include the $1/K$ scaling in Equation~(\ref{eq:main_lr-RNN}), and put the $1/K$ scaling factor post-training. No regularization was applied.
LINT-inferred connectivities in {\bf E}-{\bf L} all generated dynamics nearly identical to the quadstable attractor dynamics in {\bf A}.
}

\label{supp-figure-3}
\end{center}
\vskip -0.2in
\end{figure*}

In Equation~(\ref{eq:main_mf-lrrnn}), we have assumed that single-neuron parameters $(\boldsymbol{m}_i, \boldsymbol{n}_i, \boldsymbol{b}_i, \boldsymbol{d}_i)$ are iid samples from $p(\boldsymbol{m},\boldsymbol{n}, \boldsymbol{b},\boldsymbol{d})$. This is justified on two complementary grounds. First, in the MFT of disordered neural networks, the samples $(\boldsymbol{m}_i, \boldsymbol{n}_i, \boldsymbol{b}_i, \boldsymbol{d}_i)$ become asymptotically independent draws from the marginal distribution $p(\boldsymbol{m},\boldsymbol{n}, \boldsymbol{b},\boldsymbol{d})$ as $K\rightarrow \infty$, with finite-size corrections of order $O(1/K)$ \citep{amitgutfreund, MASTROGIUSEPPE18}. Second, and more fundamentally, a working hypothesis in systems neuroscience is that computation emerges from collective population dynamics rather than from the activity of individual neurons \citep{cunningham14, vyas20}. In this collective regime, the identity of any single neuron is irrelevant, and what matters is the statistical distribution of parameters across the population. Connector operates precisely at this regime: it infers the population-level distribution $(\boldsymbol{m}_i, \boldsymbol{n}_i, \boldsymbol{b}_i, \boldsymbol{d}_i)$, not individual neuron identities. Residual cross-neuron dependencies within a single trained lrRNN therefore do not affect the validity of the inferred distribution, as long as the population statistics are well-captured, which is exactly what the mean-field regime guarantees. We empirically verified that applying Connector to 8 independent lrRNNs trained on the same data (Figure~\ref{supp-figure-3}E-L) yields consistent distributions.

\subsection{Network architecture and hyperparameters} \label{details}

We parametrized the connectivity distribution using two neural networks: one for the marginal distribution $p(\boldsymbol{m}, \boldsymbol{b}, \boldsymbol{d})$, and one for the conditional distribution $p(\boldsymbol{n}|\boldsymbol{m},\boldsymbol{b},\boldsymbol{d})$. Both networks were trained as CNFs using flow matching \cite{lipman2023flow}.

Each network consisted of a 4-layer fully connected architecture with hidden dimension 128 and Swish activation functions \cite{ramachandran2017searchingactivationfunctions}. The marginal network took as input a sample from the $(\boldsymbol{m}, \boldsymbol{b}, \boldsymbol{d})$-space concatenated with a scalar time variable $t \in [0, 1]$, and output velocity vectors in the $(\boldsymbol{m}, \boldsymbol{b}, \boldsymbol{d})$-space. The conditional network took as input a sample from the $\boldsymbol{n}$-space, $(\boldsymbol{m}, \boldsymbol{b}, \boldsymbol{d})$-space, and time $t$, outputting velocity vectors in the $\boldsymbol{n}$-space.

Training proceeded in two stages (Algorithm~\ref{alg:connector}). For the experiments in Sections~\ref{generalized_hopfield}--\ref{cddm_rnns}, first, we trained the marginal distribution $p(\boldsymbol{m}, \boldsymbol{b}, \boldsymbol{d})$ for 1,000 epochs using batch size 64 on the loadings of the lrRNN. Second, we trained the conditional distribution $p(\boldsymbol{n}|\boldsymbol{m},\boldsymbol{b},\boldsymbol{d})$ for 1,000 epochs by: (1) sampling from the trained $p(\boldsymbol{m}, \boldsymbol{b}, \boldsymbol{d})$, and (2) computing the optimal $\hat{\boldsymbol{N}}$ via ridge regression (Equation~(\ref{eq:ridge})) on the latent dynamics with regularization strength $c=10^{-4}$ and using $K=$ 1,000, and (3) training the conditional CNF to match this target distribution. In Section~\ref{generalized_hopfield}, we set $\alpha = \Delta t / \tau = 0.1$, and in Section~\ref{cddm_rnns}, we set $\alpha = \Delta t / \tau = 0.2$, which were the time constants used for generating the latent dynamics in the ground-truth networks. For Section~\ref{luo_section}, the number of epochs, batch size, and number of samples ($K$) were the same as previous Sections, but $c=10^{-6}$ and $\alpha=0.1$. The activation function $\phi$ was always $\tanh$.

Both networks were optimized using AdamW \cite{loshchilov2019} with learning rate $10^{-3}$ and weight decay $10^{-5}$. We used affine probability path interpolation \cite{lipman2023flow} with conditional optimal transport scheduling for the flow matching objective. Sampling from the trained CNFs used the midpoint ODE solver with 10 integration steps and step size 0.05.

In our analyses, we split data into 5 folds. 4/5 of data were used in training/optimizing hyperparameters, and 1/5 were used in testing. All models were implemented in PyTorch and trained on NVIDIA GPUs. Fitting Connector to a trained lrRNN typically took less than 1 hour for all of our experiments on an NVIDIA TITAN X Pascal. We trained our CNF using flow matching, which is simulation-free, and readily scales to very high dimensions, as reported in \citet{lipman2023flow}.

\newpage

\newpage
\begin{figure*}[ht]
\vskip 0.2in
\begin{center} 
\centerline{\includegraphics[width=6.7in]{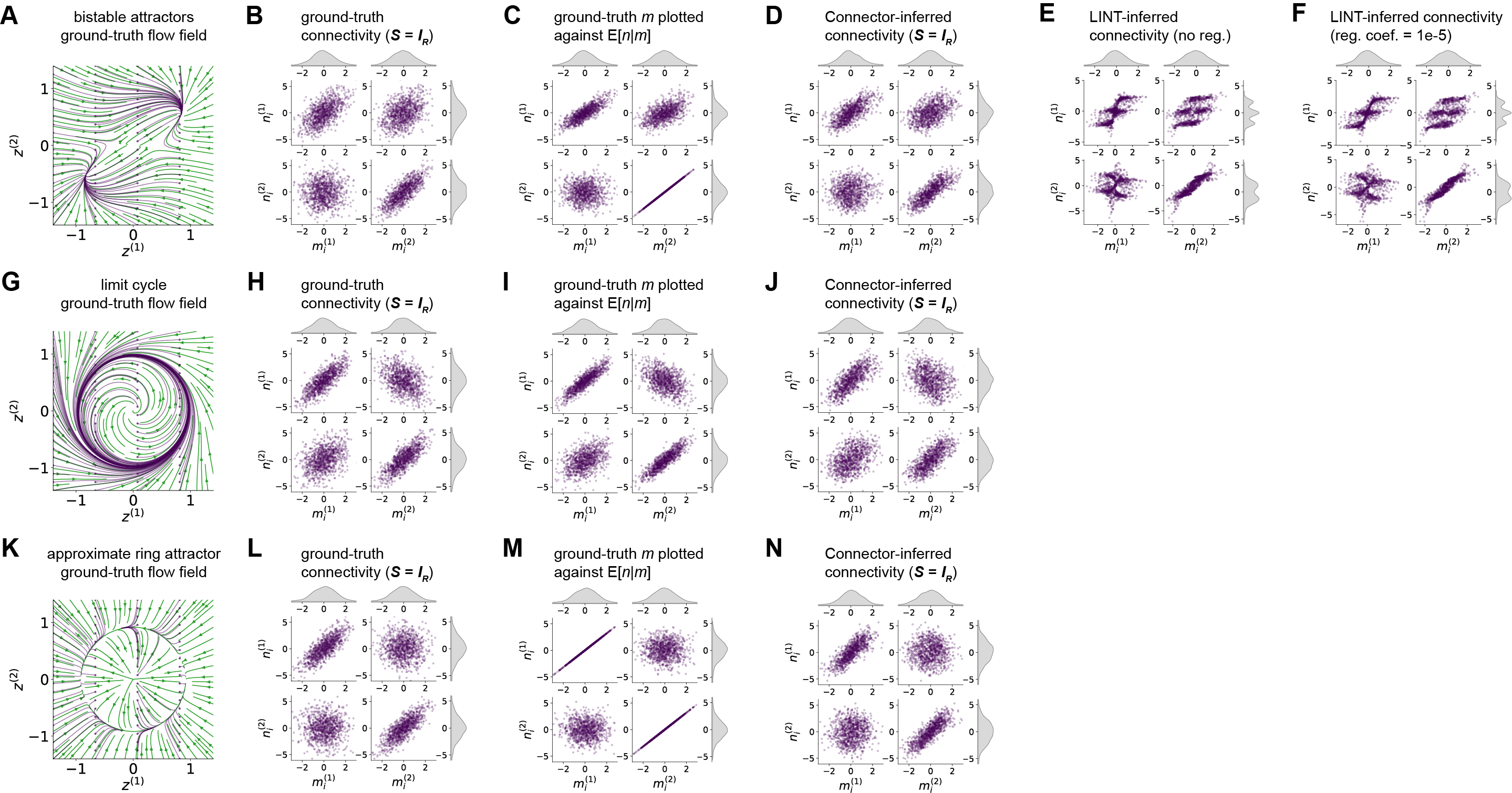}}
\caption{Connector accurately infers the connectivity distributions of bistable attractors ({\bf A}--{\bf D}), limit cycle ({\bf G}--{\bf J}), and approximate ring attractor ({\bf K}--{\bf N}). $K=$ 1,000. ({\bf A}) Ground-truth bistable attractor dynamics generated from an lrRNN with a Gaussian connectivity distribution \cite{beiran21}.
({\bf B}) Ground-truth connectivity distribution of the bistable attractors. The conditional covariance of $p(\boldsymbol{n}|\boldsymbol{m})$ is set to be $\boldsymbol{S} = \boldsymbol{I}_R$.
({\bf C}) Ground-truth bistable attractors $\boldsymbol{m}_i$'s plotted against $\mathbb{E}[\boldsymbol{n}|\boldsymbol{m}_i]$'s.
({\bf D}) Connectivity of bistable attractors inferred from Connector.
({\bf E}) Connectivity of bistable attractors inferred from LINT \cite{valente22}. No regularization was applied.
({\bf F}) Connectivity of bistable attractors inferred from LINT \cite{valente22}. Regularization with coefficient $=\textrm{1e-5}$ was applied (regularization coefficient higher than or equal to $\textrm{1e-4}$ did not converge).
({\bf G}) Ground-truth limit cycle dynamics generated from an lrRNN with a Gaussian connectivity distribution \cite{beiran21}.
({\bf H}) Ground-truth connectivity distribution of the limit cycle. The conditional covariance of $p(\boldsymbol{n}|\boldsymbol{m})$ is set to be $\boldsymbol{S} = \boldsymbol{I}_R$.
({\bf I}) Ground-truth limit cycle $\boldsymbol{m}_i$'s plotted against $\mathbb{E}[\boldsymbol{n}|\boldsymbol{m}_i]$'s.
({\bf J}), Connectivity of limit cycle inferred from Connector.
({\bf K}), Ground-truth approximate ring attractor dynamics generated from an lrRNN with a Gaussian connectivity distribution \cite{beiran21}. We have an approximate ring due to the finite size of the network ($K=$ 1,000).
({\bf L}), Ground-truth connectivity distribution of the approximate ring attractor. The conditional covariance of $p(\boldsymbol{n}|\boldsymbol{m})$ is set to be $\boldsymbol{S} = \boldsymbol{I}_R$.
({\bf M}) Ground-truth approximate ring attractor $\boldsymbol{m}_i$'s plotted against $\mathbb{E}[\boldsymbol{n}|\boldsymbol{m}_i]$'s.
({\bf N}) Connectivity of approximate ring attractor inferred from Connector.
}
\label{supp-figure-6}
\end{center}
\vskip -0.2in
\end{figure*}

\begin{figure*}[ht]
\vskip 0.2in
\begin{center} 
\centerline{\includegraphics[width=4.5in]{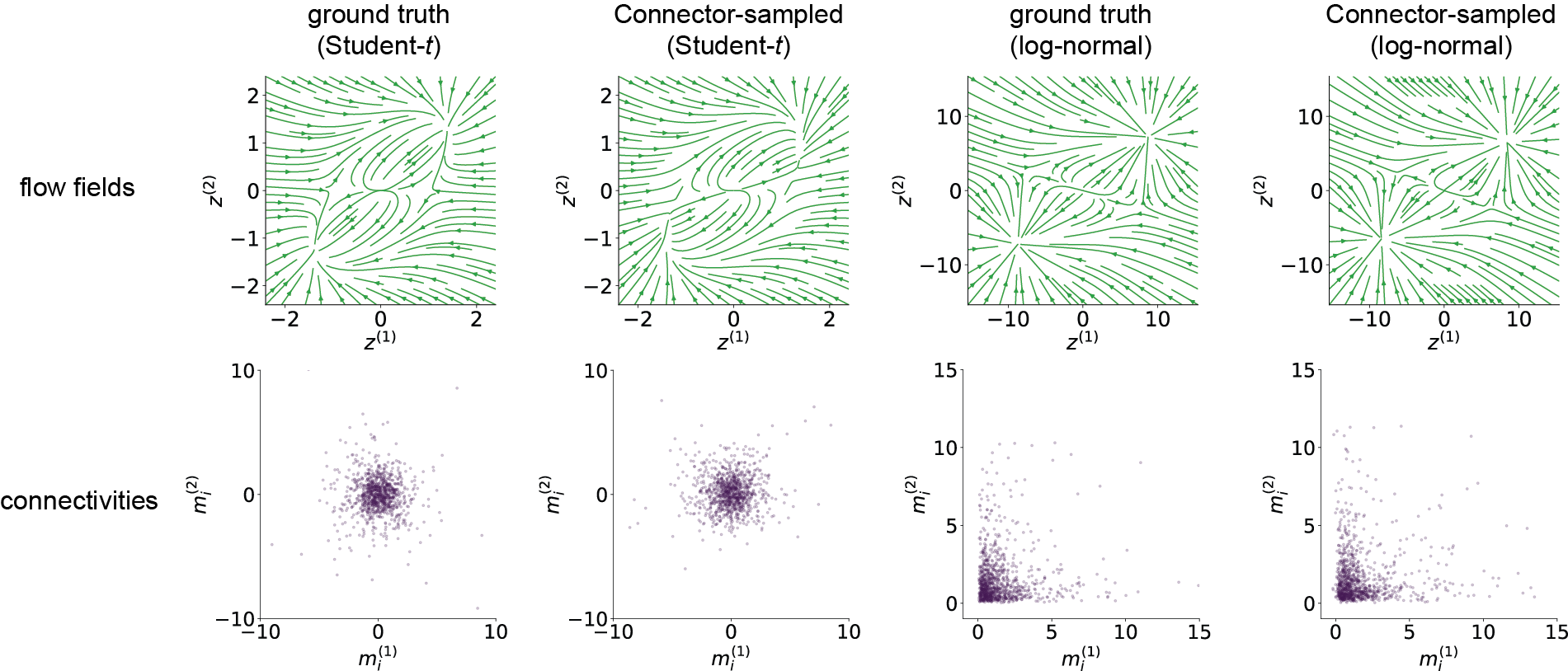}}
\caption{Connector accurately infers connectivity distributions of bistable attractors generated from a Student-$t$ distribution and a log-normal distribution. The ground-truth networks had $K_{obs}=$ 1,000. The samples from the Student-$t$ distribution were generated by using the same mean and covariance as in Figure~\ref{supp-figure-6}B, but with the number of degrees of freedom $\nu = 3$. The samples from the log-normal distribution were generated by exponentiating the samples in Figure~\ref{supp-figure-6}B. The ground-truth and Connector-sampled flow fields are similar to each other; however, Connector's $p(\boldsymbol{n}|\boldsymbol{m})$ is Gaussian, and therefore does not match the ground-truth $p(\boldsymbol{n}|\boldsymbol{m})$, which is a Student-$t$ or a log-normal.
}
\label{supp-figure-17}
\end{center}
\vskip -0.2in
\end{figure*}

\newpage
\begin{figure*}[ht]
\vskip 0.2in
\begin{center} 
\centerline{\includegraphics[width=6.5in]{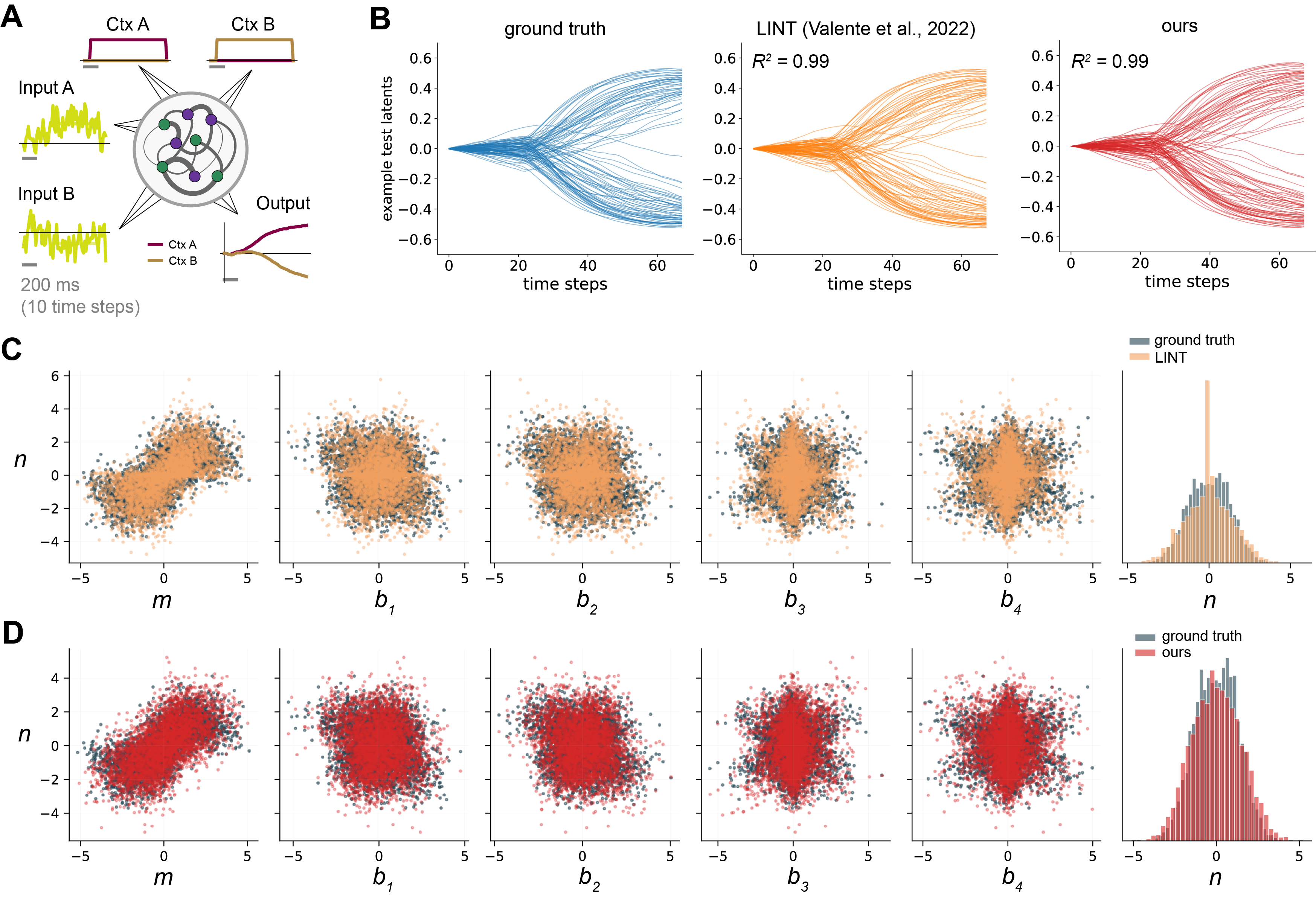}}
\caption{Identifiability in the presence of external inputs. ({\bf A}) lrRNN trained on context-dependent decision-making task in \citet{dubreuil22}. Adapted from Figure 4b of \citet{dubreuil22}.
({\bf B}) Latent trajectories of the ground-truth lrRNN (blue) and latent trajectories inferred from the neural activity of the ground-truth lrRNN (orange and red). Orange lines are from the latent trajectories inferred by LINT \cite{valente22}, and red lines are from the latent trajectories generated from an lrRNN sampled from the connectivity distribution inferred by Connector. The inferred latent trajectories matched the ground-truth trajectories well, both for LINT and Connector.
({\bf C}--{\bf D}) Despite both LINT and Connector-generated networks having similar latent trajectories, their connectivities can be different. In particular, $\boldsymbol{n}$ given $\boldsymbol{m}, \boldsymbol{b}$ can be different between the ground truth and the inferred networks due to the degeneracy {\bf (3)} discussed in Section~\ref{main_identify}. We empirically find this mismatch. In {\bf C}, notice the mismatch in the histograms approximating $p(\boldsymbol{n})$ for the ground-truth and LINT-inferred networks, with the LINT-inferred network having a sharp peak around the origin. In {\bf D}, we show the histograms approximating $p(\boldsymbol{n})$ for the ground-truth and Connector-based networks. Here for the Connector-inferred connectivity distribution, we set $\boldsymbol{S} = \boldsymbol{I}_R$. The histograms approximating $p(\boldsymbol{n})$ are similar between the ground truth and Connector, more so than between the ground truth and LINT.}
\label{figure-3}
\end{center}
\vskip -0.2in
\end{figure*}

\newpage
\begin{figure*}[ht]
\vskip 0.2in
\begin{center} 
\centerline{\includegraphics[width=6in]{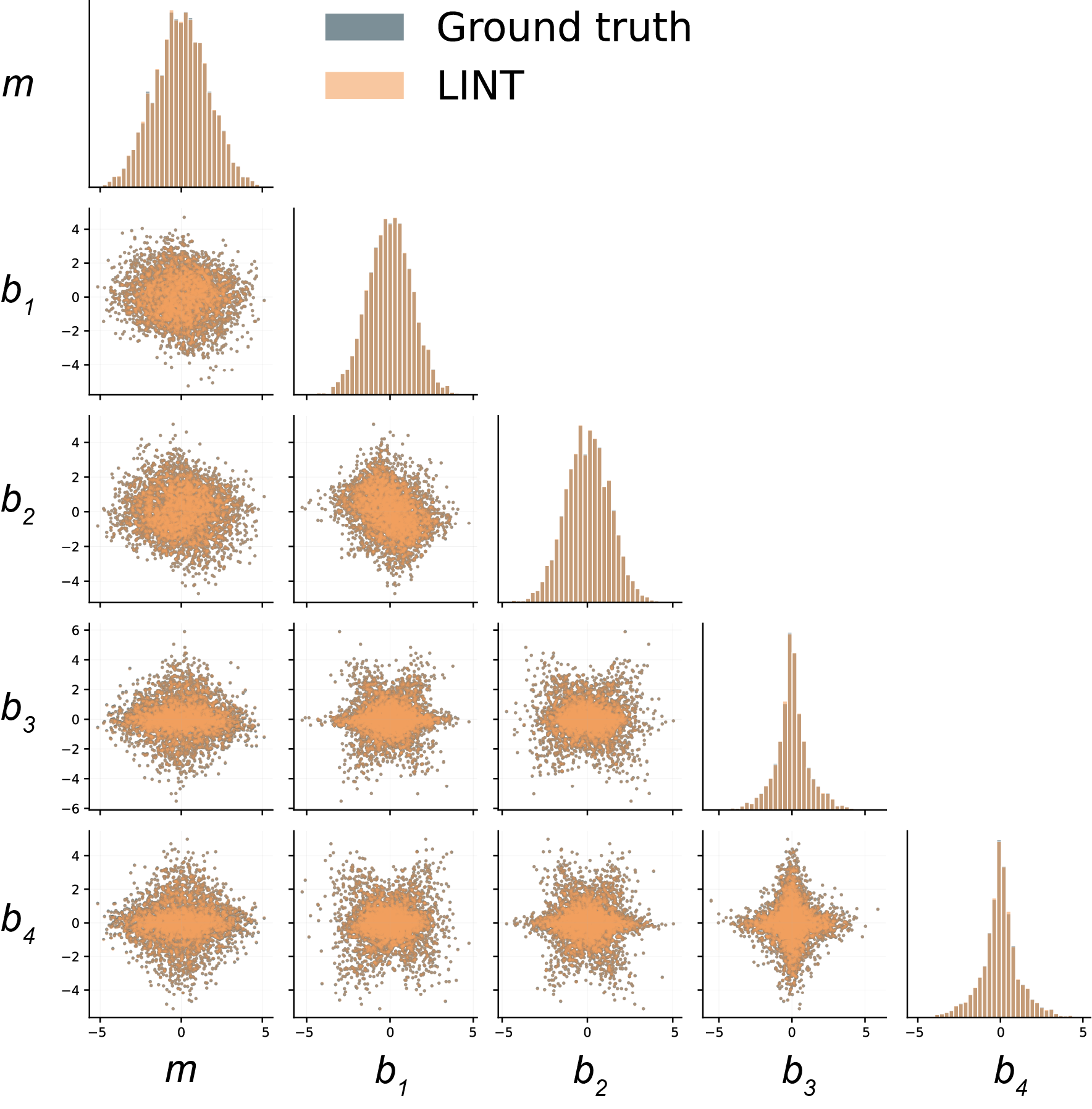}}
\caption{We trained LINT \cite{valente22} on the neural activity of the lrRNN that performs a context-dependent decision-making task (lrRNN data obtained from \citet{dubreuil22}). Consistent with our analyses in Appendix~\ref{identify}, $\boldsymbol{M}$ and $\boldsymbol{B}$ in LINT are linearly identifiable, and this leads to the ground-truth $p(\boldsymbol{m}, \boldsymbol{b})$ and our inferred $p(\boldsymbol{m}, \boldsymbol{b})$ matching each other well (after the transformation in Equation~(\ref{eq:p_transformation}) to match the inferred connectivity to the ground-truth connectivity).}
\label{supp-figure-18}
\end{center}
\vskip -0.2in
\end{figure*}

\newpage
\begin{figure*}[ht]
\vskip 0.2in
\begin{center} 
\centerline{\includegraphics[width=6in]{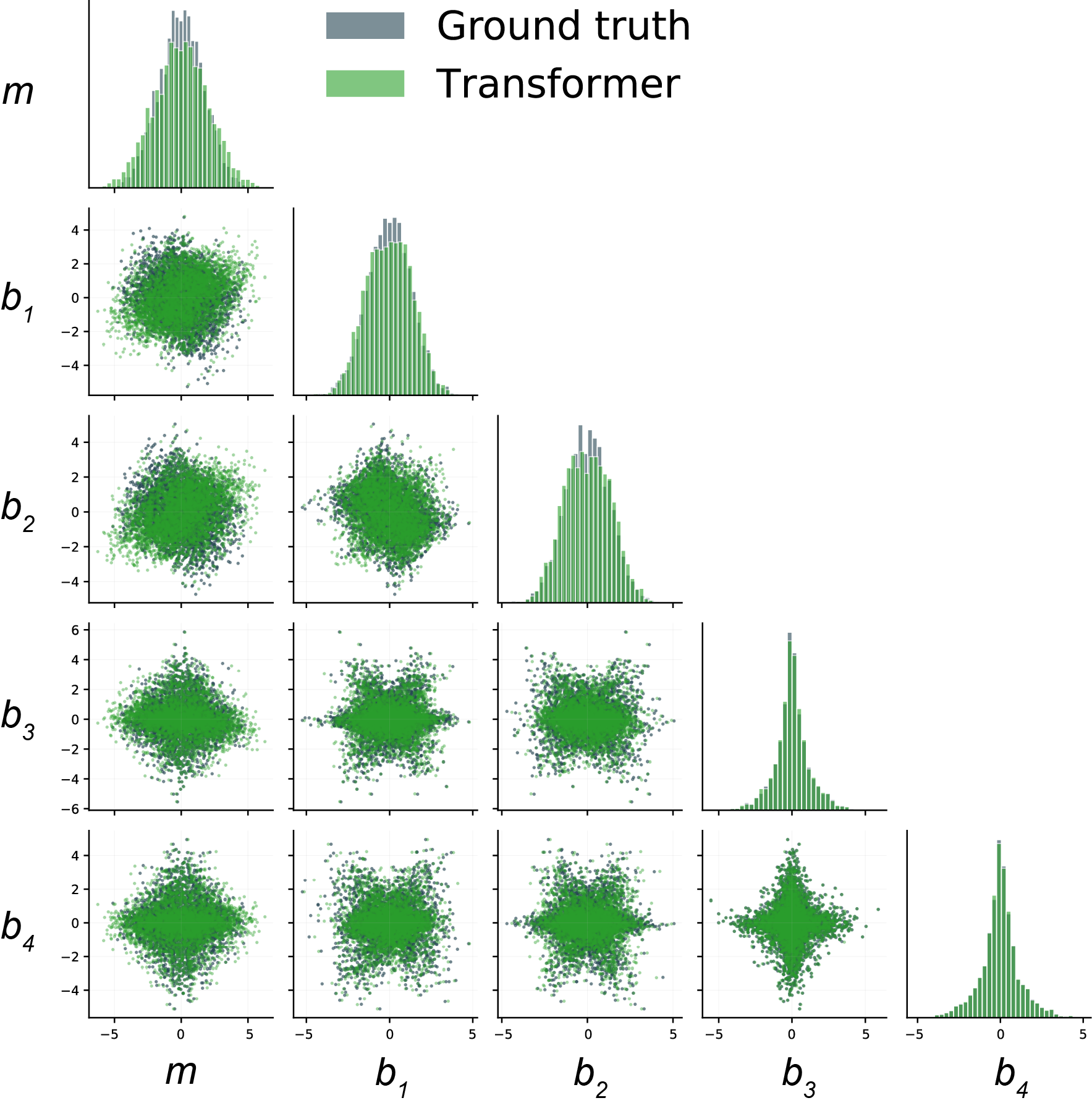}}
\caption{Same as Figure~\ref{supp-figure-18} but for a transformer-based LVM. This LVM is similar to NDT \cite{Ye_2021}, but with mapping from the latent $\boldsymbol{z}_t$ to neural activity following Equation~(\ref{eq:main_lvm}). Our analyses in Appendix~\ref{identify} suggest that $\boldsymbol{M}$ and $\boldsymbol{B}$ in LVMs are not identifiable in general. Consistent with this result, we find that the ground-truth $p(\boldsymbol{m}, \boldsymbol{b})$ and the $p(\boldsymbol{m}, \boldsymbol{b})$ inferred from the transformer-based LVM do not match well (even after linear transformation as in Figure~\ref{supp-figure-18}), despite the model producing neural activity patterns that are highly similar to the ground truth ($R^2=0.99$).}
\label{supp-figure-19}
\end{center}
\vskip -0.2in
\end{figure*}

\newpage
\begin{figure*}[t]
\vskip 0.2in
\begin{center} 
\centerline{\includegraphics[width=6.7in]{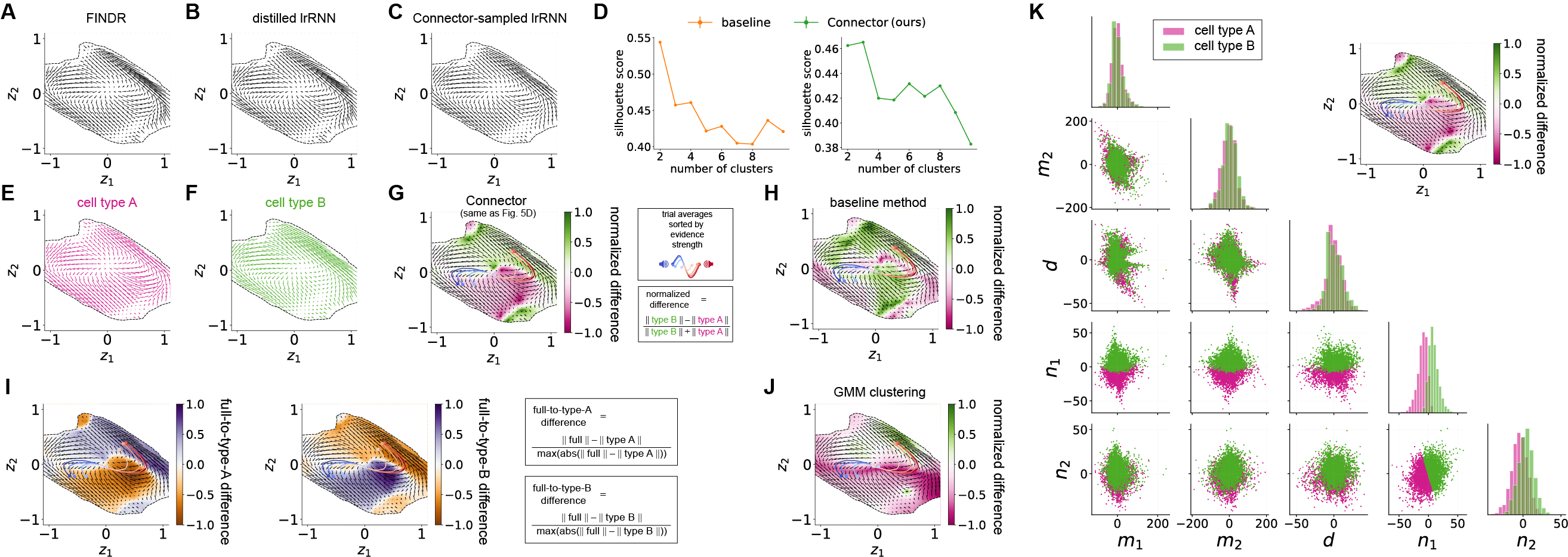}}
\caption{
Extended analyses for Figure~\ref{figure-4}. ({\bf A}) Flow field inferred by FINDR \cite{kim2025findr}.
({\bf B}) Flow field obtained from lrRNN trained via the distillation approach in Appendix~\ref{details_luo_section}.
({\bf C}) Flow field from a new lrRNN sampled from the connectivity distribution inferred by Connector ($K=$ 5,000).
({\bf D}) Silhouette scores for the connectivity of lrRNN in {\bf B} (orange; baseline), and silhouette scores for the connectivity of lrRNN in {\bf C} (green; Connector).
({\bf E}) Flow field from cell type A in the network generated in {\bf C} (Equation~(\ref{eq:ctsd2})).
({\bf F}) Flow field from cell type B in the network generated in {\bf C} (Equation~(\ref{eq:ctsd2})).
({\bf G}) Same as Figure~\ref{figure-4}D. The normalized difference index is computed with Equation~(\ref{eq:NDI}) in Section~\ref{NDI}.
({\bf H}) Instead of using {\bf C}, if we use the connectivity from {\bf B} to cluster neurons and compute the normalized difference index (Equation~(\ref{eq:NDI})), cell types A and B did not partition the state space in a way similar to Figure~\ref{figure-4}D.
({\bf I}) Instead of using the normalized difference index in Equation~(\ref{eq:NDI}), we used Equation~(\ref{eq:NDI2}) and Equation~(\ref{eq:NDI3}) for the left and right panels, respectively. We found that cell types A and B partition the state space in a way similar to Figure~\ref{figure-4}D using these indices.
({\bf J}) Instead of using $k$-means clustering as in Figure~\ref{figure-4}B, cell types A and B were clustered using GMM. The turning points of the trial-averaged latent trajectories coincide with changes in the relative contributions of the two cell types identified from GMM, similar to what we find in Figure~\ref{figure-4}D. ({\bf K}) In Figure~\ref{figure-4}B, the samples $\boldsymbol{n}_i$ were drawn from $p(\boldsymbol{n}|\boldsymbol{m},\boldsymbol{d}) = \mathcal{N}(\boldsymbol{\mu}(\boldsymbol{m},\boldsymbol{d}), \boldsymbol{S})$, assuming that $\boldsymbol{S} = \boldsymbol{0}$. Here, we let $\boldsymbol{S}$ be 1.5 times the standard deviation of $\{ \boldsymbol{\mu}(\boldsymbol{m}_i, \boldsymbol{d}_i) \}^K_{i=1}$. We then performed clustering on $\{ \boldsymbol{m}_i, \boldsymbol{n}_i, \boldsymbol{d}_i \}_{i=1}^K$ and found that cell types A and B partition the state space in a way similar to Figure~\ref{figure-4}D.
}
\label{supp-figure-15}
\end{center}
\vskip -0.2in
\end{figure*}

\newpage
\begin{figure*}[t]
\vskip 0.2in
\begin{center} 
\centerline{\includegraphics[width=6.7in]{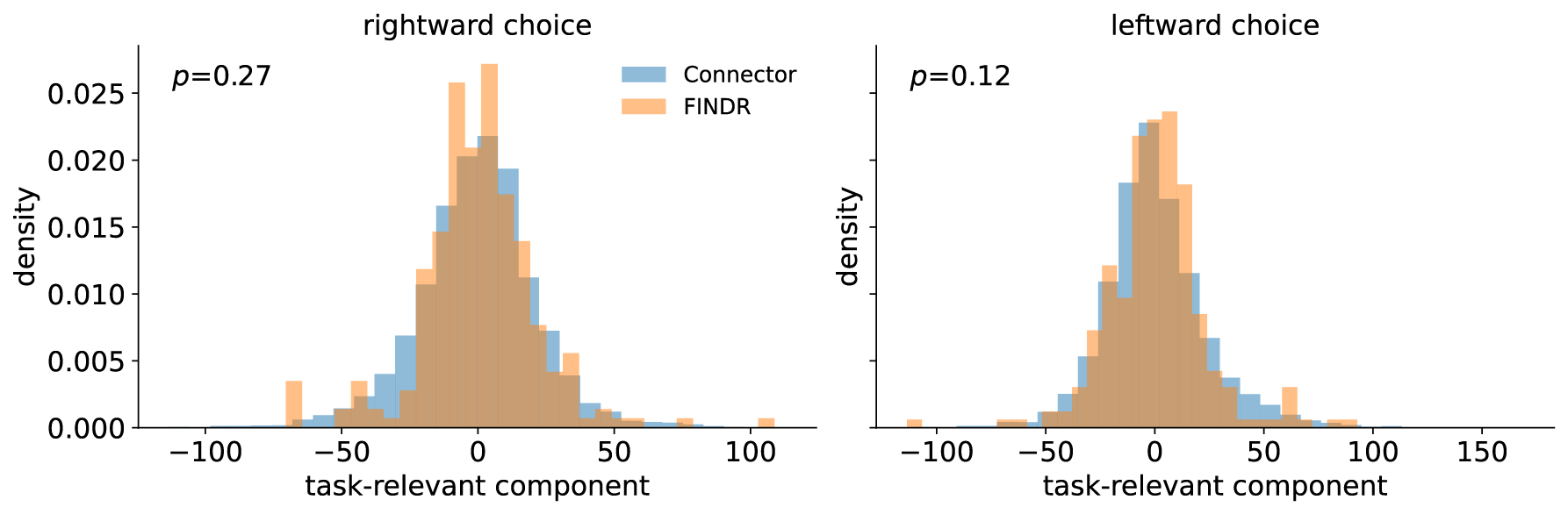}}
\caption{
Sampling new neurons from connectivity distribution inferred from data in Figure~\ref{figure-4}. We visualize the task-relevant component of neural activity $\boldsymbol{M}\boldsymbol{z}_t$ in Equation~(\ref{eq:findr_r}). FINDR was trained on $K_{obs}=240$ neurons, and the resulting task-relevant component inferred from FINDR is shown as the orange density distribution. For Connector, we sampled $K=5{,}000$ neurons, and the corresponding task-relevant component is shown as the blue density distribution. The plots show snapshots of the task-relevant component of neural activity at $t = 1\mathrm{s}$ from stimulus onset, separately for leftward and rightward choices. The $p$-values indicate two-sample two-sided Kolmogorov–Smirnov test ($p>0.1$).
}
\label{supp-figure-16}
\end{center}
\vskip -0.2in
\end{figure*}

\end{document}